\documentclass[
  journal=medium,
  manuscript=article-type,
  year=2020,
  volume=37,
]{cup-journal}

\usepackage[nopatch]{microtype}
\usepackage{booktabs}
\usepackage{amsmath,amsfonts,amssymb,amsthm,hyperref}
\usepackage{float}
\usepackage{breqn}
\usepackage[backend=biber]{biblatex}
\usepackage[english]{babel}
\usepackage{tikz-cd}
\usetikzlibrary{fit,shapes.geometric}
\tikzset{mynode/.style={
    rectangle,draw,
    inner xsep=-1pt, black,
    }}
\usetikzlibrary{calc}
\usepackage{graphicx}
\usepackage{subcaption}
\usepackage{multirow}
\usepackage{algorithm}
\usepackage{algorithmic}

\title{A Unified Bayesian Framework for Mortality Model Selection}

\author{Alex Diana}
\affiliation{School of Mathematics, Statistics, and Actuarial Science, University of Essex, Colchester, CO4 3SQ, UK}
\email[A]{ad23269@essex.ac.uk}

\author{Jackie Wong Siaw Tze}
\affiliation{School of Mathematics, Statistics, and Actuarial Science, University of Essex, Colchester, CO4 3SQ, UK}

\author{Aniketh Pittea}
\affiliation{Grant Thornton UK LLP}


\DeclareLabeldate[online]{%
  \field{date}
  \field{year}
  \field{eventdate}
  \field{origdate}
  \field{urldate}
}

\begin{document}

\begin{abstract}
 In recent years, a wide range of mortality models has been proposed to address the diverse factors influencing mortality rates, which has highlighted the need to perform model selection. Traditional mortality model selection methods, such as AIC and BIC, often require fitting multiple models independently and ranking them based on these criteria. This process can fail to account for uncertainties in model selection, which can lead to overly optimistic prediction interval, and it disregards the potential insights from combining models. To address these limitations, we propose a novel Bayesian model selection framework that integrates model selection and parameter estimation into the same process. This requires creating a model building framework that will give rise to different models by choosing different parametric forms for each term. Inference is performed using the reversible jump Markov chain Monte Carlo algorithm, which is devised to allow for transition between models of different dimensions, as is the case for the models considered here. We develop modelling frameworks for data stratified by age and period and for data stratified by age, period and product. Our results are presented in two case studies.

\end{abstract}

\section{Introduction}

Reliable mortality modelling is essential for various domains such as actuarial science, public health, and demography. Understanding and predicting mortality rates enables policymakers to make informed decisions, support financial planning in insurance and pension systems, and aid in the assessment of population health trends.  Consequently, there is a growing need for more flexible and robust methodologies to understand uncertainties around mortality modelling.

Over the past few decades, various models have been developed to enhance the accuracy of mortality forecasting, each offering unique approaches and advantages. The Age-Period-Cohort (APC) model incorporates the main effects affecting mortality, which are age, period, and cohort effect \autocite{clayton1987models}. The Lee-Carter model \autocite{lee1992modeling} also stands as a seminal method, utilising a singular time-varying factor to capture mortality trends.  Further advancements include the Cairns-Blake-Dowd (CBD) model, which simplifies the model using a linear age trend \autocite{cairns2006two}, the Renshaw-Haberman model, which adds the cohort effect to the Lee-Carter \autocite{haberman2011comparative}, and the Age-Period-Cohort-Interaction (APCI) model \autocite{richards_apci}, which extends the APC model by incorporating interactions between age and period effect for a more comprehensive analysis. These models, among others, have significantly contributed to the field by capturing various aspects of mortality, enabling actuaries to make more informed predictions and decisions.

Given the diversity of mortality forecasting models, selecting the most appropriate model can be challenging.  Mortality model selection has usually been performed using a model selection criterion such as AIC or BIC, as for example is performed in \textcite{cairns2009quantitative}. This strategy involves fitting many models and then ranking them according to one of this criteria. This process can be manual and laborious particularly if there are many models being considered. In addition, the models are fitted independently and there is no opportunity of borrowing strengths between the many models. Moreover, even if several models are equally supported, this process will ultimately involve choosing only one of them and hence disregard the information provided by the others. Furthermore, this procedure typically does not allow for joint estimation of model and parameter uncertainties, both of which are important to incorporate all sources of uncertainty and avoid producing over-optimistic prediction intervals.

A more comprehensive alternative to traditional model selection is Bayesian model selection, which offers a solution that could address some of the limitations of conventional approaches. Bayesian model selection has recently started gaining popularity in mortality modelling. \textcite{wong_2023} demonstrated  the use of marginal likelihoods, and hence Bayes factors, in comparing several stochastic mortality models for mortality forecasting. \textcite{BARIGOU2022}  illustrated the use of marginal likelihoods, stacking, and pseudo-Bayesian model averaging approaches, to compare and combine several mortality models to determine the best approach in achieving the result.  However, both these methods do not allow the competing models to interact or to borrow strengths among them. 
In this work, we propose a Bayesian probabilistic framework that views both the model choice and the parameters as part of the same parameter space.


The first step is to define a modelling framework that can accommodate a wide range of models. To achieve high flexibility, it is key to ensure that the framework has a high modelling power by encompassing a large variety of models, such as the ones already considered in the literature. Next, we perform inference using the Reversible Jump Markov chain
Monte Carlo (RJMCMC), developed by \autocite{green1995reversible}, which is an extension of the Metropolis-Hastings algorithm to propose moves between models of different dimension. Its use in the mortality modelling literature has so far been limited, which can likely be due to the challenges of slow mixing and inadequate exploration of the model space, making it difficult to efficiently perform transitions between models. This is because to achieve efficient transition between models it is necessary to propose appropriate parameters, as proposing poorly fitting parameters will result in high rejection rate. In our inference strategy, to obtain appropriate parameters, we propose to sample them from an approximation of the posterior distribution of the proposed parameters, which will require an optimisation over the new set of parameters. This optimisation can be computationally costly if the number of parameters to optimise is large. Therefore, to minimise the number of parameters required in the optimisation, it is also crucial to ensure that the framework will allow for gradual transitions between models through small, incremental changes in the model structure, which will involve a change in only a small number of parameters.  To the best of our knowledge, this is the first time that such a joint approach has been applied in the context of mortality modelling.




We apply our model selection framework to two types of data. The first involves mortality data stratified by age and period (AP data), while the second involves data stratified by age, period and another stratifying variable, which in our case is insurance product (APP data).  In the case of AP data, we define a framework aiming to  encompass a large variety of models present in the literature, such as the LC, CBD and APC, as well as new models combining different features of the models mentioned in the introduction. The framework also allows us to obtain a wide range of intermediate models bridging the transition between the models already mentioned. In the APP case, we define a model framework starting from the APCI model and extending each term of the model by allowing it to vary with product. We have chosen the APCI as our base model for two main reasons. First, the APCI is commonly used as a benchmark for fitting mortality models (e.g. \textcite{cmib_loglinear} and \textcite{cia_report2024}). Second, as we will discuss in the relevant section, the APCI allows us to define extensions that lead to fewer changes in model constraints, which will facilitate the inference.

Introducing stratification variables into mortality models, such as insurance product, has been addressed through various approaches in the literature. For example, \textcite{carter1992modeling} propose the joint-k model, which assumes that mortality rates of all groups are jointly driven by a single time-varying index. \textcite{li2005coherent} propose to model different groups by further stratifying the Lee-Carter. \textcite{villegas2014modeling} build a mortality model by modelling group-specific mortality rates with respect to a reference population modelled through a Lee-Carter with a cohort effect.

For both types of data, we apply the model to a case study. In the AP case, we consider mortality data from different countries as part of the Human Mortality Database (\textcite{hmd}). In the APP case, we consider mortality data from different insurance products provided by the Continuous Mortality Investigation (CMI).



The paper is structured as following. The general inferential strategy used for model selection, as well as notations and a short background of the statistical theory, is presented in Section \ref{sec:background}. Section \ref{sec:apmodel} will discuss the framework for AP data, as well as the inference and the applicaiton to the case study, while Section \ref{sec:appmodel} will discuss the framework for APP data and the related case study. Section \ref{sec:conclusion} summarises the paper and presents some possible future directions.


\section{Background and theory}
\label{sec:background}

In this section, we set up the notation and describe the theoretical background of our modelling and inference procedure. 

We will consider data on deaths and exposure for a specific age and period. We will denote by $d_{x,t}$ and $E_{x,t}$ the number of deaths and exposure for age $x$ and period $t$, respectively. We model deaths using the Poisson framework as proposed by \textcite{poissonlca}, $d_{x,t} \sim \text{Poisson}(E_{x,t} m_{x,t})$, where $m_{x,t}$ is the mortality rate for age $x$ and period $t$. In the case of APP data, we add the subscript $p$ to all the previous quantity to denote product, e.g. $d_{x,t,p}$ denotes the number of deaths of age $x$ and period $t$ and product $p$ for APP data. We also denote the total number of ages, years and products by $X$, $T$ and $P$, respectively.

We perform model selection in a Bayesian framework. To set up some notations, we denote the data as $d$, the parameters as $\theta$ and the model as $M$, and we denote the posterior distribution of the model $M$ and the parameters $\theta$ conditional on the data $d$ by $p(\theta, M | d)$, which is proportional to the likelihood, $p(d | \theta, M)$, and the prior, $p(\theta, M)$. Since we are performing model selection, we assume to have a set of models $(M_1,\dots,M_k)$, each with a set of parameter $\theta_k$, and we denote by $n_k$ the dimension of model $M_k$. We denote by $p(\theta_k, M_k) = p(\theta_k | M_k) p(M_k)$ the joint prior on the model and the corresponding model parameters, and by $L_k(\theta)$ the likelihood of the data $d$ given the parameters and the model $M_k$.  We note that choosing a prior of the form $p(M_k) \propto \frac{1}{n_k}$ corresponds to a posterior distribution with the same form as BIC, while choosing $p(M_k) \propto \text{exp}(-n_k)$ corresponds to AIC.

 When proposing a move between models of different dimensions, the standard Metropolis-Hastings algorithm cannot be used anymore because of the difference in dimensions between models, and instead we resort to the RJMCMC algorithm \autocite{green1995reversible}, which we briefly review.  The RJMCMC relies on defining two proposal distributions:
 
 \begin{itemize}
     \item The move from model $k$ to $k'$ happens with probability $q(k \rightarrow k')$.
     \item To define the parameter of the new proposed model, for each proposed model transition from $k$ to $k'$, we sample a latent variable $u$ of dimension $d_{k \rightarrow k'}$ from a distribution $q_{k \rightarrow k'}(u)$, such that the new set of proposed parameters $\theta_{k'}$ of model $k'$ can be obtained as $\theta_{k'} = g_{k \rightarrow k'}(\theta_{k},u)$.
 \end{itemize} 
Finally, the new model is accepted with probability
\begin{equation}
\label{eq:rjratio}
\alpha[(k, \theta_k),(k', \theta_{k'})] = \text{min}\left\{ 1, \frac{L_{k'}(\theta_{k'}) p(\theta_{k'} | k') p(k') q(k' \rightarrow k) q_{k' \rightarrow k}(u')}{L_{k}(\theta_{k}) p(\theta | k) p(k) q(k \rightarrow k') q_{k \rightarrow k'}(u)} \left| \frac{\partial g_{k \rightarrow k'}(\theta_k, u)}{\partial(u,k)} \right| \right\}    
\end{equation}
where $q(k' \rightarrow k)$ and $q_{k' \rightarrow k}(u')$ are the transition probabilities of the reverse move. 
 A requirement of RJMCMC is the \textit{dimension matching condition}: $n_k + d_{k \rightarrow k'} = n_{k'} + d_{k' \rightarrow k}$. It follows that in the case where $n_k + d_{k \rightarrow k'} = n_{k'}$, we have $d_{k' \rightarrow k} = 0$, with the last expression meaning that the reverse move to the smaller model is deterministic.  For example, when models are nested, if we can find a proposal (from the smaller to the larger model) with dimension equal to the different in the model dimensions, the proposal to the smaller model will be deterministic. 
 
 Generally, the closer the proposal distribution $q_{k \rightarrow k'}(u)$ to the posterior distribution of the variables $u$, the more efficient the sampling scheme will be. To achieve this, we choose as proposal distributions $q(u)$ a Laplace approximation of the posterior distribution. The Laplace approximation is given by a normal distribution $N(\mu^{\star}, \Sigma^{\star})$, where $\mu^{\star}$ is the set of a parameters corresponding to the maximum of the posterior distribution, and $\Sigma^{\star}$ is the inverse of the negative Hessian  of the posterior distribution computed in $\mu_{\star}$. We note that this corresponds to solving an optimisation tasks for each update.  It follows that, when possible, it is preferable to choose moves with the smallest possible dimensions changes $d_{k \rightarrow k'}$ and  $d_{k' \rightarrow k}$, since this will require performing fewer optimisation task and over smaller number of variables.



We give a simple example of how RJMCMC works in practice. Let us consider two models for age only data: model $1$, having mortality as a constant $\text{log}(m_{x}) = a$, and model $2$, having a age-varying mortality $\text{log}(m_{x}) = a_x$. We could propose the move in two different ways. In the first case, when going from model $1$ to model $2$, we can propose $a_x$ from a proposal distribution $q_{1\rightarrow 2}(a_x)$ (of dimension $X$), and for the reverse move (from model $2$ to model $1$), we propose $a$ from a proposal distribution $q_{2\rightarrow 1}(a)$ (of dimension $1$). The dimension matching condition is satisfied as $n_1 (= 1) + d_{1 \rightarrow 2} (= X) = n_2 (= X) + d_{2 \rightarrow 1} (= 1)$. Alternatively, we could design the move from model $1$ to model $2$ by only proposing the increments $\epsilon_x$ from $a$ to $a_x$, such that $a_x = a + \epsilon_x$, where $\sum_x \epsilon_x = 0$ (and therefore $\epsilon_x$ is of dimension $X - 1$).  The dimension matching condition is again satisfied since $n_1 (= 1) + d_{1 \rightarrow 2} (= X - 1) = n_2 (= X)$, and the reverse move is deterministic, where the proposed $a$ is the mean of the current $a_x$. As mentioned, the second strategy is likely to be more efficient, since it only requires finding the posterior distribution on a smaller dimensional space ($X -1$) against finding two posterior distributions (of dimension $1$ and $X$). 


\section{Mortality model selection for age-period data}
\label{sec:apmodel}

In this section, we will define a modelling framework for building mortality models for age and period data and will next use this framework to perform model selection. We define our framework by separating the log-mortality rate into four terms: age-effect, period effect, age-period effect and cohort effect, which have already been considered in a wide variety of models \autocite{cairns2009quantitative}.

We assume that the log-mortality rate can be expressed as
\begin{equation}
\label{eq:ap}
    \text{log}(m_{x,t}) = f_1(x) + f_2(t, f_3(x)) + f_4(t-x)
\end{equation}
where $f_1(x)$ is the age-effect, $f_2(t, f_3(x))$ is the age-period effect and $f_4(t-x)$ is the cohort effect. We model each term in the following way:

\begin{itemize}
\item \textit{Age effect:} we assume that age effect can be modelled linearly ($f_1(x) = a + b (x - \bar{x})$) or nonlinearly ($f_1(x) = a_x$). We introduce the variable $\delta_1$, which has value $1$ if the age effect is not linear and $2$ if the effect is linear.
\item \textit{Period effect:} we assume that the period effect can be modelled in three ways and introduce the variable $\delta_2$ to represent each outcome: independent from age $(f_2(t,f_3(x)) =  k^1_t)$ ($\delta_2 = 1$), modelled through an interaction with age  $(f_2(t,f_3(x)) = k^2_t f_3(x))$ ($\delta_2 = 2$), and a sum of the previous two $(f_2(t,f_3(x)) =  k^1_t + k^2_t f_3(x))$,  ($\delta_2 = 3$).
\item  \textit{Age-Period effect:} if we have chosen an interaction between age and period ($\delta_2 = 2$ or $3$), we introduce the variable $\delta_3$ that is $1$ if the age term in the interaction is modelled linearly with respect to age ($f_3(x) = \bar{b}(x - \bar{x})$, where $\bar{b}$ is the slope) and $2$ if modelled non-linearly ($f_3(x) = b_x$).
\item \textit{Cohort effect:} we introduce the variable $\delta_4$ that is $1$ if there is a cohort effect $\gamma_{c}$ in the model ($f_4(c) = \gamma_c$), where $c = t - x$, and $2$ otherwise ($f_4(c) = 0$). 
\end{itemize}

After reparametrisation, the log-mortality rate can be expressed as
\begin{equation}
\label{eq:ap}
    \text{log}(m_{x,t}) = f_1(x) + f_2(t, f_3(x)) + f_4(t-x)
\end{equation}
where
$$
f_1(x) = \begin{cases}
    a_x  \hspace{1.8cm} \text{if } \delta_1 = 1 \\
    a + b (x - \bar{x}) \quad \text{if } \delta_1 = 2
\end{cases}
\hspace{1cm}
f_2(t, f_3(x)) = \begin{cases}
    k^1_t  \hspace{2cm} \text{if } \delta_2 = 1 \\
    k^2_t (1 + f_3(x)) \hspace{.4cm} \text{if } \delta_2 = 2 \\
     k^1_t + k^2_t f_3(x) \hspace{.5cm} \text{if } \delta_2 = 3
\end{cases}
$$
$$
f_3(x) = \begin{cases}
    \bar{b} (x - \bar{x})  \hspace{.6cm} \text{if } \delta_3 = 1 \\
    b_x  \hspace{1.6cm} \text{if } \delta_3 = 2 
\end{cases}
\hspace{1cm}
f_4(t - x) = \begin{cases}
    0  \hspace{1.7cm} \text{if } \delta_4 = 1 \\
   \gamma_{t-x} \hspace{1.2cm} \text{if } \delta_4 = 2 
\end{cases}
$$
The complete list of models that can be obtained as a result of the different choices is summarised in Table \ref{fig:allmodels}. It can be seen that models such as the APC, LC, and CBD, are also included. However, as mentioned, we are also able to achieve intermediate models between the one already considered.

\begin{table}[h]
\centering
\hspace{-1.525cm}
\begin{tabular}{|l|l|l|l|l|l|}
\hline
Model & $\delta_1$  & $\delta_2$  & $\delta_3$  & $\delta_4$  &  Id. Constraints \\ \hline

$a_x + k^1_t$ & 1 & 1  & 1  & 1   &  $\sum k^1_t = 0$  \\ \hline

$a + b (x-\bar{x}) + k^1_t 
$ & 2 & 1  & 1  & 1     & $\sum k^1_t = 0$   \\ \hline

$a_x + k^2_t (1 + \bar{b} (x-\bar{x}))$ & 1 & 2  & 1  & 1    & $\sum k^2_t = 0$  \\ \hline

$a + b (x-\bar{x}) + k^2_t (1 + \bar{b} (x-\bar{x}))$ & 2 & 2  & 1  & 1    &  $\sum k^2_t = 0$  \\ \hline

$ a_x + k^2_t (1 + b_x)$ (M1- Lee Carter)  & 1 & 2  & 2  & 1   &  $\sum {k}^2_t = \sum {b}_x = 0$ \\ \hline

$a + b (x-\bar{x}) + k^2_t (1 + b_x)$ & 2 & 2  & 2  & 1   &  $\sum {k}^2_t$ =  $\sum b_x = 0$  \\ \hline

$ a_x + k^1_t + k^2_t (x-\bar{x})$ & 1 & 3  & 1  & 1    &   $\sum k^1_t = \sum k^2_t = 0$  \\ \hline

$ a + b (x-\bar{x})  + k^1_t + k^2_t (x-\bar{x}) $ (M5) & 2 & 3  & 1  & 1   &  $\sum k^1_t = \sum k^2_t = 0$  \\ \hline

$ a_x + k^1_t + k^2_t b_x$ & 1 & 3  & 2  & 1    &  $\sum {k}^1_t = \sum k^2_t =  \sum b_x = 0$, $b_1 = -1$  \\ \hline

$ a + b (x-\bar{x}) + k^1_t + k^2_t b_x $ & 2 & 3  & 2  & 1   & $\sum {k}^1_t = \sum {k}^2_t = \sum b_x = 0$, $b_1 = -1$    \\ \hline

$a_x + k^1_t + \gamma_{t-x}$ (M3) & 1 & 1  & 1  & 2   &  $\sum k^1_t = \sum_{c} \gamma_{c} = \gamma_{1-X} = 0$  \\ \hline

$ a + b (x-\bar{x}) + k^1_t + \gamma_{t-x}$ & 2 & 1  & 1  & 2   &  $\sum k^1_t = \sum_{c} \gamma_{c} = \gamma_{1-X} = 0$  \\ \hline

$a_x + k^2_t (1 + \bar{b} (x-\bar{x})) + \gamma_{t- x}$ & 1 & 2  & 1  & 2    &  $\sum k^2_t = \sum_{c} \gamma_{c} = \gamma_{1-X} = 0$   \\ \hline

$a + b (x-\bar{x}) + k^2_t (1 + \bar{b} (x-\bar{x})) + \gamma_{t- x}$ & 2 & 2  & 1  & 2    &  $\sum k^2_t = \sum_{c} \gamma_{c} = \gamma_{1-X} = 0$  \\ \hline

$ a_x + k^2_t (1 + b_x)  + \gamma_{t-x}$ ($\approx$ M2)  & 1 & 2  & 2  & 2  &  $\sum k^2_t = \sum b_x = \sum_{c} \gamma_{c} = \gamma_{1-X} = 0$  \\ \hline

$a + b (x-\bar{x}) + k^2_t (1 + b_x)  + \gamma_{t-x}$ & 2 & 2  & 2  & 2   &    $\sum k^2_t = \sum b_x =  \sum_{c} \gamma_{c} = \gamma_{1-X} = 0$  \\ \hline

$ a_x + k^1_t + k^2_t (x-\bar{x}) + \gamma_{t-x}$ & 1 & 3  & 1  & 2    &   $\sum k^1_t = \sum k^2_t =  \sum_{c} \gamma_{c} = \gamma_{1-X} = 0$  \\ \hline

$ a + b (x-\bar{x})  + k^1_t + k^2_t (x-\bar{x}) + \gamma_{t-x} $ (M6) & 2 & 3  & 1  & 2   &  $\sum k^1_t = \sum k^2_t =  \sum_{c} \gamma_{c} = \gamma_{1-X} =  0$  \\ \hline

$ a_x + k^1_t + k^2_t b_x + \gamma_{t-x}$ & 1 & 3  & 2  & 2   &  $\sum {k}^1_t = \sum k^2_t = \sum b_x =  \sum_{c} \gamma_{c} = \gamma_{1-X} = 0$, $b_1 = -1$  \\ \hline

$ a + b (x-\bar{x}) + k^1_t + k^2_t b_x + \gamma_{t-x} $ & 2 & 3  & 2  & 2  & $\sum {k}^1_t =  {\sum k^2_t = } \sum b_x =  \sum_{c} \gamma_{c} = \gamma_{1-X} = 0$, $b_1 = -1$    \\ \hline

\end{tabular}
\caption{Complete set of models that can be obtained in the framework for each choice of $\delta_1$, $\delta_2$, $\delta_3$ and $\delta_4$. We have also denoted in brackets the models defined in Table 1 of \autocite{cairns2009quantitative}.}
\label{fig:allmodels}
\end{table}

The set of identifiability constraints are:

\begin{itemize}
    \item $\sum_t k^1_t = 0$, since any constant added to $k^1_t$ can be absorbed into $f_1(x)$.
    \item $\sum_t k^2_t = 0$, since we can remove a constant $C$ from $k^2_t$ and absorb $C f_3(x)$ into $f_1(x)$ . 
    \item $\sum_{x} b_x = 0$. If $\delta_2 = 3$, we also need on $b_x$ the additional constraint $b_1 = -1$, since we can otherwise replace $b_x$ with $C b_x$ such and absorb the rest of the constants into the other parameters.
    \item $\sum_{c} \gamma_{c} = 0$, since any additive constant can be absorbed into $f_1(x)$. Also, following \textcite{cairns2009quantitative}, another constraint is needed since we can remove $\delta ( (t - \bar{t}) - (x - \bar{x}))$ from $\gamma_{t-x}$ and add the same contribution to $a_x$ and $k_t$. We choose to impose $\gamma_{1-X} = 0$.
\end{itemize}

We note that our constraints are slightly different than the one usually used in the literature because of the different parametrisation that we have used. 

It is worth pointing out that the constraints vary according to the modelling choice of each of the other terms. This means that the constraints for one term cannot be considered in isolation. Specifically:
\begin{itemize}
    \item If $\delta_2 = \delta_3 = 2$, $\sum_t k^2_t = 0$ does not apply since the nonlinear term $C b_x$ cannot be absorbed into the linear term $a + b (x - \bar{x})$.
    \item If $\delta_2 = 2$, $\gamma_1 = 0$ is not needed as we cannot anymore absorb the term $\delta (t - \bar
    {t})$ since we do not have a term with $k_t$ only.
\end{itemize}


\subsection{Inference}

We sample from the posterior distribution of $(\theta_1,\delta_1,\theta_2,\delta_2,\theta_3,\delta_3,\theta_4,\delta_4)$, where $\theta_i$ is the set of parameters of $f_i$, using an MCMC algorithm. The algorithm will alternate between sampling each set of parameters $\theta_i$ and the model configuration $\delta_i$ of each term in \ref{eq:ap}, conditional on the rest of the parameters and the current model configuration. 


\subsubsection{Parameters update}

For the parameter $\theta_i$, we sample from the posterior distribution of the parameters with a Metropolis-Hastings algorithm using as proposal the Laplace approximation of the posterior distribution. To find the Laplace approximation, we optimise the the log-posterior distribution $\text{log}(p(\theta_i | \theta_{-i}, \delta))$ to find the maximum $\hat{\theta_i}$ and the hessian $\hat{H}$ at the maximum. When constraints are present, we reparametrise and optimise in term of the free variables. We use an explicit expression of the solution $\hat{\theta_i}$ and of the Hessian $\hat{H}$ when analytically available, and resort to using an optimisation procedure such as the \textit{optim} function in \textsf{R} when analytical expressions are not available. To achieve better performances, we also provide \textit{optim} with the explicit expression of the gradient $\nabla_{\theta_i} \text{log}(p(\theta_i | \theta_{-i}, \delta))$. Next, we propose a new value $\theta_i^*$ from $N(\hat{\theta_i}, \hat{H})$ and accept it with probability 
$$
\min\left\{1, \frac{p(d | \theta_i^*, \theta_{-i}) N(\theta_i | \hat{\theta_i}, \hat{H})}{p(d | \theta_i, \theta_{-i})N(\theta_i^{\star} | \hat{\theta_i}, \hat{H})}\right\}.
$$

\subsubsection{Model space update}

To explore the model space, we propose to change each variable $\delta_i$ at a time. For $\delta_1$, $\delta_3$ and $\delta_4$, we simply propose the value $2$ if we are currently in the value $1$ and viceversa. For $\delta_2$, we propose the following transitions: $(\delta_2 = 1) \Rightarrow (\delta_2^{\star} = 2, \delta_3^{\star} = 1)$, $(\delta_2 = 2) \Rightarrow (\delta_2^{\star} = 1)$, $(\delta_2 = 2) \Rightarrow (\delta_2^{\star} = 3)$ and $(\delta_2 = 3) \Rightarrow (\delta_2^{\star} = 2)$. 

To set up some notation, we assume that the current model $\delta$ has parameter $(\theta_1,\theta_2)$ and the new proposed model $\delta^{\star}$ has parameter $(\theta_1,\theta_2^{\star})$ (noting that $\theta_2$ or $\theta_2^{\star}$ can also be empty). For example, if the current model is nested within the proposed model, we have $\theta_2 = \emptyset$. 


The moves we will consider will fall into the following cases:

\begin{enumerate}
    \item $\theta_2 = \emptyset$, $\theta_2^* \neq \emptyset$. This is the case where we are proposing to add new parameters to the model. In this case, we can simply choose as auxiliary variables $u$ the new set of parameters $\theta_2^{\star}$. Specifically, we generate $u = \theta_2^{\star}$ from their Laplace approximation to the posterior $p(\theta_2 | \theta_1, \delta^*, d)$, $N(\hat{\theta_2}, \hat{H})$, and accept the proposed model with the new set of parameters, $((\theta_1,\theta_2^*),\delta^*)$  with probability
    $$
\text{min}\left\{ 1, \frac{p(d | (\theta_1, \theta_2^*),\delta^*) p(\theta_2 | \delta^*) p(\delta^*) q(\delta^* \rightarrow \delta) }{p(d | \theta_1, \delta)  p(\delta) q(\delta \rightarrow \delta^*) N(\theta_2^* | \hat{\theta_2}, \hat{H})} \right\}.
$$
 \item $\theta_2^* = \emptyset$, $\theta_2 \neq \emptyset$. This is the case where we are proposing to remove parameters of the model. In this case, the acceptance ratio is the inverse of the ratio in the previous move.
 \item $\theta_2 \neq \theta_2^*$ both not empty. This is the case where we are proposing to change parameters of the model. In this case, we choose as auxiliary variables $u = \delta_2^*$ and $u' = \delta_2$. The new model is accepted with probability as in (\ref{eq:rjratio}).
\end{enumerate}

However, we will need to pay particular attention to the cases where the move will cause changes in the identifiability constraints (ICs) on $\theta_1$. In the cases where this happens, we cannot straightforwardly use the updates defined above because the dimension matching condition is not satisfied anymore. In that case, we could either reparametrise the model such that the additional ICs are simply reflected in a reduction in the number of parameters or, more simply, impose the ICs also for the model where these are not theoretically needed. We note that it is also possible in principle to propose in addition to $\theta_2^{\star}$ the set of parameters $\theta_1$ for which the ICs change, using essentially the third type of update. However, we avoid this as it would imply optimising over a large number of parameters.

We now consider each different update of $\delta_i$:

\begin{itemize}
    \item \textit{Update $\delta_1$}

To propose the move from $\delta_1$ to $\delta^*_1$, we use the third type of update by choosing as auxiliary variables $u = a_x$ (if $\delta^{*}_1 = 1$) and $u = (a,b)$ (if $\delta^{*}_1 = 2$).

We note that in the case $\delta_2 = 2$ or $\delta_2 = 3$, changing $\delta_1$ causes a change in the ICs on $k^2_t$, since $\sum_t k^2_t = 0$ is required when $\delta_1 = 1$ but it is not in the case $\delta_1 = 2$. We therefore choose to keep the identifiability constraint $\sum_t k^2_t = 0$ throughout.

We also note that the move from $a + b(x - \bar{x})$ to $a_x$ could also be seen as simply an addition of new parameters, falling into the first type of update, by parameterising, (for example) the model $\delta_1 = 1$ as $a + b (x - \bar{x}) + \epsilon_x$, with $\sum_x \epsilon_x = \sum_x \epsilon^2_x = 0$. However, solving the optimisation with respect to $\epsilon_x$ is more computationally intensive than with respect to $a_x$, and therefore we choose not to propose the move in this fashion.

    \item \textit{Update $\delta_2$}

We note that this move causes several changes in the ICs according to the model specification for the other parameters. For example, in the case $\delta_4 = 2$, when going from $\delta_2 = 1$ to $\delta^{\star}_2 = 2$ (or from $\delta_2 = 3$ to $\delta^{\star}_2 = 2$), the ICs on $\delta_4$ change since $\gamma_{1-X} = 0$ is not required when $\delta_2 = 2$. Moreover, when proposing moves between $\delta_2 = 3$ to $\delta^{\star}_2 = 2$ in the case $\delta_3 = 2$, the ICs on $b_x$ change since $b_x = - 1$ is not required when $\delta_2 = 2$. 



We now discuss each move separately.

In the case of $(\delta_2 = 1) \Rightarrow (\delta_2^{\star} = 2, \delta_3^{\star} = 1)$, we choose the first type of update by setting as auxiliary variables $u = \bar{b}$. The reverse move, $(\delta_2 = 2) \Rightarrow (\delta_2^{\star} = 1)$, follows the second type of update.



In the case of $(\delta_2 = 2) \Rightarrow (\delta_2^{\star} = 3)$ (and viceversa), we choose to keep the identifiability constraint $\gamma_{1-X} = 0$ also for $\delta_2 = 2$ even though it is not required. With this choice, the case of $\delta_3 = 1$ poses no changes in ICs since we are only adding the new parameters $k^2_t$ to the model with no change in ICs in the rest of the parameters. Therefore, we can use the first type of update with $u = k^2_t$ in the one move and the second type of update in the opposite move. In the case $\delta_3 = 2$, as mentioned above, the ICs on $\delta_3$ change, which need to be accounted for in the move. We propose a change in the model reparametrisation such that the changes in ICs is absorbed by the addition of new parameters. We notice that the model for $\delta_2 = 3$:
    $$
    f_1(x) + k_t + k_t b_x + f_4(t-x) \qquad 
    $$ 
    can be reparametrised as
    $$
    f_1(x) + k_t + k_t \bar{b} b_x + f_4(t-x)
    $$
    with the additional constraint $b_x$ = -1, as in the second model. Therefore, to propose to move to the previous model, starting from the model for  $\delta_2 = 3$,
    $$
    f_1(x) + k^1_t + k^2_t b_x + f_4(t-x)  
    $$ we can now choose the third type of update, where the proposed parameter is $\theta_2^* = (k_t^1,k_t^2)$ in the move $(\delta_2 = 2) \Rightarrow (\delta_2^{\star} = 3)$ and $\theta_2^* = (k_t,\bar{b})$ in the reverse move $(\delta_2 = 3) \Rightarrow (\delta_2^{\star} = 2)$.




    \item \textit{Update $\delta_3$}

We distinguish two cases. 

In the case $\delta_2 = 2$, we are moving between the models $f_2(t,f_3(x)) = k^2_t (1 + b_x)$ and $k^2_t (1 + \bar{b}(x- \bar{x}))$. In this case, use the third type of update, using the auxiliary variables $u = b_x$ in the move from $\delta_3 = 1$ to $\delta_3^* = 2$ and $u = \bar{b}$ in the reverse move.

In the case $\delta_2 = 3$, we are moving between the models $f_2(t,f_3(x)) = k_t^1 + k^2_t (x - \bar{x})$ to $k^1_t + k^2_t b_x)$. Since one model is nested within the other, we can use the first type of update with $u = \bar{b}_x$ when moving from $\delta_3 = 1$ to $\delta_3^* = 2$ and the second type of update in the reverse move.
    
    \item \textit{Update $\delta_4$}

The move from $\delta_4 = 1$ to $\delta_4^* = 2$ falls into the first scheme, while the move from $\delta_4 = 2$ to $\delta_4^* = 1$ falls into the second scheme. In the first case, we simply choose $u = \gamma_{c}$.
\end{itemize}

\subsection{Simulation study}

To test the accuracy of our model selection procedure, we apply our method to several models and assess the ability of the method in recovering the true model.

We consider two simulation studies. In the first, we evaluate the ability of the framework to distinguish between linear and non-linear age effect and to detect the presence of a cohort effect. In the second, we evaluate the ability to detect an age-dependent period effect and nonlinearity in this age-dependent effect. More specifically: 

\begin{itemize}
    \item We investigate the ability to identify between a linear or nonlinear effect in $f_1(x)$ ($\delta_1 = 1$ or $\delta_1 = 2$) and a cohort effect ($\delta_4 = 1$ or $\delta_4 = 2$), for simplicity by taking an independent period effect. We simulate data from the model with both effects present, $a_x + k_t + \gamma_{t-x}$ (which corresponds to the model $(\delta_1, \delta_2, \delta_3, \delta_4) = (1,1,1,2)$), and increasingly deviate this model from the base model $a + b(x - \bar{x}) + k_t$ (which corresponds to the model $(\delta_1, \delta_2, \delta_3, \delta_4) = (2,1,1,1)$). To achieve this, we simulate $a_x = a + b (x - \bar{x}) + \epsilon_x$, where $\epsilon_x \sim \text{N}(0, \sigma^2_a)$ and $\gamma_{t-x} \sim \text{N}(0, \sigma^2_g)$. We varied $\sigma_a$ in the set $\{0.025,0.05,0.075\}$ and $\sigma_g$ in the set $\{0.05,0.07,0.09\}$, for a total of $9$ combinations of simulation parameters.
    
    \item We investigate the ability of the method to pick up an independent period effect ($\delta_2 = 1$) versus an interaction ($\delta_2 = 2$), and the ability to detect nonlinearity in the age term ($\delta_3 = 1$ or $\delta_3 = 2$). We simulate data from the model $a_x + k_t (1 + \bar{b} b_x)$ (which corresponds to the model $(\delta_1, \delta_2, \delta_3, \delta_4) = (2,2,2,1)$), where $b_x \sim \text{N}(0, \sigma^2_b)$, by varying $\bar{b}\in \{0.3,0.4,0.5\}$ and $\sigma_b\in \{0.05,0.10,0.15\}$. Lower values of $\bar{b}$ will push the model towards the setting $\delta_2 = 1$; while lower values of $\sigma_b$ will push the model towards the setting $\delta_3 = 1$.
\end{itemize}

For each of the $9$ combinations of simulation parameters, we simulated $10$ datasets and apply our model selection procedure to each dataset. To simulate the datasets, we have used $X = 20$ different ages and $Y = 30$ years, and simulated the exposures from a Poisson$(1000)$.

Suppose $\delta=(\delta_1,\delta_2,\delta_3,\delta_4)$ denotes the configuration of a specific model, we define as $p^{(i)}_\delta$ the posterior probability that $p(\delta | d^{(i)})$, where $d^{(i)}$ is the $i$-th simulated dataset. We report the average posterior probability across the $10$ simulations, $\frac{1}{10} \sum_{i=1}^{10} p^{(i)}_\delta$ for each configuration $\delta\in\{(1,1,1,1),(1,1,1,2),\ldots\}$.

Results of the first case study are presented in Table \ref{tab:sim1}. The trend of the posterior model probabilities is as expected, with the posterior probabilities of the models including the term $a_x$ (instead of $a - b(x - \bar{x})$ increasing as the parameter $\sigma_a$ increases. A similar pattern applies for the posterior probabilities of the model including the cohort effect $\gamma_{t-x}$, which increase as $\sigma_g$ increases.
We also note that even if the only models considered should be one of the following: 
\begin{itemize}
    \item $\delta=(2,1,1,1)$: $a + b(x - \bar{x})$ ;
    \item $\delta=(1,1,1,1)$: $a_x$ ;
    \item $\delta=(1,1,1,2)$: $a + b(x - \bar{x}) + \gamma_{t-x}$ ;
    \item $\delta=(2,1,1,2)$: $a_x + \gamma_{t-x}$ ;                      
\end{itemize}
the model selection procedure sometimes explores also additional models. For example, the model corresponding to $\delta=(1,2,1,2)$: $a_x + k_t (1 + \bar{b} (x - \bar{x})) + \gamma_{t-x}$, which is a slight variation from the model $a_x + k_t + \gamma_{t-x}$ is often explored.

\begin{table}[]
\centering 
\hspace*{-1.1cm} 
\tiny
\begin{tabular}{|l|l|l|l|l|l|l|l|l|l|l|l|l|}
\hline
Parameter Values & (1,1,1,1) & (1,1,1,2) & (2,1,1,1) & (2,1,1,2) & (2,2,1,1) & (1,2,1,1) & (1,2,1,2) & (1,2,2,1) & (1,2,2,2) & (1,3,2,2) & (2,2,1,2) & (2,2,2,1) \\ \hline
$\sigma_a = .025$, $\sigma_g = .05$ & 0.000 & 0.000 & 0.685 & 0.000 & 0.304 & 0.000 & 0.000 & 0.000 & 0.000 & 0.000 & 0.011 & 0.000 \\ \hline
$\sigma_a = .05$, $\sigma_g = .05$ & 0.747 & 0.000 & 0.096 & 0.000 & 0.018 & 0.138 & 0.000 & 0.001 & 0.000 & 0.000 & 0.000 & 0.000 \\ \hline
$\sigma_a = .075$, $\sigma_g = .05$ & 0.833 & 0.000 & 0.000 & 0.000 & 0.000 & 0.166 & 0.000 & 0.001 & 0.000 & 0.000 & 0.000 & 0.000 \\ \hline
$\sigma_a = .025$, $\sigma_g = .07$ & 0.000 & 0.000 & 0.346 & 0.418 & 0.061 & 0.000 & 0.000 & 0.000 & 0.000 & 0.000 & 0.165 & 0.012 \\ \hline
$\sigma_a = .05$, $\sigma_g = .07$ & 0.191 & 0.154 & 0.000 & 0.005 & 0.000 & 0.204 & 0.427 & 0.003 & 0.011 & 0.002 & 0.003 & 0.002 \\ \hline
$\sigma_a = .075$, $\sigma_g = .07$ & 0.165 & 0.374 & 0.000 & 0.003 & 0.000 & 0.038 & 0.400 & 0.001 & 0.009 & 0.009 & 0.001 & 0.002 \\ \hline
$\sigma_a = .025$, $\sigma_g = .09$ & 0.000 & 0.000 & 0.002 & 0.790 & 0.002 & 0.000 & 0.000 & 0.000 & 0.000 & 0.000 & 0.204 & 0.002 \\ \hline
$\sigma_a = .05$, $\sigma_g = .09$ & 0.000 & 0.189 & 0.000 & 0.470 & 0.000 & 0.000 & 0.193 & 0.000 & 0.003 & 0.000 & 0.144 & 0.001 \\ \hline
$\sigma_a = .075$, $\sigma_g = .09$ & 0.000 & 0.439 & 0.000 & 0.003 & 0.000 & 0.000 & 0.534 & 0.000 & 0.024 & 0.000 & 0.000 & 0.001 \\ \hline
\end{tabular}
\caption{First simulation study: Posterior probabilities of each model configuration, averaged across $10$ simulations.}
\label{tab:sim1}
\end{table}

Results of the second simulation study are presented in Table \ref{tab:sim2}. As expected the posterior probability of the configurations with $\delta_2 = 2$ increase as $\bar{b}$ increases; while the posterior probability of the configuration $(2,2,2,1)$ increases when both $\bar{b}$ and $\sigma_b$ increase. This suggests that the inference framework can correctly identify the true model.

\begin{table}[]
\scriptsize
\begin{tabular}{|l|l|l|l|l|l|l|}
\hline
Parameter Values & $(2,1,1,1)$ & $(2,2,1,1)$ & $(2,2,2,1)$ & $(2,3,1,1)$  \\ \hline
 $\bar{b} = .3$,   $\sigma_{b} = .05$    &   0.874 & 0.075 & 0.051 & 0.000   \\ \hline
   $\bar{b} = .4$,   $\sigma_{b} = .05$  &  0.756 & 0.148 & 0.097 & 0.000  \\ \hline
   $\bar{b} = .5$,   $\sigma_{b} = .05$  &   0.871 & 0.074 & 0.053 & 0.002    \\ \hline
  $\bar{b} = .3$,   $\sigma_{b} = .1$    &   0.797 & 0.117 & 0.086 & 0.000    \\ \hline
  $\bar{b} = .4$,   $\sigma_{b} = .1$    &   0.620 & 0.198 & 0.182 & 0.000    \\ \hline
  $\bar{b} = .5$,   $\sigma_{b} = .1$    &   0.254 & 0.179 & 0.567 & 0.000 \\ \hline
  $\bar{b} = .3$,   $\sigma_{b} = .15$   &   0.654 & 0.179 & 0.167 & 0.000  \\ \hline
  $\bar{b} = .4$,   $\sigma_{b} = .15$   &   0.191 & 0.041 & 0.768 & 0.000   \\ \hline
  $\bar{b} = .5$,   $\sigma_{b} = .15$   &   0.000 & 0.000 & 1.000 & 0.000  \\ \hline
\end{tabular}
\caption{Second simulation study: Posterior probabilities of each model average across the $10$ simulations}
\label{tab:sim2}
\end{table}

\subsection{Case study: Human Mortality Database}

We apply the models defined in the previous section to mortality data of $35$ countries as part of the Human Mortality Database (\textcite{hmd}). For each country, we selected a subset of data comprising individuals aged between $60$ and $90$, and covering the period from $1990$ to $2022$.

\begin{figure}[h!]
    \centering
    \begin{tabular}{ccc}
        \begin{subfigure}[b]{0.3\textwidth}
            \centering
            \includegraphics[width=\textwidth]{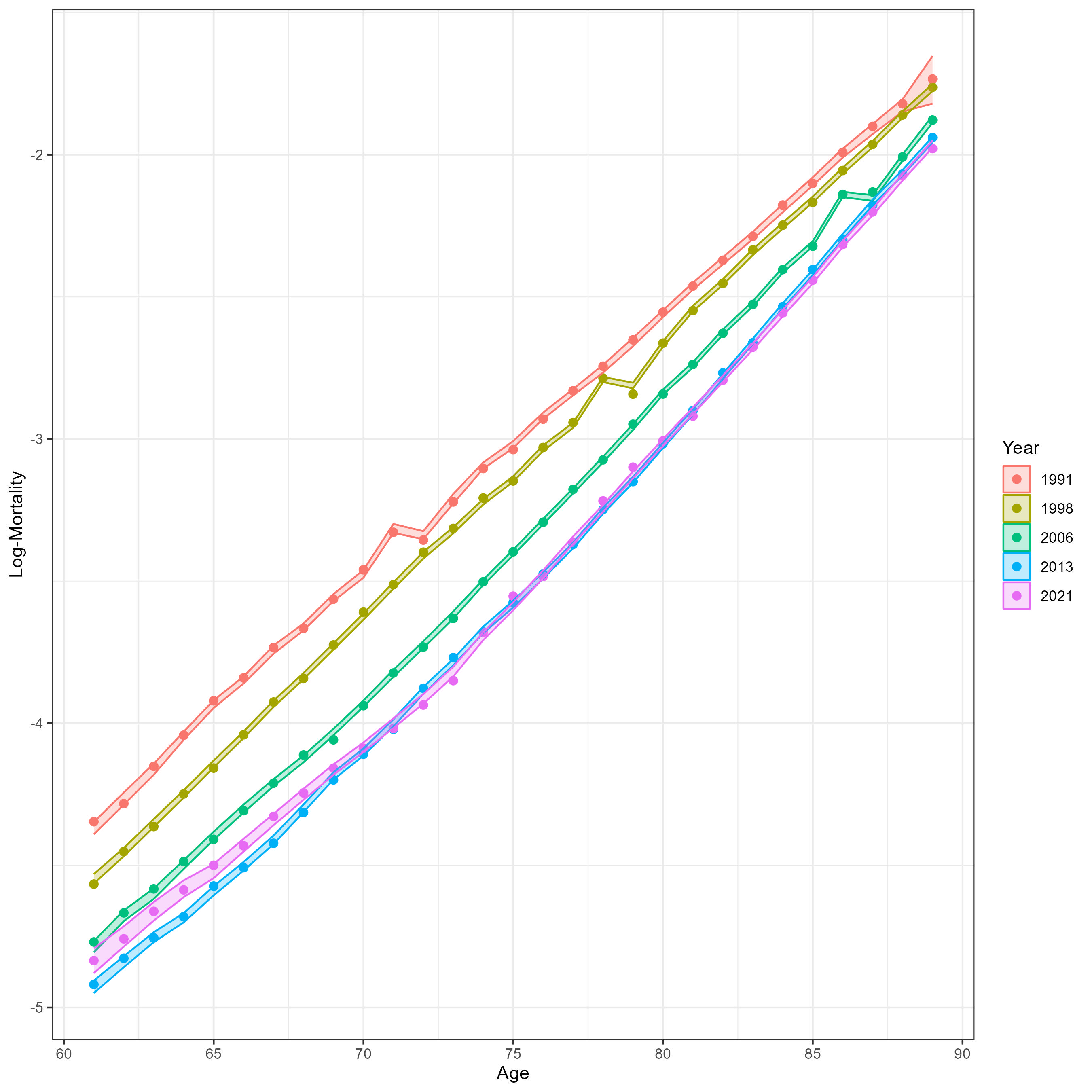} 
            \caption{UK}
        \end{subfigure} &
        \begin{subfigure}[b]{0.3\textwidth}
            \centering
            \includegraphics[width=\textwidth]{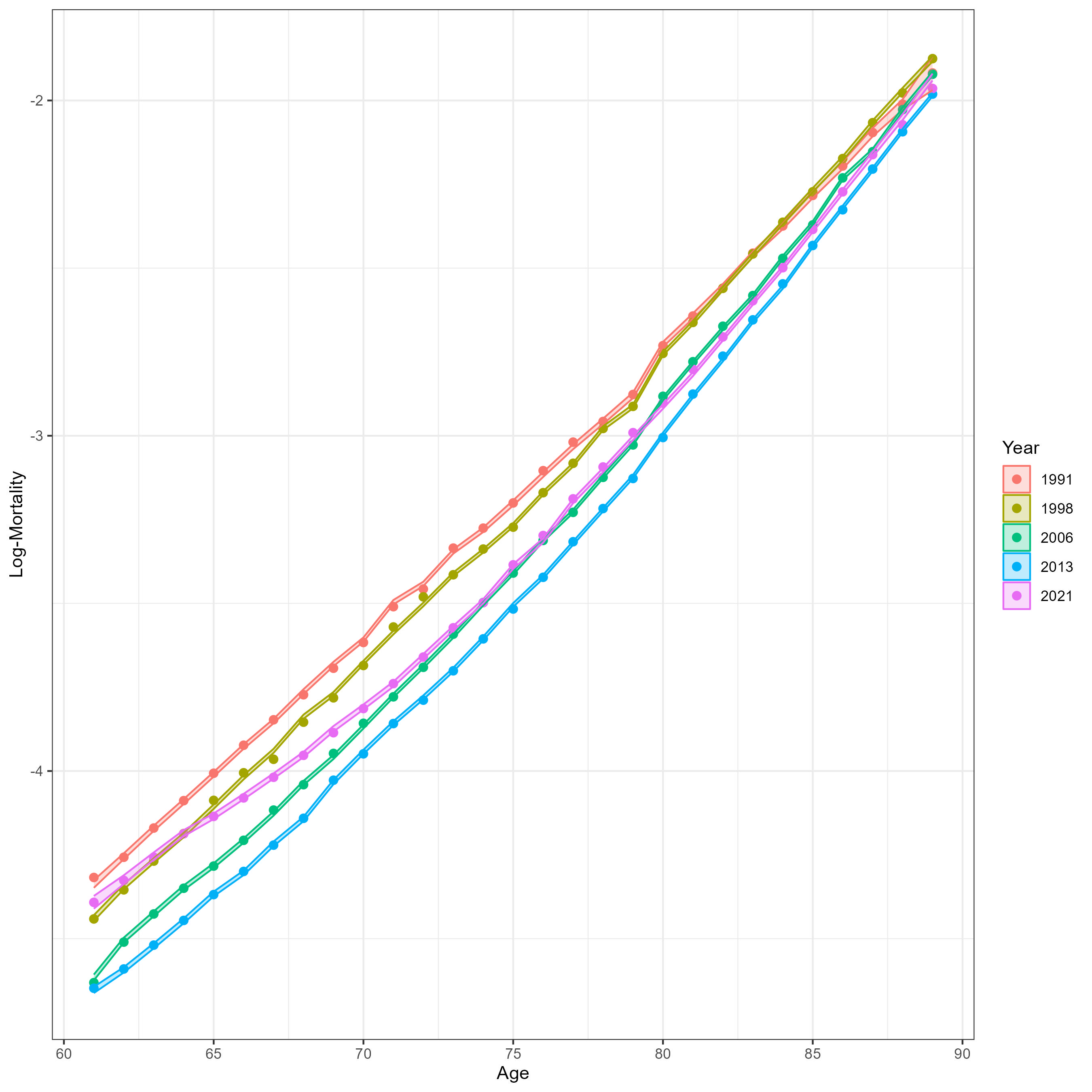} 
            \caption{USA}
        \end{subfigure} &
        \begin{subfigure}[b]{0.3\textwidth}
            \centering
            \includegraphics[width=\textwidth]{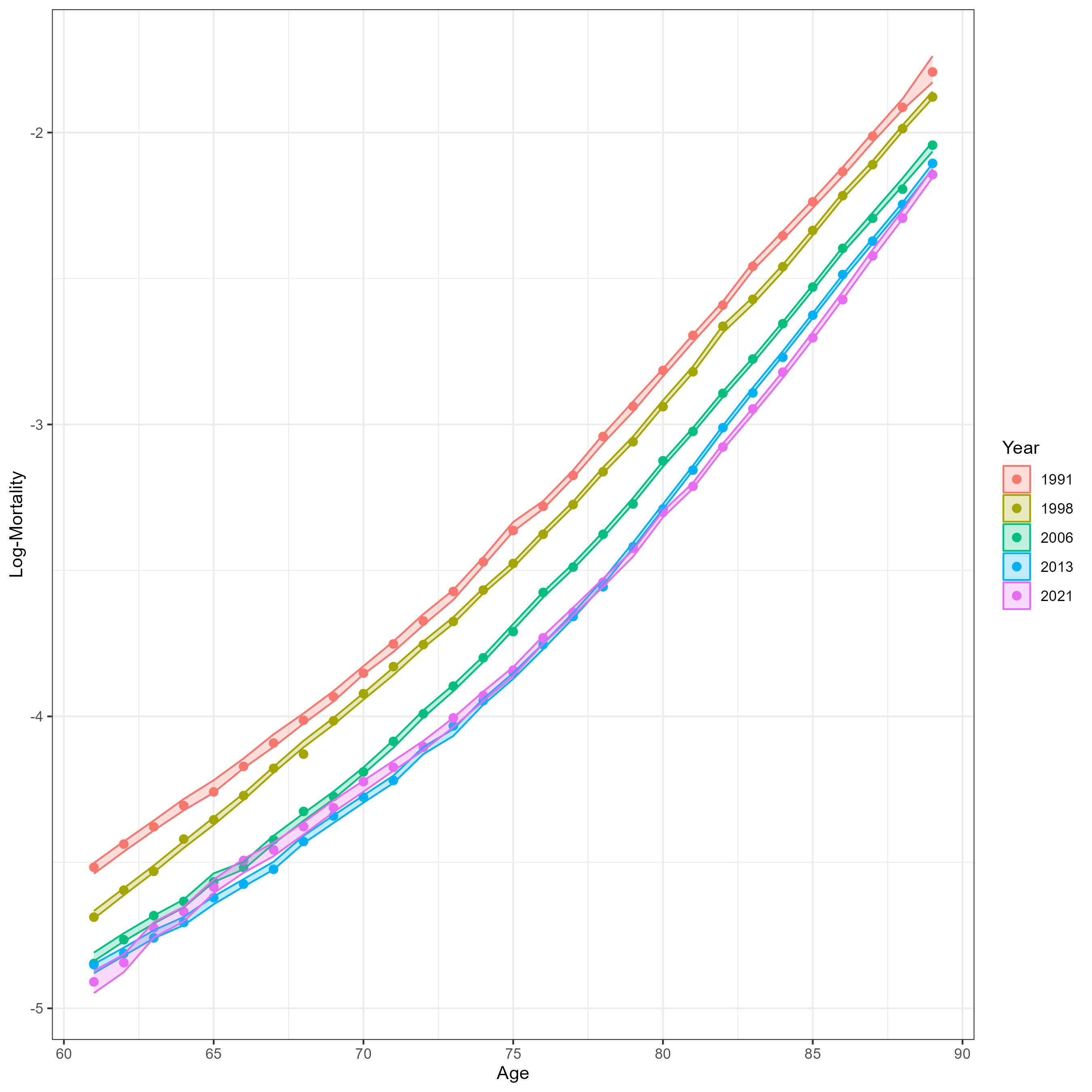} 
            \caption{France}
        \end{subfigure} \\
        
        \begin{subfigure}[b]{0.3\textwidth}
            \centering
            \includegraphics[width=\textwidth]{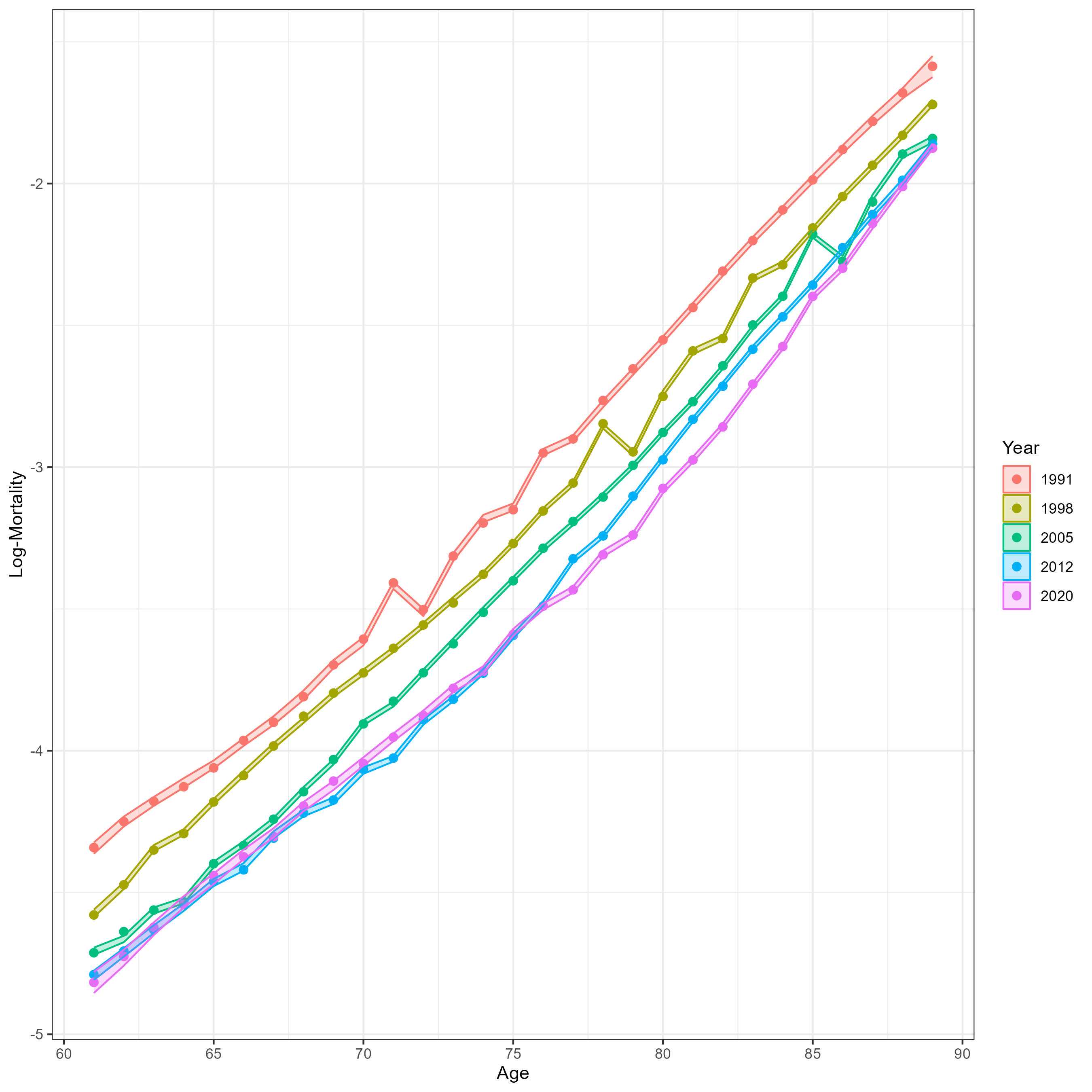} 
            \caption{Germany}
        \end{subfigure} &
        \begin{subfigure}[b]{0.3\textwidth}
            \centering
            \includegraphics[width=\textwidth]{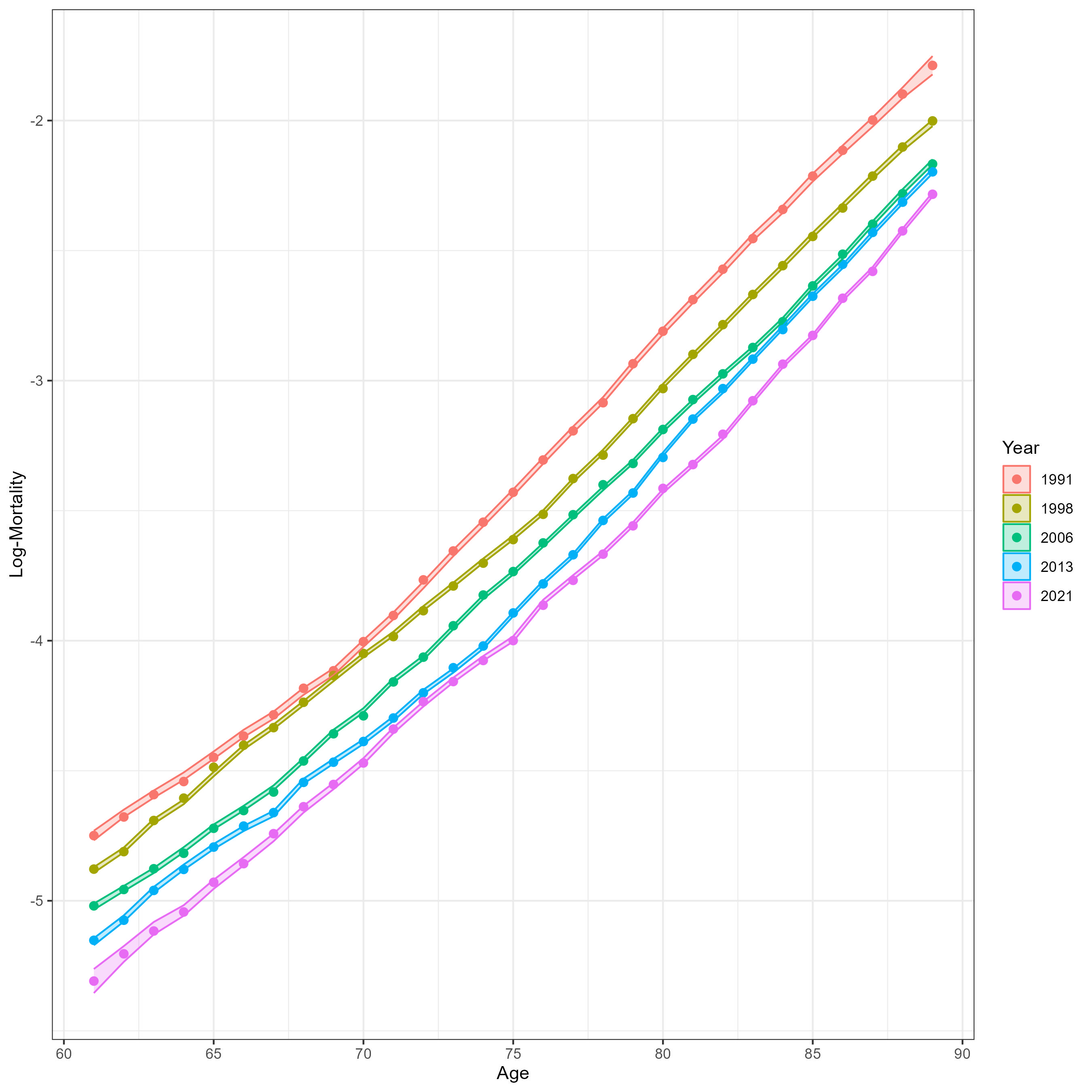} 
            \caption{Japan}
        \end{subfigure} &
        \begin{subfigure}[b]{0.3\textwidth}
            \centering
            \includegraphics[width=\textwidth]{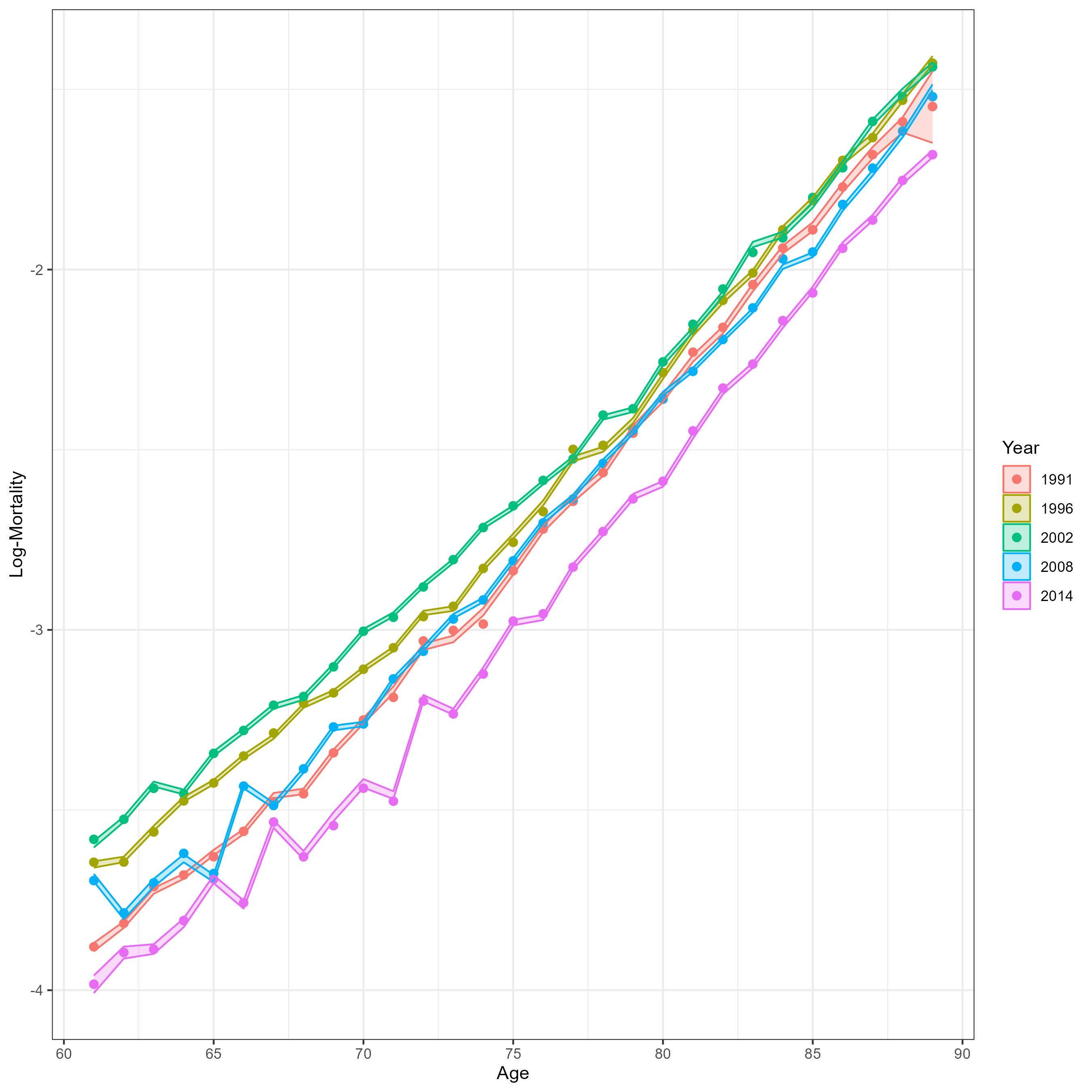} 
            \caption{Russia}
        \end{subfigure} \\
        
        \begin{subfigure}[b]{0.3\textwidth}
            \centering
            \includegraphics[width=\textwidth]{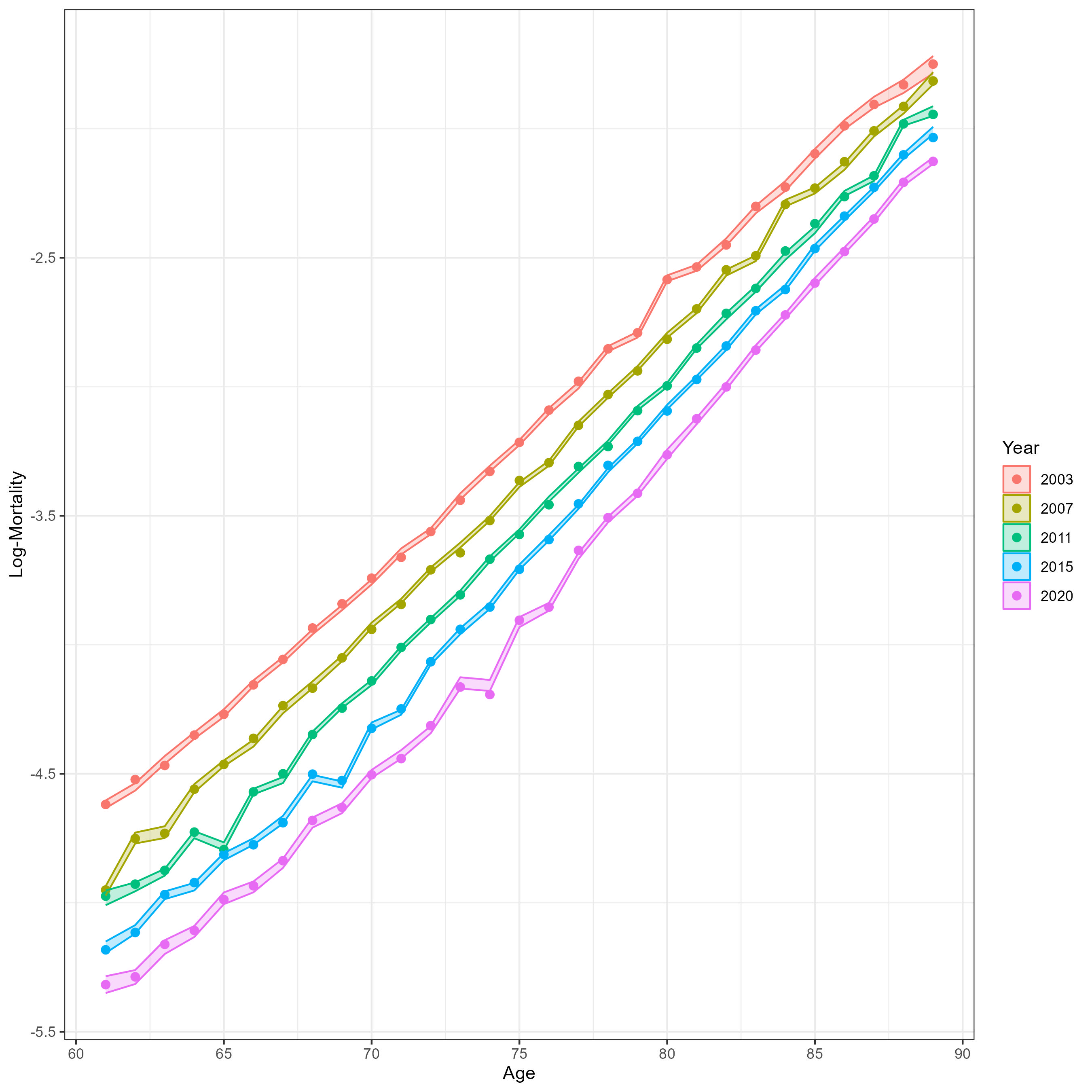} 
            \caption{Korea}
        \end{subfigure} &
        \begin{subfigure}[b]{0.3\textwidth}
            \centering
            \includegraphics[width=\textwidth]{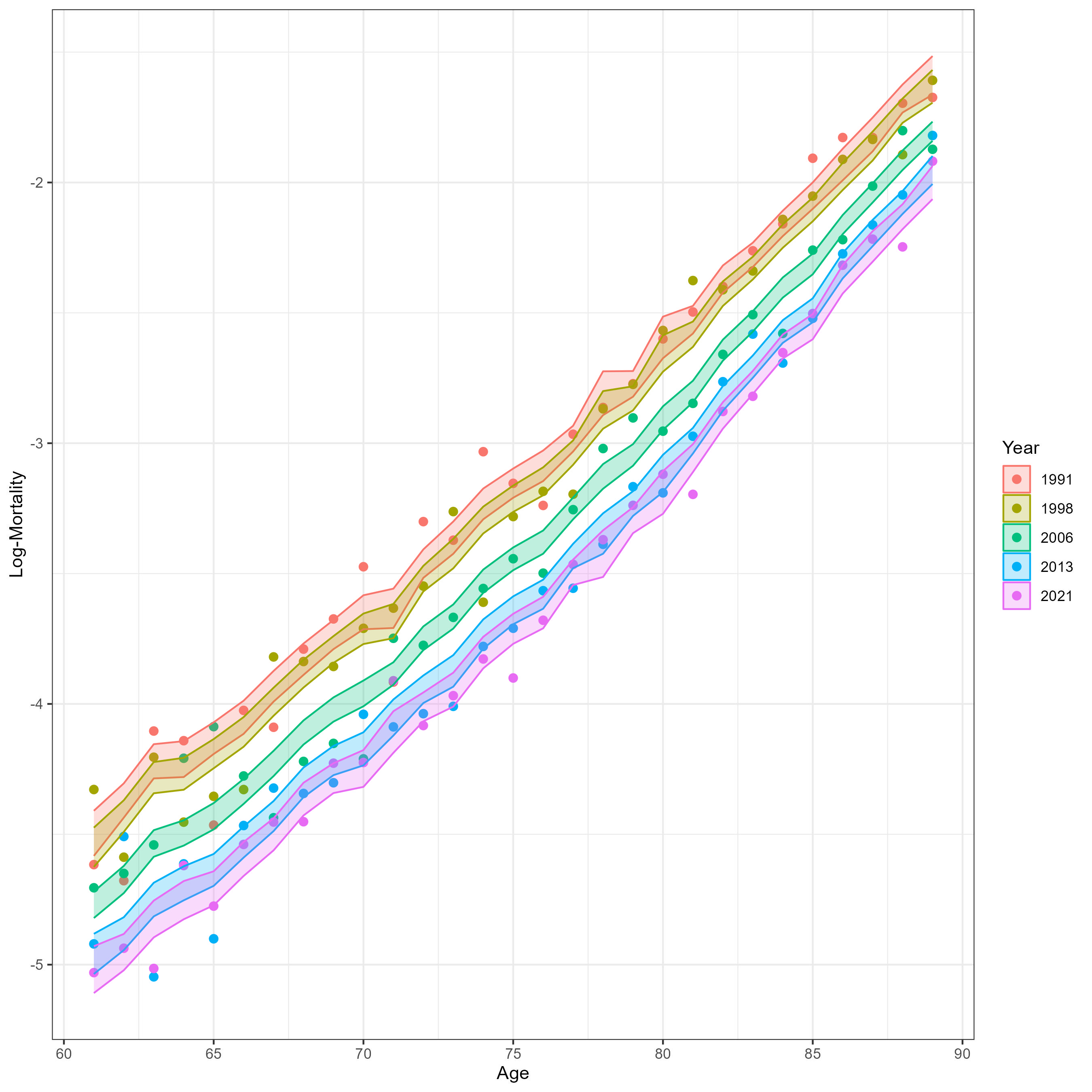} %
            \caption{Luxembourg}
        \end{subfigure} &
        \begin{subfigure}[b]{0.3\textwidth}
            \centering
            \includegraphics[width=\textwidth]{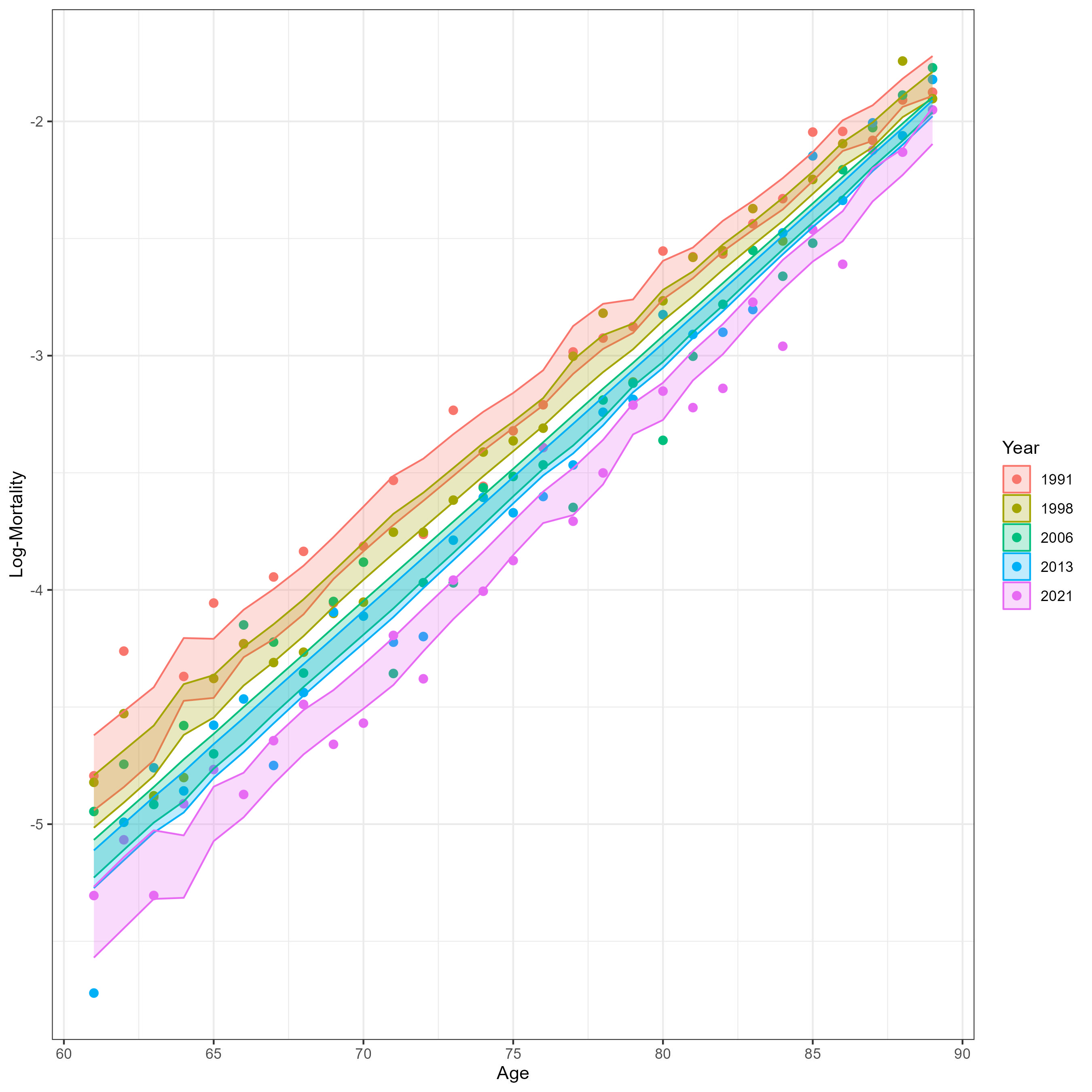} 
            \caption{Iceland}
        \end{subfigure}
    \end{tabular}
    \caption{HMD data results: 95\% PCI of the log-mortality rates $m_{x,t}$. The dots represent the crude rates.}
\label{fig:ap_mortalityrates}
\end{figure}

For each country, we ran the MCMC for $2$ chains of $5000$ iterations, with $5000$ burn-in iterations. We have assumed flat prior on the model parameters. 
The posterior credible intervals (PCI) of the estimated log mortality rates are presented in Fig. \ref{fig:ap_mortalityrates}. For simplicity, we only show here the results for a few countries (the full results are presented in Supplementary material). We also present the posterior probabilities of the model selection procedure in Table \ref{fig:ap_delta}. In many countries, it can be seen that the evidence for the selection of one term is strong enough that the posterior probability is $1$, while for other countries multiple models are selected. Few notable observations are that the cohort term ($\delta_4 = 2$) is almost always selected for all countries. Also, most of the countries do not select a linear effect for $a_x$ ($\delta_1 = 2$), suggesting that this effect is mostly non-linear.


From a model selection perspective, countries with large exposures, such as the UK, USA, Germany and Japan, generally tend to support with strong evidence only one model. Countries with lower exposures, such as Iceland and Luxembourg, will result in higher uncertainty in both the credible intervals of the mortality rates and in the model selection. Interestingly, the model selection for Korea seems to exhibit two local modes at the models $a_x + k_t b_x + \gamma_{t-x}$ and $a + b(x - \bar{x}) + k^1_t + k^2_t b_x + \gamma_{t-x}$, with each of the two chains converging to each model, which shows that both models equally support the data.

\begin{table}[]
\centering
\begin{tabular}{|l|ll|lll|ll|ll|}
\hline
\multirow{2}{*}{} & \multicolumn{2}{c|}{$\delta_1$} & \multicolumn{3}{c|}{$\delta_2$}                     & \multicolumn{2}{c|}{$\delta_3$} & \multicolumn{2}{c|}{$\delta_4$} \\ \hline \cline{2-10} 
                         & \multicolumn{1}{l|}{1}    & 2   & \multicolumn{1}{l|}{1} & \multicolumn{1}{l|}{2} & 3 & \multicolumn{1}{l|}{1}    & 2   & \multicolumn{1}{l|}{1}    & 2   \\ \hline
Australia                & \multicolumn{1}{l|}{1.0000} & 0.0000 & \multicolumn{1}{l|}{0.0000} & \multicolumn{1}{l|}{1.0000} & 0.0000 & \multicolumn{1}{l|}{1.0000} & 0.0000 & \multicolumn{1}{l|}{0.0000} & 1.0000 \\ \hline
Austria                  & \multicolumn{1}{l|}{1.0000} & 0.0000 & \multicolumn{1}{l|}{0.0000} & \multicolumn{1}{l|}{1.0000} & 0.0000 & \multicolumn{1}{l|}{0.8233} & 0.1767 & \multicolumn{1}{l|}{0.0000} & 1.0000 \\ \hline
Belgium                  & \multicolumn{1}{l|}{1.0000} & 0.0000 & \multicolumn{1}{l|}{0.0000} & \multicolumn{1}{l|}{1.0000} & 0.0000 & \multicolumn{1}{l|}{0.9999} & 0.0001 & \multicolumn{1}{l|}{0.0000} & 1.0000 \\ \hline
Bulgaria                 & \multicolumn{1}{l|}{1.0000} & 0.0000 & \multicolumn{1}{l|}{0.0000} & \multicolumn{1}{l|}{1.0000} & 0.0000 & \multicolumn{1}{l|}{0.0000} & 1.0000 & \multicolumn{1}{l|}{0.0000} & 1.0000 \\ \hline
Canada                   & \multicolumn{1}{l|}{1.0000} & 0.0000 & \multicolumn{1}{l|}{0.3829} & \multicolumn{1}{l|}{0.6166} & 0.0004 & \multicolumn{1}{l|}{0.9696} & 0.0304 & \multicolumn{1}{l|}{0.0000} & 1.0000 \\ \hline
Chile                    & \multicolumn{1}{l|}{0.4999} & 0.5001 & \multicolumn{1}{l|}{0.0000} & \multicolumn{1}{l|}{0.0000} & 1.0000 & \multicolumn{1}{l|}{0.0001} & 0.9999 & \multicolumn{1}{l|}{0.0000} & 1.0000 \\ \hline
Croatia                  & \multicolumn{1}{l|}{0.9999} & 0.0001 & \multicolumn{1}{l|}{0.6904} & \multicolumn{1}{l|}{0.3051} & 0.0044 & \multicolumn{1}{l|}{0.9810} & 0.0190 & \multicolumn{1}{l|}{0.0000} & 1.0000 \\ \hline
Czechia                  & \multicolumn{1}{l|}{1.0000} & 0.0000 & \multicolumn{1}{l|}{0.0000} & \multicolumn{1}{l|}{1.0000} & 0.0000 & \multicolumn{1}{l|}{0.0000} & 1.0000 & \multicolumn{1}{l|}{0.0000} & 1.0000 \\ \hline
Denmark                  & \multicolumn{1}{l|}{1.0000} & 0.0000 & \multicolumn{1}{l|}{0.0000} & \multicolumn{1}{l|}{1.0000} & 0.0000 & \multicolumn{1}{l|}{1.0000} & 0.0000 & \multicolumn{1}{l|}{0.0000} & 1.0000 \\ \hline
Estonia                  & \multicolumn{1}{l|}{1.0000} & 0.0000 & \multicolumn{1}{l|}{0.0000} & \multicolumn{1}{l|}{1.0000} & 0.0000 & \multicolumn{1}{l|}{0.5924} & 0.4076 & \multicolumn{1}{l|}{1.0000} & 0.0000 \\ \hline
Finland                  & \multicolumn{1}{l|}{1.0000} & 0.0000 & \multicolumn{1}{l|}{0.0000} & \multicolumn{1}{l|}{1.0000} & 0.0000 & \multicolumn{1}{l|}{0.5000} & 0.5000 & \multicolumn{1}{l|}{0.5000} & 0.5000 \\ \hline
France                   & \multicolumn{1}{l|}{1.0000} & 0.0000 & \multicolumn{1}{l|}{0.0000} & \multicolumn{1}{l|}{0.4423} & 0.5577 & \multicolumn{1}{l|}{0.0000} & 1.0000 & \multicolumn{1}{l|}{0.0000} & 1.0000 \\ \hline
Germany                  & \multicolumn{1}{l|}{1.0000} & 0.0000 & \multicolumn{1}{l|}{0.0000} & \multicolumn{1}{l|}{0.0000} & 1.0000 & \multicolumn{1}{l|}{0.0000} & 1.0000 & \multicolumn{1}{l|}{0.0000} & 1.0000 \\ \hline
Greece                   & \multicolumn{1}{l|}{1.0000} & 0.0000 & \multicolumn{1}{l|}{0.0000} & \multicolumn{1}{l|}{1.0000} & 0.0000 & \multicolumn{1}{l|}{0.0000} & 1.0000 & \multicolumn{1}{l|}{0.0000} & 1.0000 \\ \hline
HK                       & \multicolumn{1}{l|}{0.9997} & 0.0003 & \multicolumn{1}{l|}{0.0000} & \multicolumn{1}{l|}{0.9115} & 0.0885 & \multicolumn{1}{l|}{0.0000} & 1.0000 & \multicolumn{1}{l|}{0.0000} & 1.0000 \\ \hline
Hungary                  & \multicolumn{1}{l|}{1.0000} & 0.0000 & \multicolumn{1}{l|}{0.0000} & \multicolumn{1}{l|}{1.0000} & 0.0000 & \multicolumn{1}{l|}{0.9860} & 0.0140 & \multicolumn{1}{l|}{0.0000} & 1.0000 \\ \hline
Iceland                  & \multicolumn{1}{l|}{0.0000} & 1.0000 & \multicolumn{1}{l|}{0.0000} & \multicolumn{1}{l|}{1.0000} & 0.0000 & \multicolumn{1}{l|}{0.6606} & 0.3394 & \multicolumn{1}{l|}{1.0000} & 0.0000 \\ \hline
Ireland                  & \multicolumn{1}{l|}{0.0000} & 1.0000 & \multicolumn{1}{l|}{0.0000} & \multicolumn{1}{l|}{1.0000} & 0.0000 & \multicolumn{1}{l|}{0.0000} & 1.0000 & \multicolumn{1}{l|}{1.0000} & 0.0000 \\ \hline
Israel                   & \multicolumn{1}{l|}{0.0000} & 1.0000 & \multicolumn{1}{l|}{0.0000} & \multicolumn{1}{l|}{1.0000} & 0.0000 & \multicolumn{1}{l|}{0.8577} & 0.1423 & \multicolumn{1}{l|}{0.0000} & 1.0000 \\ \hline
Italy                    & \multicolumn{1}{l|}{1.0000} & 0.0000 & \multicolumn{1}{l|}{0.0000} & \multicolumn{1}{l|}{0.0000} & 1.0000 & \multicolumn{1}{l|}{0.7693} & 0.2307 & \multicolumn{1}{l|}{0.0000} & 1.0000 \\ \hline
Japan                    & \multicolumn{1}{l|}{1.0000} & 0.0000 & \multicolumn{1}{l|}{0.0000} & \multicolumn{1}{l|}{0.0000} & 1.0000 & \multicolumn{1}{l|}{0.0000} & 1.0000 & \multicolumn{1}{l|}{0.0000} & 1.0000 \\ \hline
Korea                    & \multicolumn{1}{l|}{0.5000} & 0.5000 & \multicolumn{1}{l|}{0.0000} & \multicolumn{1}{l|}{0.5000} & 0.5000 & \multicolumn{1}{l|}{0.5000} & 0.5000 & \multicolumn{1}{l|}{0.0000} & 1.0000 \\ \hline
Latvia                   & \multicolumn{1}{l|}{1.0000} & 0.0000 & \multicolumn{1}{l|}{0.6783} & \multicolumn{1}{l|}{0.3217} & 0.0000 & \multicolumn{1}{l|}{0.8613} & 0.1387 & \multicolumn{1}{l|}{1.0000} & 0.0000 \\ \hline
Lithuania               & \multicolumn{1}{l|}{1.0000} & 0.0000 & \multicolumn{1}{l|}{0.0000} & \multicolumn{1}{l|}{0.9998} & 0.0002 & \multicolumn{1}{l|}{1.0000} & 0.0000 & \multicolumn{1}{l|}{0.0000} & 1.0000 \\ \hline
Luxembourg              & \multicolumn{1}{l|}{1.0000} & 0.0000 & \multicolumn{1}{l|}{0.0765} & \multicolumn{1}{l|}{0.9235} & 0.0000 & \multicolumn{1}{l|}{0.6049} & 0.3951 & \multicolumn{1}{l|}{1.0000} & 0.0000 \\ \hline
Netherlands             & \multicolumn{1}{l|}{0.0000} & 1.0000 & \multicolumn{1}{l|}{0.0000} & \multicolumn{1}{l|}{0.0000} & 1.0000 & \multicolumn{1}{l|}{1.0000} & 0.0000 & \multicolumn{1}{l|}{0.0000} & 1.0000 \\ \hline
Poland                  & \multicolumn{1}{l|}{1.0000} & 0.0000 & \multicolumn{1}{l|}{0.0000} & \multicolumn{1}{l|}{1.0000} & 0.0000 & \multicolumn{1}{l|}{0.0000} & 1.0000 & \multicolumn{1}{l|}{0.0000} & 1.0000 \\ \hline
Portugal                & \multicolumn{1}{l|}{1.0000} & 0.0000 & \multicolumn{1}{l|}{0.0000} & \multicolumn{1}{l|}{1.0000} & 0.0000 & \multicolumn{1}{l|}{0.0000} & 1.0000 & \multicolumn{1}{l|}{0.0000} & 1.0000 \\ \hline
Russia                  & \multicolumn{1}{l|}{1.0000} & 0.0000 & \multicolumn{1}{l|}{0.0000} & \multicolumn{1}{l|}{0.0000} & 1.0000 & \multicolumn{1}{l|}{0.0000} & 1.0000 & \multicolumn{1}{l|}{0.0000} & 1.0000 \\ \hline
Slovakia                & \multicolumn{1}{l|}{1.0000} & 0.0000 & \multicolumn{1}{l|}{0.0026} & \multicolumn{1}{l|}{0.9974} & 0.0000 & \multicolumn{1}{l|}{0.0421} & 0.9579 & \multicolumn{1}{l|}{0.0000} & 1.0000 \\ \hline
Slovenia                & \multicolumn{1}{l|}{0.0000} & 1.0000 & \multicolumn{1}{l|}{0.0000} & \multicolumn{1}{l|}{1.0000} & 0.0000 & \multicolumn{1}{l|}{0.9666} & 0.0334 & \multicolumn{1}{l|}{0.0000} & 1.0000 \\ \hline
Taiwan                  & \multicolumn{1}{l|}{1.0000} & 0.0000 & \multicolumn{1}{l|}{0.0000} & \multicolumn{1}{l|}{1.0000} & 0.0000 & \multicolumn{1}{l|}{0.0000} & 1.0000 & \multicolumn{1}{l|}{0.0000} & 1.0000 \\ \hline
UK                       & \multicolumn{1}{l|}{1.0000} & 0.0000 & \multicolumn{1}{l|}{0.0000} & \multicolumn{1}{l|}{0.0000} & 1.0000 & \multicolumn{1}{l|}{0.0001} & 0.9999 & \multicolumn{1}{l|}{0.0000} & 1.0000 \\ \hline
Ukraine                 & \multicolumn{1}{l|}{1.0000} & 0.0000 & \multicolumn{1}{l|}{0.0000} & \multicolumn{1}{l|}{0.0000} & 1.0000 & \multicolumn{1}{l|}{0.0014} & 0.9986 & \multicolumn{1}{l|}{0.0000} & 1.0000 \\ \hline
USA                      & \multicolumn{1}{l|}{1.0000} & 0.0000 & \multicolumn{1}{l|}{0.0000} & \multicolumn{1}{l|}{0.0000} & 1.0000 & \multicolumn{1}{l|}{0.0000} & 1.0000 & \multicolumn{1}{l|}{0.0000} & 1.0000 \\ \hline
\end{tabular}
\caption{HMD data results: Posterior probabilities of the model variables $\delta_i$}
\label{fig:ap_delta}
\end{table}

\section{APCI model stratified by product}
\label{sec:appmodel}

In this section, we demonstrate how to define mortality models for data stratified by an additional variable, i.e. the APP data. For example, the additional stratifying variable could represent  different insurance products, sexes, or countries. For simplicity, we will refer to this additional stratifying variable as ``product'' throughout this section.

We build the model by extending the APCI model \autocite{richards_apci}. As mentioned in the introduction, we have chosen this base model for two main reasons. First, the APCI model is one of the most popular models used to fit mortality data and is currently used by the CMI \autocite{cmib_2023}. Second, we aim to define an extension of the base model that involves fewer changes in the ICs when each model changes according to product, which, similar to the previous section, will help the design of the RJMCMC framework.

According to the APCI model, log-mortality is modelled as 
$$
\text{log}(m_{x,t}) = a_x + b_x (t - \bar{t}) +  k_t + \gamma_{t-x} \ ,
$$
where $a_x$ represents the age effect, $b_x$ represents the change in period effect by age, $k_t$ represents the period effect and $\gamma_{t-x}$ represents the cohort effect. The ICs that we use are $\sum_x b_x = \sum_t k_t = \sum_{c} \gamma_{c} = \gamma_{1-X} = \gamma_{Y-1} = 0$. Even though it is more common to assume $\sum_c c \gamma_c = \sum_c c^2 \gamma_c = 0$, we have instead assumed that the first and last element of $\gamma_c$ ($\gamma_{1-X}$ and $\gamma_{Y-1}$, respectively) are 0, as it leads to easier MCMC updates. We then extend the base APCI model to allow for product effect by proposing different functional forms for the terms $a_x$, $b_x$ and $k_t$. 

\begin{itemize}
    \item For $a_x$, we propose the product forms: a) $a_x$ (no change) b) $a_x + c^1_p$ and c) $a_{x,p}$, and we denote each choice with the latent variable $\delta_1 = 1$,$2$ and $3$, respectively. The ICs are $\sum c^1_p = 0$.
    \item For $b_x$, we propose the product forms: a) $b_x$  (no change) and b) $b_x c^2_p$, modelled with the latent variable $\delta_2 = 1$ and $2$, respectively. The ICs are $\sum c^2_p = P$, since $c^2_p$ appears in a product with $b_x$.
    \item   For $k_t$, we propose the forms: a) $k_t$ (no change) and b) $k_t + k^2_t c^3_p$, modelled with the latent variable $\delta_3 = 1$ and $2$, respectively. The ICs are the following. $\sum k^2_t = 0$, since any constant $K c^3_p$ can be incorporated into $ c^1_p$ (or into $a_{x,p}$), $\sum c^3_p = 0$ since any additive constant can be incorporated into $k_t$. $c^3_1 = 1$, Since $c_p$ can still be multiplied by a constant which can be incorporated into $k^2_t$.
\end{itemize}Similarly to the previous section, note that some constraints depend on the modelling choices. Specifically, $\sum k^2_t = 0$ only holds if $\delta_1 \neq 1$, and $\sum_{x} b_x = 0$ only holds in the case $\delta_2 = 1$. We note that we still assume $\sum_x b_x = 0$ also in the case $\delta_2 = 2$ since, in our experiments, the model was very close to being unidentifiable without the constraint $\sum_x b_x = 0$. Moreover, similarly to the previous section, to help the design of the RJMCMC, we choose to still impose the constraint $\sum k^2_t = 0$ even when $\delta_1 = 1$.

The model for age-period-product data can be summarised as
$$
\text{log}(m_{x,t,p}) = a_{\delta_1}(x,p) + b_{\delta_2}(x,p) (t - \bar{t}) + k_{\delta_3}(t,p) + \gamma_{t- x}
$$
where 

\hspace{-1cm}$a_{\delta_1}(x,p) = \begin{cases}  a_x \hspace{.9cm} \text{ if } \delta_1 = 1 \\ a_x + c^1_p \hspace{.2cm}\text{ if } \delta_1 = 2 \\ a_{x,p} \hspace{.7cm} \text{ if } \delta_1 = 3 \\ \end{cases},$
$ b_{\delta_2}(x,p) = \begin{cases} b_{x} \hspace{.5cm} \text{ if } \delta_2 = 1 \\ b_x c^2_p \hspace{.2cm }\text{ if } \delta_2 = 2 \\ \end{cases}$, $ k_{\delta_3}(t,p) = \begin{cases} k_{t} \hspace{1.3cm} \text{ if }  \delta_3 = 1 \\ k_t + k^2_t c^3_p \hspace{.3cm} \text{ if } \delta_3 = 2 \\ \end{cases}$.
The full set of models is presented in Table \ref{fig:allmodels_app}.

\begin{table}[H]
\hspace{-2.5cm}
\begin{tabular}{|l|l|l|l|l|l|}
\hline
Model & $\delta_1$  & $\delta_2$  & $\delta_3$ & Identifiability constraint \\ \hline

$a_x + b_x (t - \bar{t}) + k_t + \gamma_{t-x}$ & 1 & 1  & 1  & {\tiny $\sum b_x = \sum k_t = \sum \gamma_c = \gamma_{1-X} = \gamma_{Y - 1} = 0$} \\ \hline

$a_x + c^1_p + b_x (t - \bar{t}) + k_t + \gamma_{t-x} 
$ & 2 & 1  & 1 &  {\tiny $\sum b_x = \sum c^1_p = \sum k_t = \sum \gamma_c = \gamma_{1-X} = \gamma_{Y - 1} = 0$  }\\ \hline

$a_{x,p} + b_x (t - \bar{t}) + k_t + \gamma_{t-x} 
$ & 3 & 1  & 1  &  {\tiny $\sum b_x = \sum k_t = \sum \gamma_c = \gamma_{1-X} = \gamma_{Y - 1} = 0$ } \\ \hline

$a_x + (b_x  c^2_p) (t - \bar{t}) + k_t + \gamma_{t-x}$ & 1 & 2  & 1 &  {\tiny $\sum b_x = \sum k_t = \sum \gamma_c = \gamma_{1-X} = \gamma_{Y - 1} = 0$ , $\sum c^2_p = P$ }  \\ \hline

$a_x + c^1_p + (b_x  c^2_p) (t - \bar{t}) + k_t + \gamma_{t-x} 
$ & 2 & 2  & 1  & {\tiny $\sum b_x = \sum c^1_p = \sum k_t = \sum \gamma_c = \gamma_{1-X} = \gamma_{Y - 1} = 0$ , $\sum c^2_p = P$ }  \\ \hline

$a_{x,p} + (b_x  c^2_p) (t - \bar{t}) + k_t + \gamma_{t-x} 
$ & 3 & 2  & 1   &  {\tiny $\sum b_x = \sum k_t = \sum \gamma_c = \gamma_{1-X} = \gamma_{Y - 1} = 0$ , $\sum c^2_p = P$}   \\ \hline

$a_x + b_x (t - \bar{t}) + (k_t + k^2_t c^3_p) + \gamma_{t-x}$ & 1 & 1  & 2  & {\tiny $\sum b_x = \sum k_t = \sum k^2_t =  \sum \gamma_c = \gamma_{1-X} = \gamma_{Y - 1} = 0$, $\sum c^3_p = P$,  $c^3_1 = 1$ } \\ \hline

$a_x + c^1_p + b_x (t - \bar{t}) + (k_t + k^2_t c^3_p) + \gamma_{t-x} 
$ & 2 & 1  & 2  & {\tiny $\sum b_x = \sum c^1_p = \sum b_x = \sum k_t = \sum k^2_t = \sum \gamma_c = \gamma_{1-X} = \gamma_{Y - 1} = 0$, $\sum c^3_p = P$,  $c^3_1 = 1$ } \\ \hline

$a_{x,p} + b_x (t - \bar{t}) + (k_t + k^2_t c^3_p) + \gamma_{t-x} 
$ & 3 & 1  & 2 &  {\tiny $ \sum b_x = \sum k_t = \sum k^2_t = \sum \gamma_c = \gamma_{1-X} = \gamma_{Y - 1} = 0$, $\sum c^3_p = P$,  $c^3_1 = 1$   } \\ \hline

$a_x + (b_x  c^2_p) (t - \bar{t}) + (k_t + k^2_t c^3_p) + \gamma_{t-x}$ & 1 & 2  & 2  &  {\tiny$ \sum b_x = \sum k_t = \sum k^2_t =  \sum \gamma_c = \gamma_{1-X} = \gamma_{Y - 1} = 0$, $\sum c^3_p = P$,  $c^3_1 = 1$} \\ \hline

$a_x + c^1_p + (b_x  c^2_p) (t - \bar{t}) + (k_t + k^2_t c^3_p) + \gamma_{t-x} 
$ & 2 & 2  & 2 & {\tiny $ \sum b_x = \sum c^1_p = \sum k_t = \sum k^2_t =  \sum \gamma_c = \gamma_{1-X} = \gamma_{Y - 1} = 0$, $\sum c^3_p = P$,  $c^3_1 = 1$  } \\ \hline

$a_{x,p} + (b_x  c^2_p) (t - \bar{t}) + (k_t + k^2_t c^3_p) + \gamma_{t-x} 
$ & 3 & 2  & 2  &  {\tiny $ \sum b_x = \sum k_t = \sum k^2_t =  \sum \gamma_c = \gamma_{1-X} = \gamma_{Y - 1} = 0$, $\sum c^3_p = P$,  $c^3_1 = 1$  }  \\ \hline

\end{tabular}
\caption{Complete set of models for the APP data} 
\label{fig:allmodels_app}
\end{table}

\subsection{Inference}

We sample from the posterior distribution of $(\theta_1,\delta_1,\theta_2,\delta_2,\theta_3,\delta_3,\gamma_{t-x})$, where $\theta_1$ are the parameter of $a_{\delta_1}(x,p)$, $\theta_2$ are the parameter of $b_{\delta_2}(x,p)$ and $\theta_3$ are the parameter of $k_{\delta_3}(t,p)$, with an MCMC algorithm. We iteratively sample the parameters $\theta_1$,$\theta_2$,$\theta_3$ and $\gamma_{t-x}$ and the model choices $\delta_1$, $\delta_2$ and $\delta_3$.

The scheme follows the same ideas used for the AP model for both the parameter updates and the model space, where we use a Metropolis-Hastings update with Laplace approximation to the proposal for the parameter updates, and explore the model space by proposing to change each variable $\delta_i$ at a time. We separately discuss each proposal.

\begin{itemize}
    \item Updating $\delta_1$

When proposing to move from model $1$ ($a_x$) to model $2$ ($a_x + c^1_p$), we propose to add the variables $c^1_p$. As proposal distribution for $c^1_p$, we choose the Laplace approximation to the posterior of $(c^1_1,\dots,c^1_{P-1})$, since $c^1_P = - \sum_{p=1}^{P-1} c^1_p$. The reverse move is deterministic.

When proposing to move from model $2$ ($a_x + c^1_p$) to model $3$ ($a_{x,p}$), we note that the difference in the models dimenion is $X P - X - (P-1) = (X-1)(P-1)$, to which the dimension of the proposal needs to be equal to. To build a proposal of this dimension, we note that $a_{x,p}$ can be parametrised as $a_x + c^1_p + \tilde{c}^1_{x,p}$, where $\sum_p c^1_p = \sum_p c^2_{x,p} (\text{ for each $x$}) = \sum_x c^2_{x,p} (\text{ for each $p$}) = 0$, from which it follows that the dimension of $c^2_{x,p}$ is $(X-1)(P-1)$. Therefore, we choose as auxiliary variables $u$ of the RJMCMC the variables $c^2_{x,p}$ and as proposal distribution we choose the Laplace approximation to the posterior. The reverse move is deterministic.
    
    \item Updating $\delta_2$

To design the proposal from model $1$ ($b_x (t - \bar{t})$) to model $2$ ($b_x c^2_p (t - \bar{t})$), we propose the variables $c^2_p$, from the Laplace approximation to the posterior distribution, while the reverse move is deterministic.

    
    \item Updating $\delta_3$ 

To propose from model $1$ ($k_t$) to model $2$ ($k_t + k^2_t c^3_p$), we propose the variables $k^2_t$ and $c^3_p$ from the Laplace approximation to the posterior distribution, while the reverse move is deterministic.
    
\end{itemize}

\subsection{Case study: Insurance products}

\begin{figure}[H]
    \centering
    \includegraphics[scale=.1]{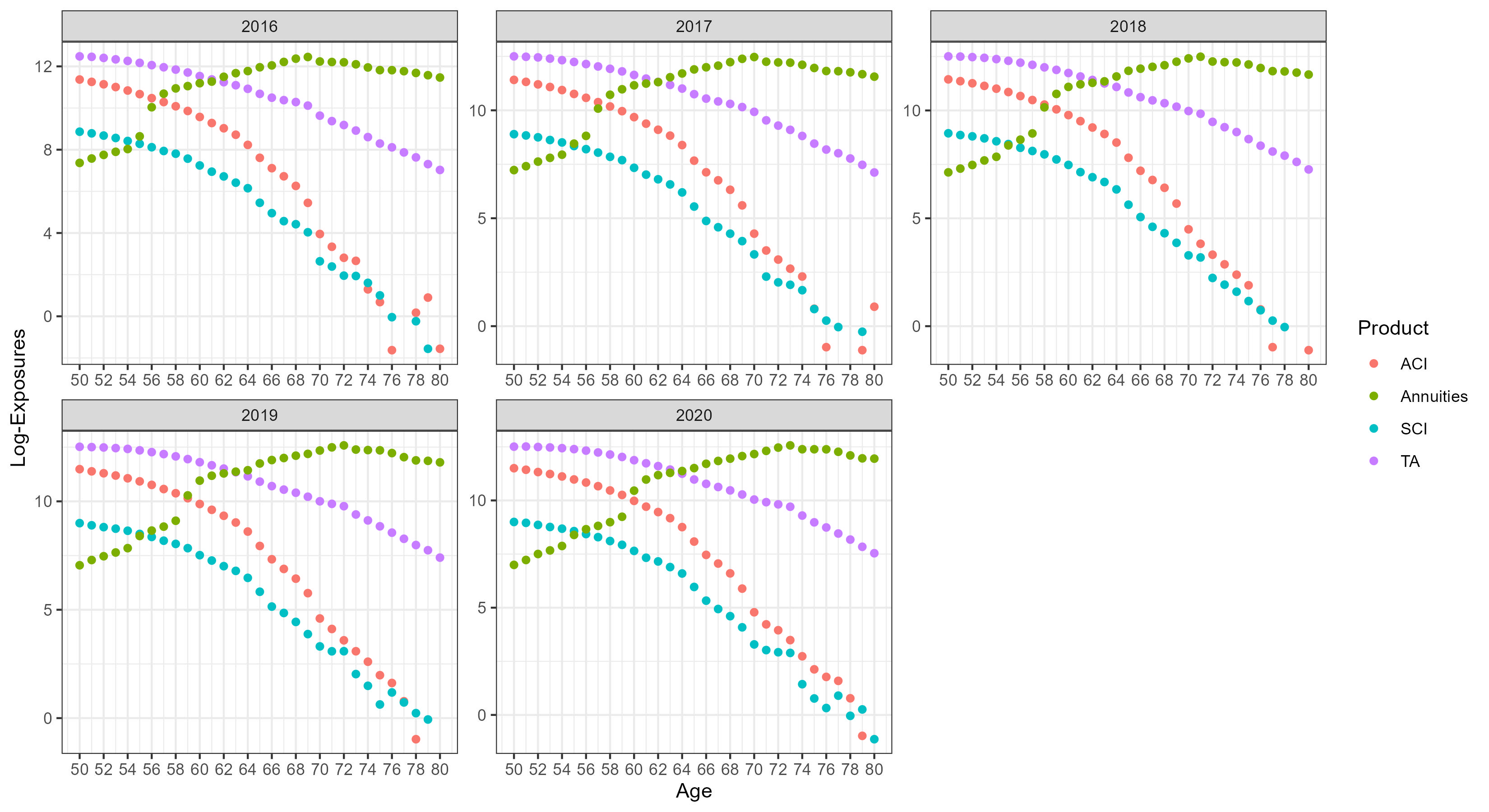}
    \caption{Insurance product Case study: Log(Exposures)}.
        \label{fig:app_exposure}
\end{figure}

We apply the previously described framework on mortality data stratified by different product. The data were sourced by the CMI and were provided by different life insurance providers. The available products are pension annuities in payment (Annuities), Term Assurances (TA), Accelerated Critical Illness (ACI) and Standalone Critical Illness (SCI). The data ranges from $2016$ to $2020$ and covers both males and females and across varying commencement years, durations and sum assured bands. In our example, we have used ages from $50$ to $80$. We note that the data comes with several limitations, such as very low exposure for some combination of age, year and product. as shown in Fig. \ref{fig:app_exposure}. However, working in a Bayesian framework naturally allows the model to take the low sample size into account in the estimation. For more details on the data, we refer the reader to the CMI Working Paper 117 and 162 \autocite{cmib_2019, cmib_2022}.

\hspace{-1cm}

\begin{figure}
    \centering
    \includegraphics[scale=.115]{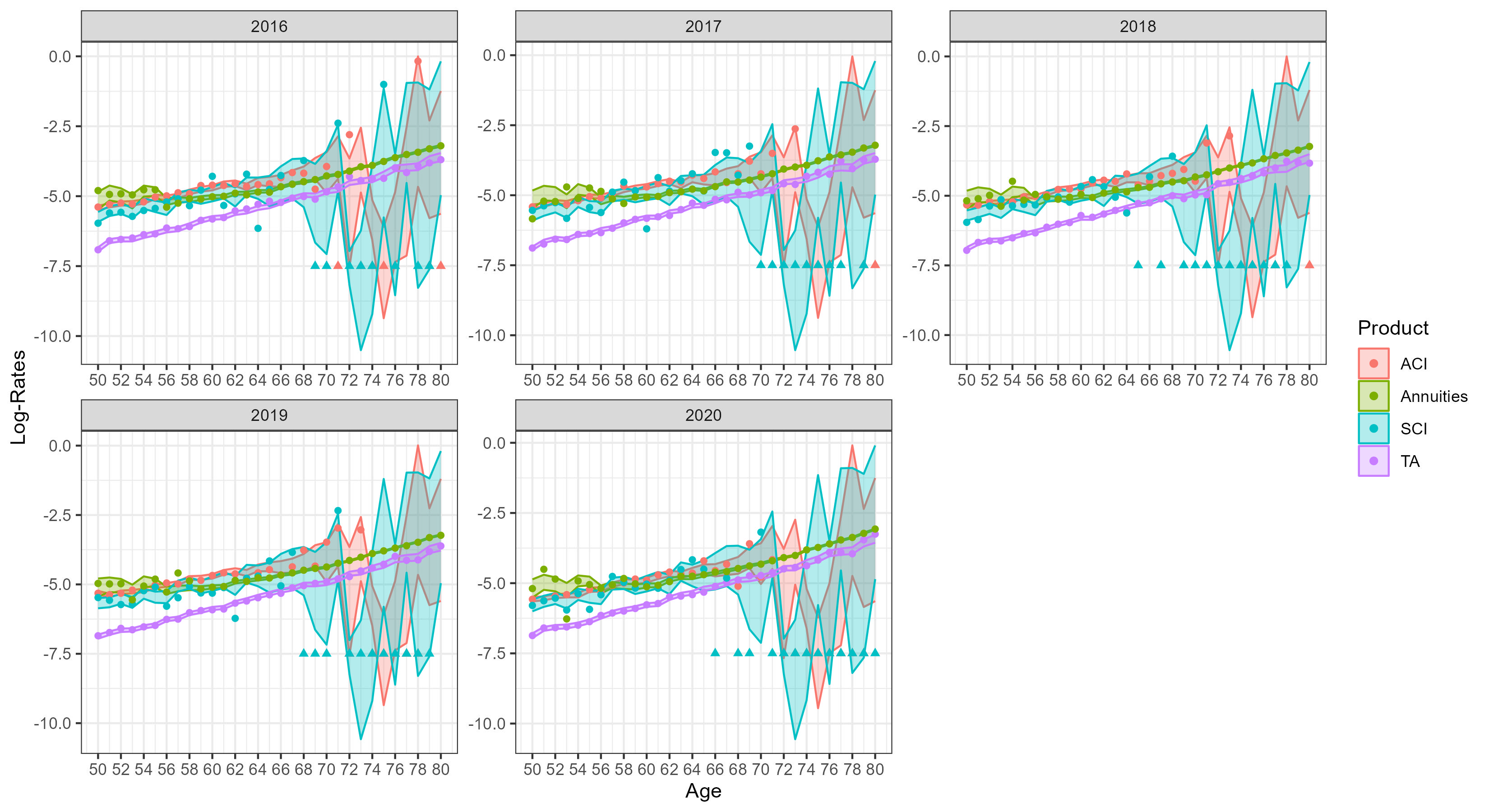}
    \caption{Insurance product Case study: $95$ \% PCI of the log-mortality rates. The crude mortality rates are represented by the dots. Data points where the crude mortality rates are $0$ are represented by a triangle.}.
        \label{fig:app_results}
\end{figure}

The credible intervals of the log-mortality curves are shown in Fig. \ref{fig:app_results}. 
It is clear that individuals associated with term assurance products exhibit lower mortality than the other products. Surprisingly, mortality rates for the Term Assurance product are lower than the ones for the Annuities product, which is counter-intuitive since individuals signing up to Annuities product are generally expected to live longer. For older ages, the conclusion for the products ACI and SCI is less clear since the variability of their estimates considerably increases in older ages, which results from the very low exposure as visible in Fig. \ref{fig:app_exposure}. 

The posterior probabilities of the model choice variables are shown in Table \ref{tab:productmodel}. The most supported model has the form $$a_{x,p} + b_x (t - \bar{t}) + (k_t + k^2_t c^3_p) + \gamma_{t-x}$$ with a posterior probability of $0.96727$, followed by the model $$a_{x,p} + b_x c^2_p (t - \bar{t}) + (k_t + k^2_t c^3_p) + \gamma_{t-x}$$ with a lower posterior probability of $0.03273$. We note that this result highlights the advantage of using our model selection framework, where the preferred model shares (with high probability) the term $b_x (t- \bar{t})$ across all the different products. By contrast, fitting the model separately for each product would involve specifying a separate term $b_{x,p} (t - \bar{t})$ for each $p$, which may not be the most ideal choice (as in the case here).

\begin{figure}[htbp]
    \centering
    \begin{subfigure}[b]{0.45\textwidth}
        \centering
        \includegraphics[width=\textwidth]{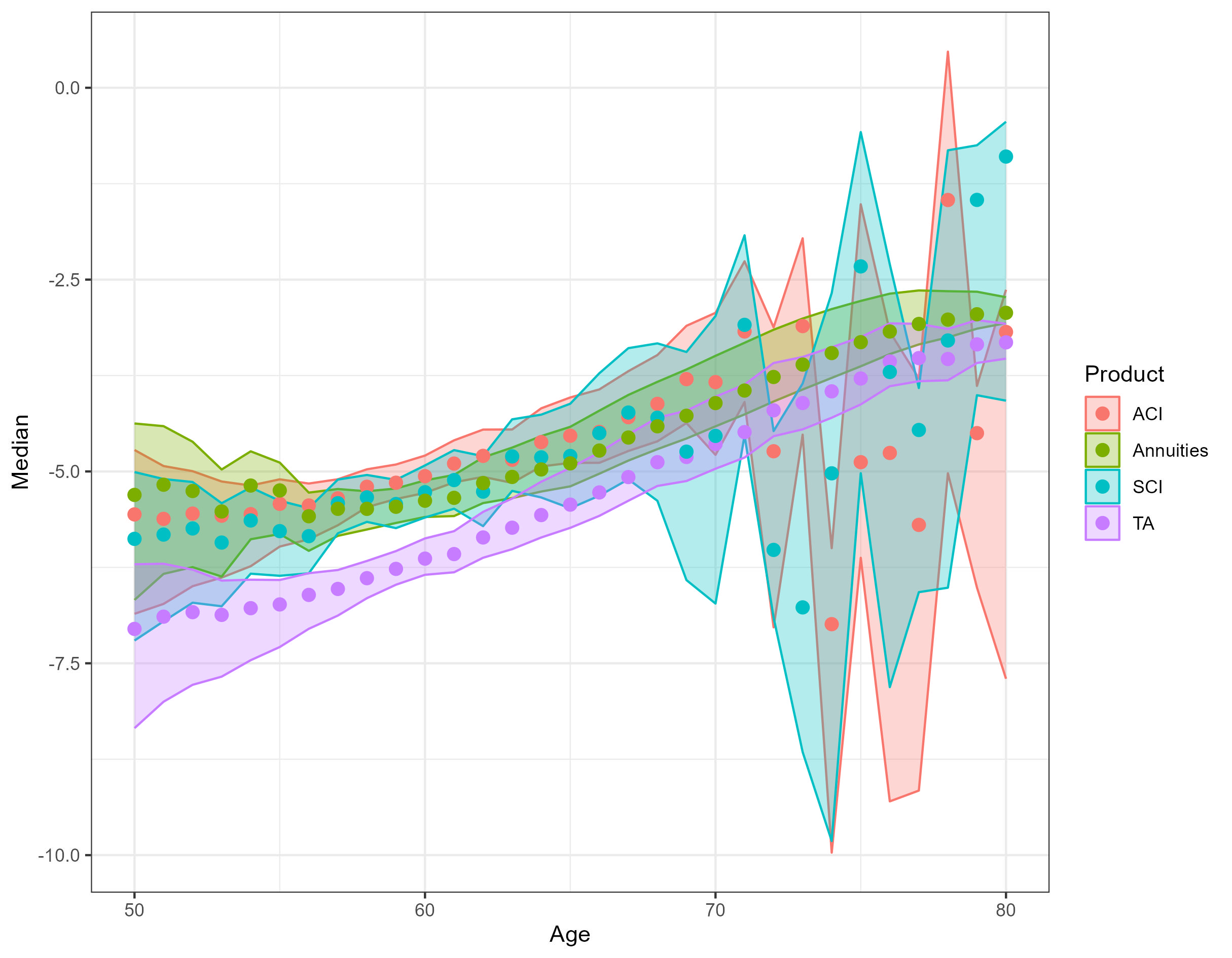} 
        \caption{$95$ \% PCI of the age effets $a_{x,p}$.}
        \label{fig:app_a}
    \end{subfigure}
    \hfill
    \begin{subfigure}[b]{0.45\textwidth}
        \centering
        \includegraphics[width=\textwidth]{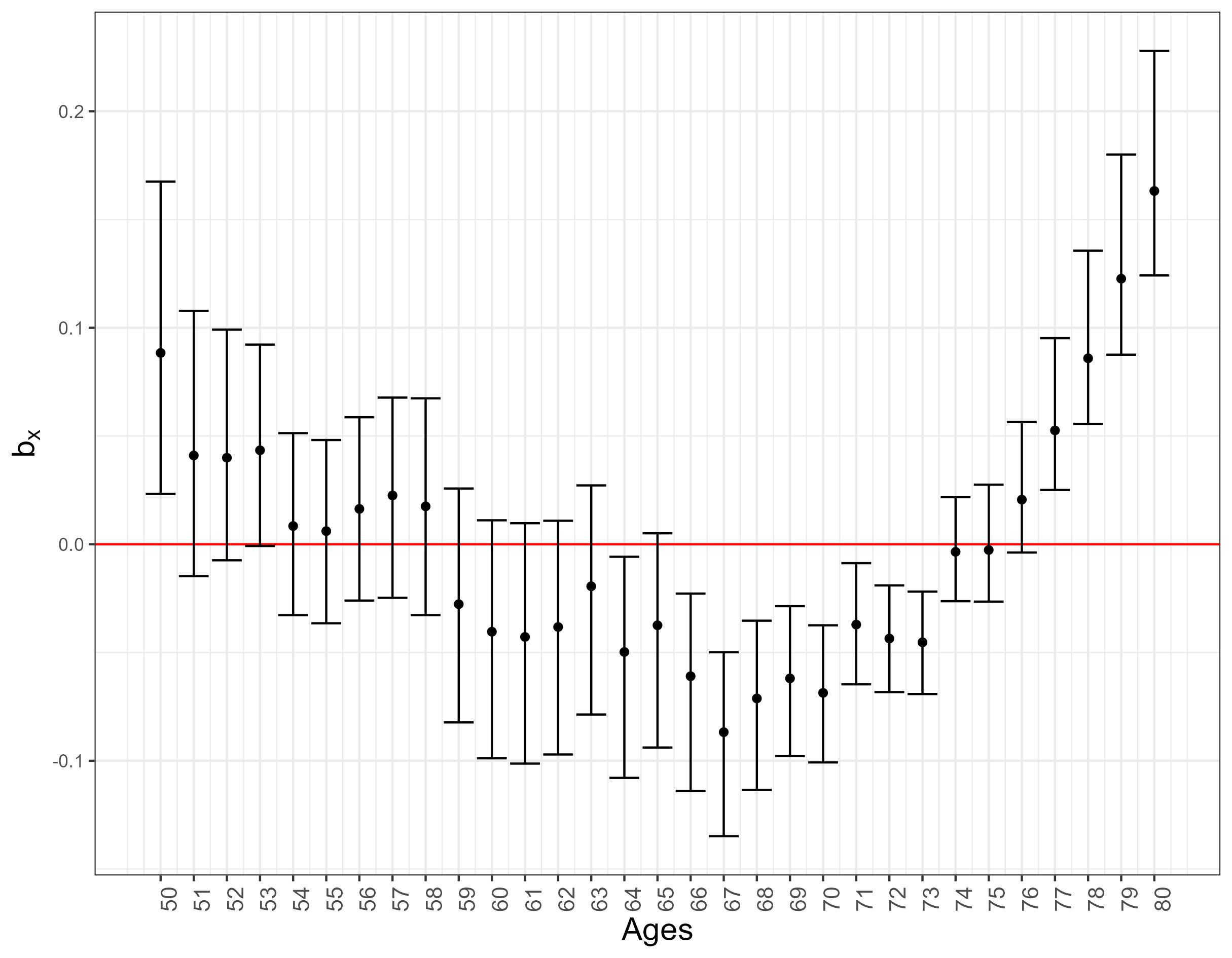} 
        \caption{$95$ \% PCI of the age-year effets $b_{x}$.}
        \label{fig:app_bk}
    \end{subfigure}
    
    \vspace{1em} 
    
    \begin{subfigure}[b]{0.45\textwidth}
        \centering
        \includegraphics[width=\textwidth]{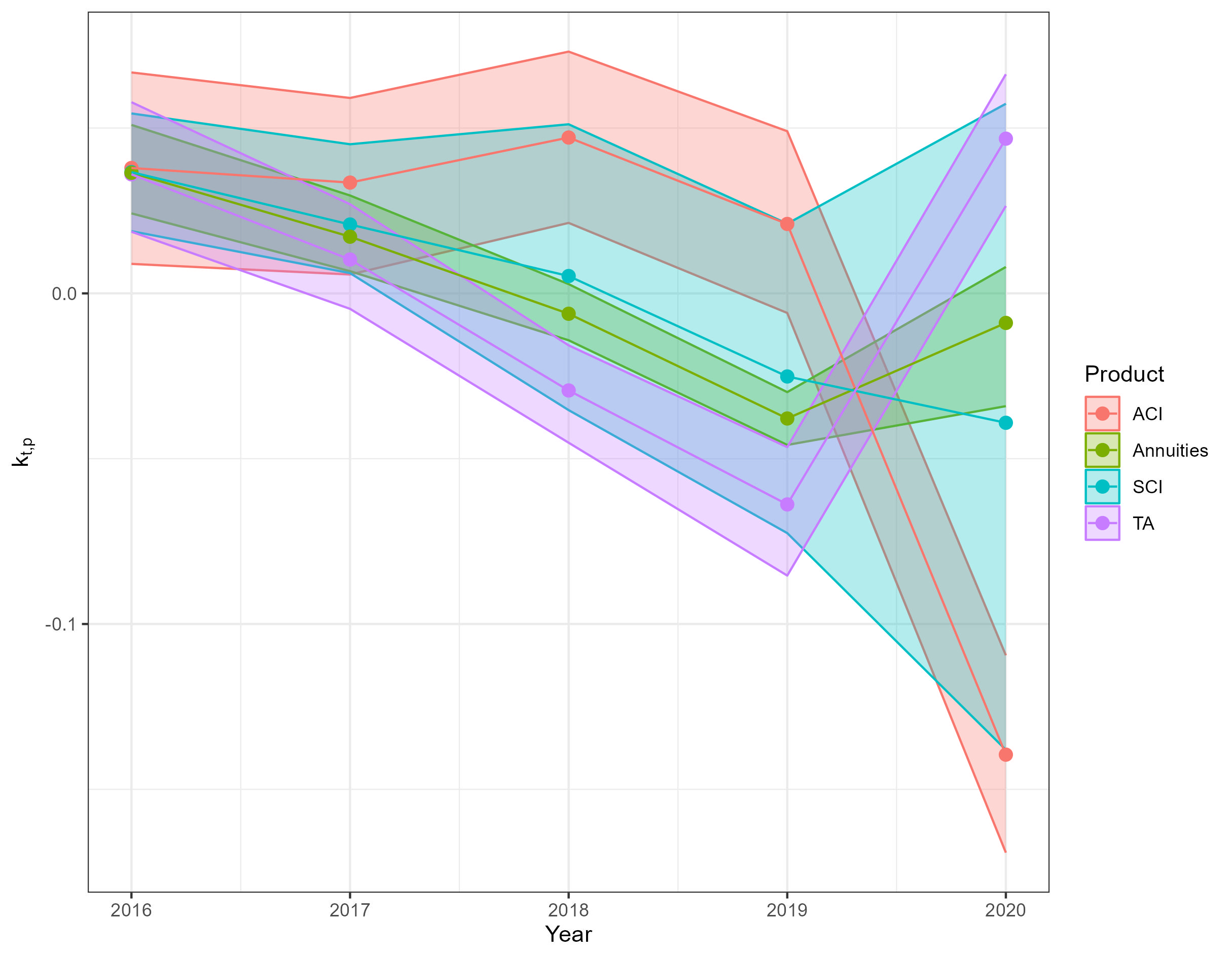} 
        \caption{$95$ \% PCI of the year effect $k({t,p})$.}
        \label{fig:app_kt}
    \end{subfigure}
    \hfill
    \begin{subfigure}[b]{0.45\textwidth}
        \centering
        \includegraphics[width=\textwidth]{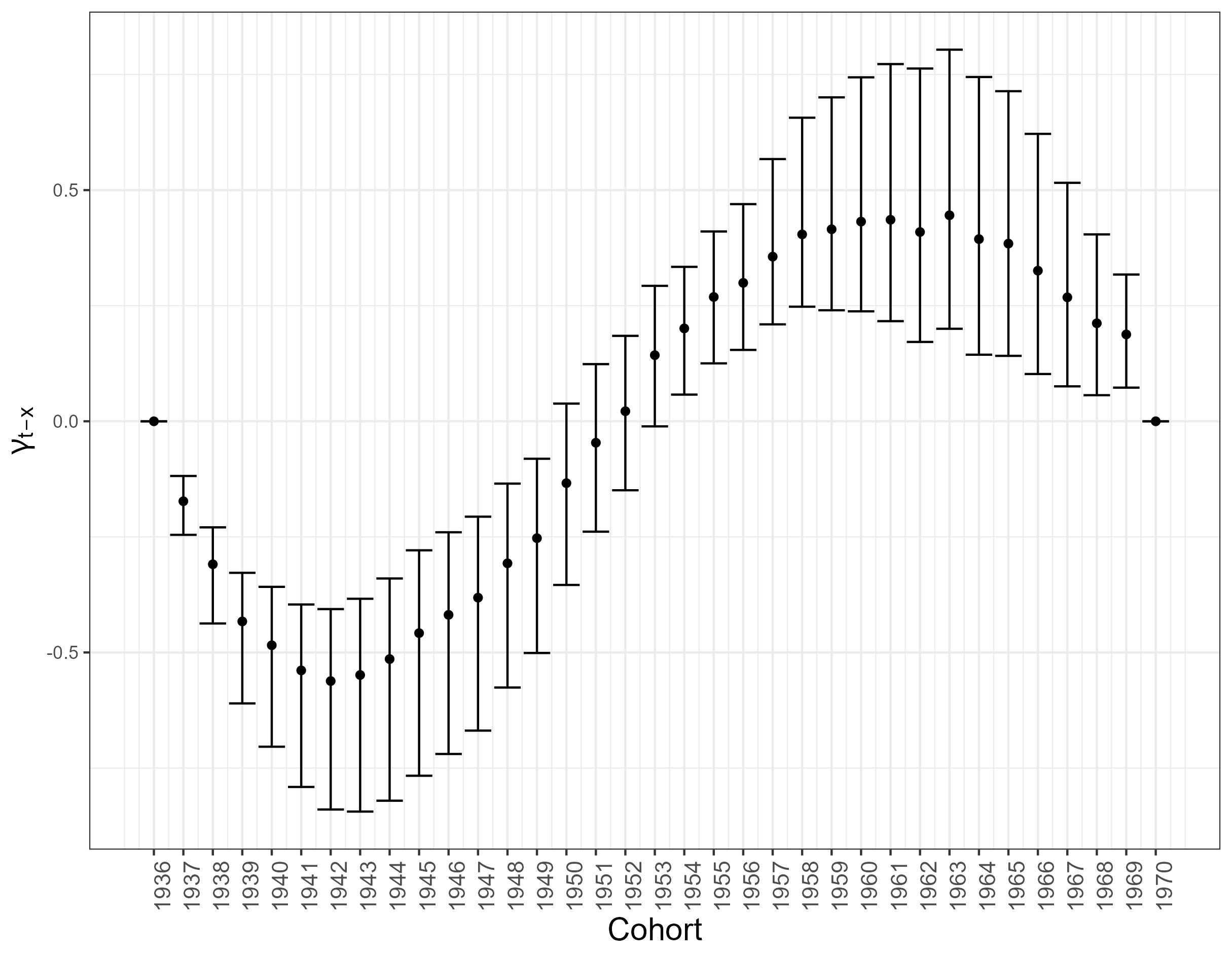} 
        \caption{$95$ \% PCI of the cohort effets $\gamma_{t-x}$.}
        \label{fig:app_gtx}
    \end{subfigure}
    
    \caption{Product insurance data: results}
    \label{fig:app_results2}
\end{figure}

We also present in Figure \ref{fig:app_results2} the posterior distributions for the parameters of the model. In Fig. \ref{fig:app_kt}, we report the posterior credible interval of $k(t,p)$. It can be seen that the the period effect is generally negative, since mortality tends to decrease with time. But for some products the trend is not always present, potentially as a result of the small sample size (only $5$ years of data). In Fig. \ref{fig:app_bk} we present the posterior credible intervals of $b_x$, which represents the interaction between age and period effect, after accounting for age and period effect individually. Lower values of $b_x$ represent large improvement in mortality as period increases, and vice versa. In our results, it can be seen that for older ages, the improvement in mortality is lower. The result seems to suggest that ages between $60$ and $75$ are the ones who are likely to experience the largest improvement in mortality in the long run. However, we suggest analysing the results with care because of the very small number of years used (only $5$ years).

\hspace{-1cm}

\begin{table}[]
\begin{tabular}{|lll|ll|ll|}
\hline
\multicolumn{3}{|l|}{$\delta_1$ }                     & \multicolumn{2}{l|}{$\delta_2$}    & \multicolumn{2}{l|}{$\delta_3$}    \\ \hline
\multicolumn{1}{|l|}{1 $(a_x)$} & \multicolumn{1}{l|}{2 $(a_x + c^1_p)$ } & 3 $(a_{x,p})$ & \multicolumn{1}{l|}{1 ($b_x (t - \bar{t})$)} & 2 ($b_x c^2_p (t - \bar{t})$) & \multicolumn{1}{l|}{1 ($k_t$)} & 2 ($k_t + k^2_t c^3_p$) \\ \hline
\multicolumn{1}{|l|}{0} & \multicolumn{1}{l|}{0} & 1 & \multicolumn{1}{l|}{0.96727} & 0.03273   & \multicolumn{1}{l|}{0} &  1 \\ \hline
\end{tabular}
\caption{Product data: Posterior probabilities on the model choice variables $\delta_i$}
\label{tab:productmodel}
\end{table}

\section{Conclusion}
\label{sec:conclusion}

In this paper, we introduced a novel Bayesian probabilistic framework for mortality model selection using Reversible Jump Markov Chain Monte Carlo (RJMCMC). Our approach addresses key limitations of traditional model selection methods by incorporating model uncertainty directly into the inference process. To apply Bayesian model selection, we have defined two modelling frameworks: for AP data, our model building process is inspired by the different models available in the literature; while for APP data we have defined a framework extending the APCI model.

For AP data, we applied the framework to mortality data of 35 countries in the HMD, focusing on individuals aged 60 to 90 from years 1990 to 2022. The results show that for a substantial number of countries more than one model is supported, reinforcing the importance of taking multiple models into account.  For APP data, we applied the framework to mortality data stratified by different insurance products. The results show large posterior variances for specific combinations of age and years because of data sparsity. This highlights the benefits of a Bayesian approach where uncertainty is properly accounted for when the sample size is low. However, this also suggests that it would be ideal to take full advantage of the Bayesian approach and define a joint prior over different ages and years in order to borrow strengths when data are sparse. 

We emphasise that the ideas laid out in this paper need not be restricted to model selection of AP or APP based data, but can be extended to other types of data, provided a framework encompassing several model is defined. Future extension could include extending the framework for AP data to more complex models. For APP data, since we have worked under the assumption that the stratifying variable is a categorical variable, it would be interesting to consider continuous variables, such as duration.




\begin{acknowledgement}
We thank the CMI for providing the data on the insurance products. We also thank Charlotte Midford who has supported us in promoting this work.
\end{acknowledgement}

\paragraph{Funding Statement}


\paragraph{Competing Interests}



\printbibliography

@article{villegas2014modeling,
  title={On the modeling and forecasting of socioeconomic mortality differentials: An application to deprivation and mortality in England},
  author={Villegas, Andr{\'e}s M and Haberman, Steven},
  journal={North American Actuarial Journal},
  volume={18},
  number={1},
  pages={168--193},
  year={2014},
  publisher={Taylor \& Francis}
}

@article{carter1992modeling,
  title={Modeling and forecasting US sex differentials in mortality},
  author={Carter, Lawrence R and Lee, Ronald D},
  journal={International Journal of forecasting},
  volume={8},
  number={3},
  pages={393--411},
  year={1992},
  publisher={Elsevier}
}

@article{li2005coherent,
  title={Coherent mortality forecasts for a group of populations: An extension of the Lee-Carter method},
  author={Li, Nan and Lee, Ronald},
  journal={Demography},
  volume={42},
  pages={575--594},
  year={2005},
  publisher={Springer}
}

@article{haberman2011comparative,
  title={A comparative study of parametric mortality projection models},
  author={Haberman, Steven and Renshaw, Arthur},
  journal={Insurance: Mathematics and Economics},
  volume={48},
  number={1},
  pages={35--55},
  year={2011},
  publisher={Elsevier}
}

@article{green1995reversible,
  title={Reversible jump Markov chain Monte Carlo computation and Bayesian model determination},
  author={Green, Peter J},
  journal={Biometrika},
  volume={82},
  number={4},
  pages={711--732},
  year={1995},
  publisher={Oxford University Press}
}

@article{cairns2006two,
  title={A two-factor model for stochastic mortality with parameter uncertainty: theory and calibration},
  author={Cairns, Andrew JG and Blake, David and Dowd, Kevin},
  journal={Journal of Risk and Insurance},
  volume={73},
  number={4},
  pages={687--718},
  year={2006},
  publisher={Wiley Online Library}
}

@article{clayton1987models,
  title={Models for temporal variation in cancer rates. II: age--period--cohort models},
  author={Clayton, David and Schifflers, E},
  journal={Statistics in medicine},
  volume={6},
  number={4},
  pages={469--481},
  year={1987},
  publisher={Wiley Online Library}
}

@article{lee1992modeling,
  title={Modeling and forecasting US mortality},
  author={Lee, Ronald D and Carter, Lawrence R},
  journal={Journal of the American statistical association},
  volume={87},
  number={419},
  pages={659--671},
  year={1992},
  publisher={Taylor \& Francis}
}

@article{BARIGOU2022,
title = {Bayesian model averaging for mortality forecasting using leave-future-out validation},
journal = {International Journal of Forecasting},
year = {2022},
doi = {https://doi.org/10.1016/j.ijforecast.2022.01.011},
author = {Karim Barigou and Pierre-Olivier Goffard and Stéphane Loisel and Yahia Salhi}
}

@TECHREPORT{cia_report2024,
    author = {Kai Kaufhold and Martina Brück and Claus Neidhardt},
    title = {Mortality {I}mprovements
{R}esearch},
    type={{Report to the Canadian Institute of
Actuaries’ Project Oversight Group}},
    institution = {Canadian Institute of Actuaries},
    year = {2024}
}

@TECHREPORT{cmib_loglinear,
    author = {{Continuous Mortality Investigation Bureau}},
    title = {{CMI} Mortality Projections Model consultation},
    type={{CMI} {W}orking {P}aper no. 90},
    institution = {London: Institute and Faculty of Actuaries},
    year = {2016}
}

@TECHREPORT{cmib_2019,
    author = {{Continuous Mortality Investigation Bureau}},
    title = {{“All offices” experience of pension
annuities in payment, 2011-2016}},
    type={{CMI} {W}orking {P}aper no. 117},
    institution = {London: Institute and Faculty of Actuaries},
    year = {2019}
}

@TECHREPORT{cmib_2022,
    author = {{Continuous Mortality Investigation Bureau}},
    title = {{“All offices” experience of term assurances in 2020}},
    type={{CMI} {W}orking {P}aper no. 162},
    institution = {London: Institute and Faculty of Actuaries},
    year = {2022}
}

@TECHREPORT{cmib_2023,
    author = {{Continuous Mortality Investigation Bureau}},
    title = {{CMI Mortality Projections Model: CMI\_2023}},
    type={{CMI} {W}orking {P}aper no. 189},
    institution = {London: Institute and Faculty of Actuaries},
    year = {2024}
}

@Article{poissonlca,
  author =       {Brouhns,~N. and Denuit,~M. and Vermunt,~J.~K.},
  title =        {A {P}oisson log-bilinear regression approach to the construction of projected lifetables},
  journal =      IME,
  year =         {2002},
  OPTkey =       {},
  volume =       {{\bf 31}},
  number =       {3},
  OPTmonth =     {},
  pages =        {373-393},
  OPTnote =      {},
  OPTannote =    {}
}

@article{cairns2009quantitative,
  title={A quantitative comparison of stochastic mortality models using data from England and Wales and the United States},
  author={Cairns, Andrew JG and Blake, David and Dowd, Kevin and Coughlan, Guy D and Epstein, David and Ong, Alen and Balevich, Igor},
  journal={North American Actuarial Journal},
  volume={13},
  number={1},
  pages={1--35},
  year={2009},
  publisher={Taylor \& Francis}
}

@Article{richards_apci,
  author =       {Richards,~S.~J. and Currie,~I.~D. and Kleinow,~T. and Ritchie,~G.~P.},
  title =        {{A stochastic implementation of the APCI model for mortality projections}},
  journal =      {British Actuarial Journal,},
	publisher =    {Cambridge University Press},
  year =         {2019},
  OPTkey =       {},
  volume =       {{\bf 24}},
  OPTnumber =       {},
  OPTmonth =     {},
  pages =        {e13},
  OPTnote =      {},
  OPTannote =    {}
}

@article{wong_2023,
    author = {Wong, Jackie S T and Forster, Jonathan J and Smith, Peter W F},
    title = "{Bayesian model comparison for mortality forecasting}",
    journal = {Journal of the Royal Statistical Society Series C: Applied Statistics},
    volume = {72},
    number = {3},
    pages = {566-586},
    year = {2023},
    doi = {10.1093/jrsssc/qlad021}
}

@online{hmd,
  author = {HMD},
  title = {Human Mortality Database. Max Planck Institute for Demographic Research (Germany), University of California, Berkeley (USA), and French Institute for Demographic Studies (France)},
  url = {http://www.mortality.org},
year = {2024},
  note = {Available at www.mortality.org.}
}


\end{document}


\section{Additional plots for AP data results}

\begin{figure}[H]
\label{fig:ap_mortalityrates}
    \centering
    \begin{tabular}{ccc}
        \begin{subfigure}[b]{0.3\textwidth}
            \centering
            \includegraphics[width=\textwidth]{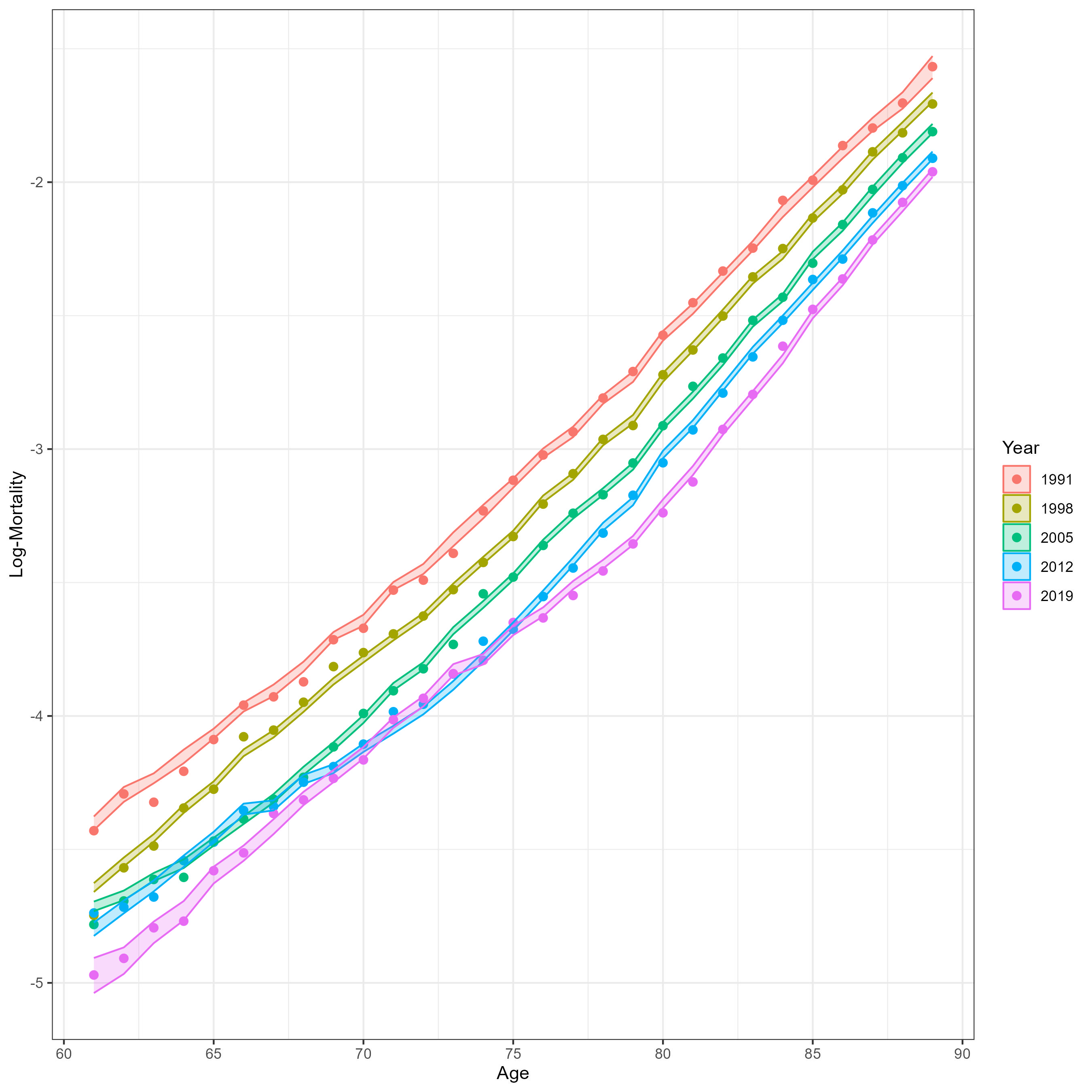} 
            \caption{Austria}
        \end{subfigure} &
        \begin{subfigure}[b]{0.3\textwidth}
            \centering
            \includegraphics[width=\textwidth]{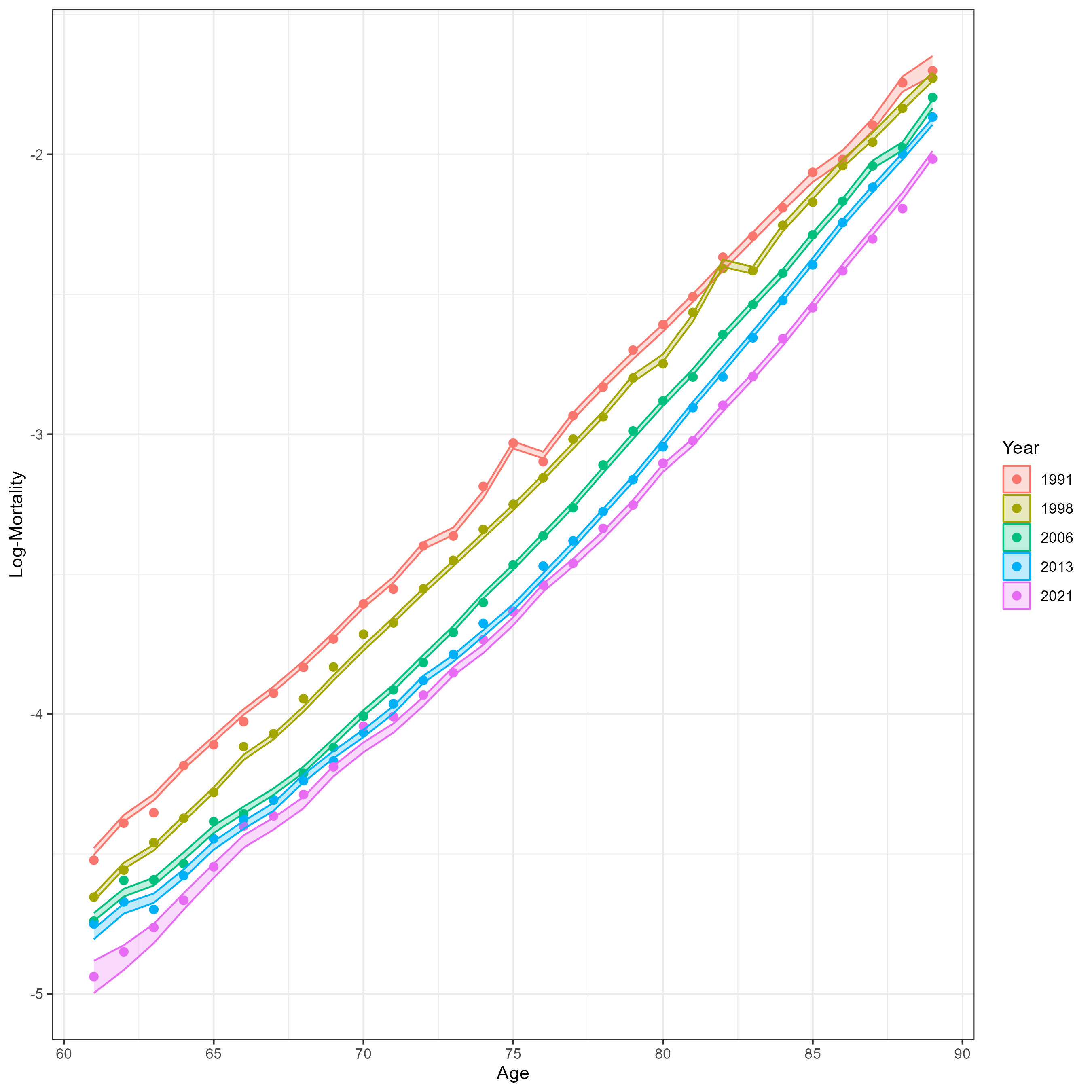} 
            \caption{Belgium}
        \end{subfigure} &
        \begin{subfigure}[b]{0.3\textwidth}
            \centering
            \includegraphics[width=\textwidth]{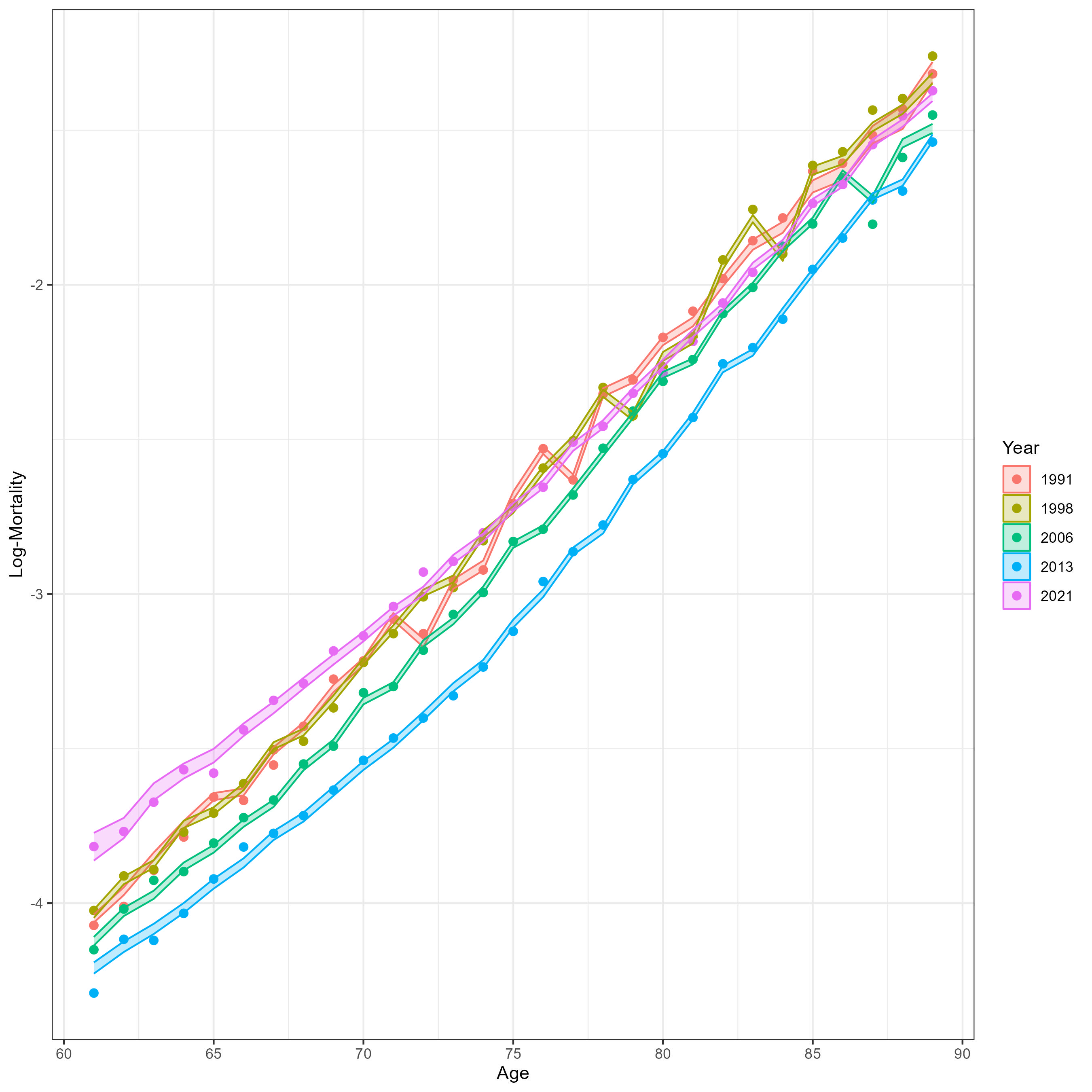} 
            \caption{Bulgaria}
        \end{subfigure} \\
        
        \begin{subfigure}[b]{0.3\textwidth}
            \centering
            \includegraphics[width=\textwidth]{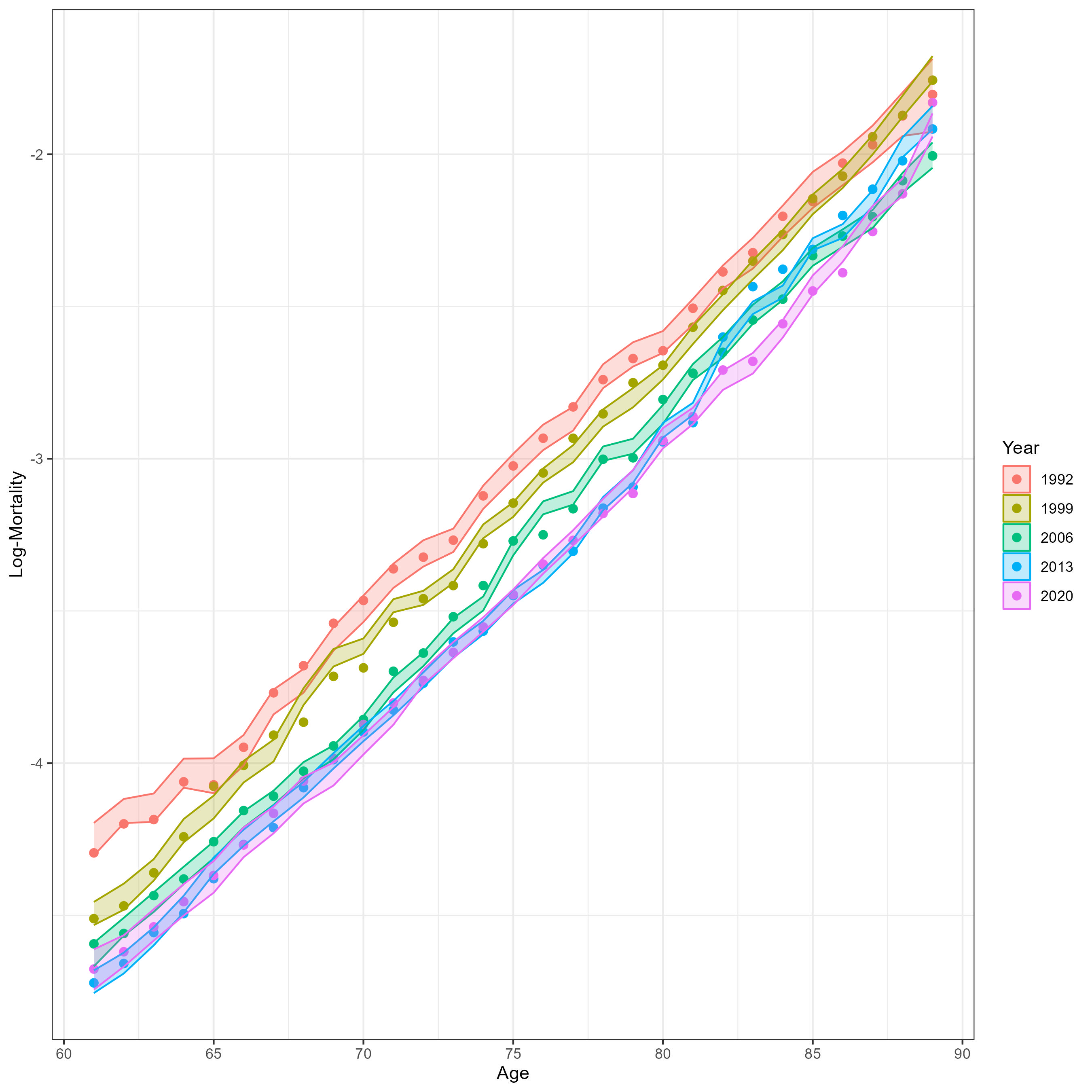} 
            \caption{Chile}
        \end{subfigure} &
        \begin{subfigure}[b]{0.3\textwidth}
            \centering
            \includegraphics[width=\textwidth]{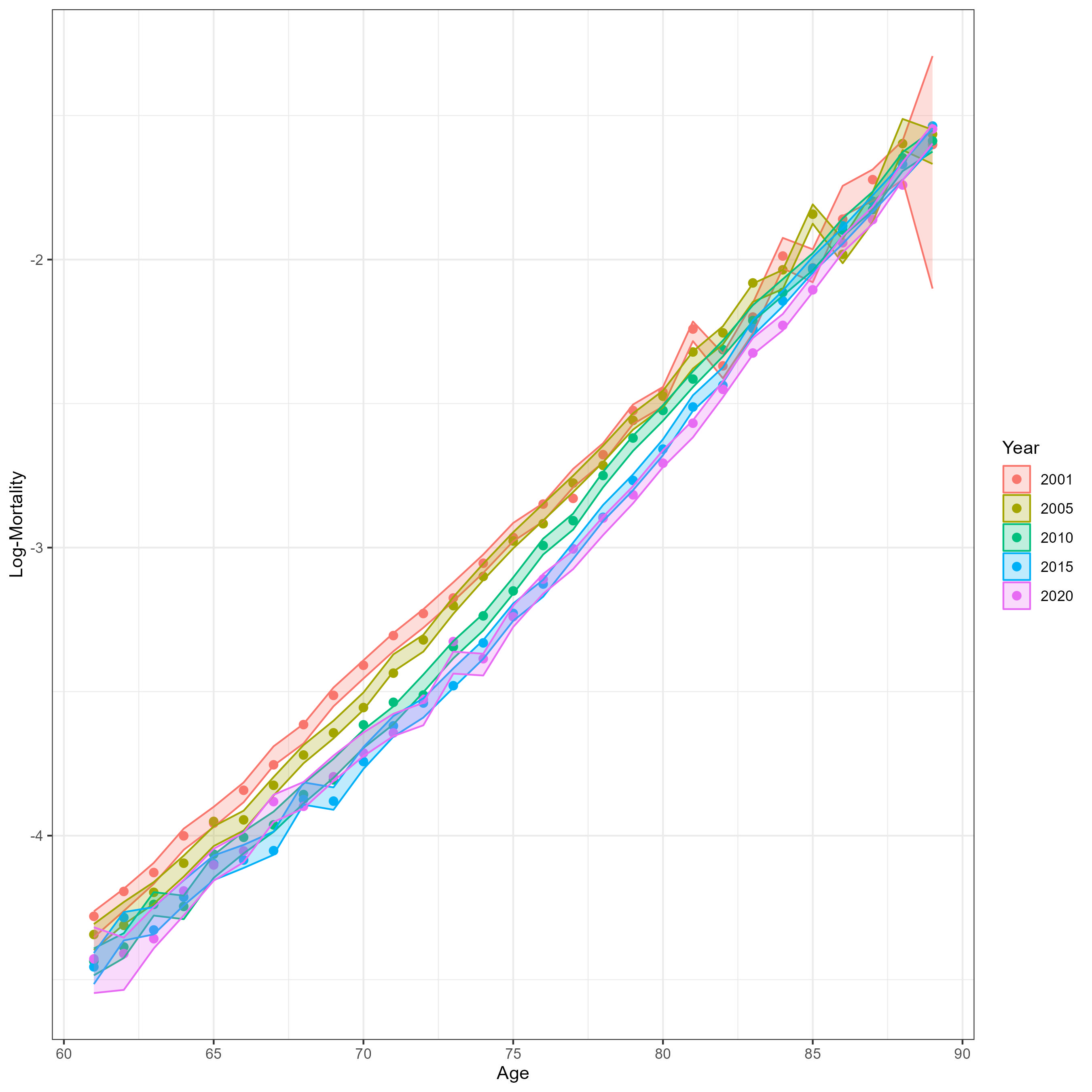} 
            \caption{Croatia}
        \end{subfigure} &
        \begin{subfigure}[b]{0.3\textwidth}
            \centering
            \includegraphics[width=\textwidth]{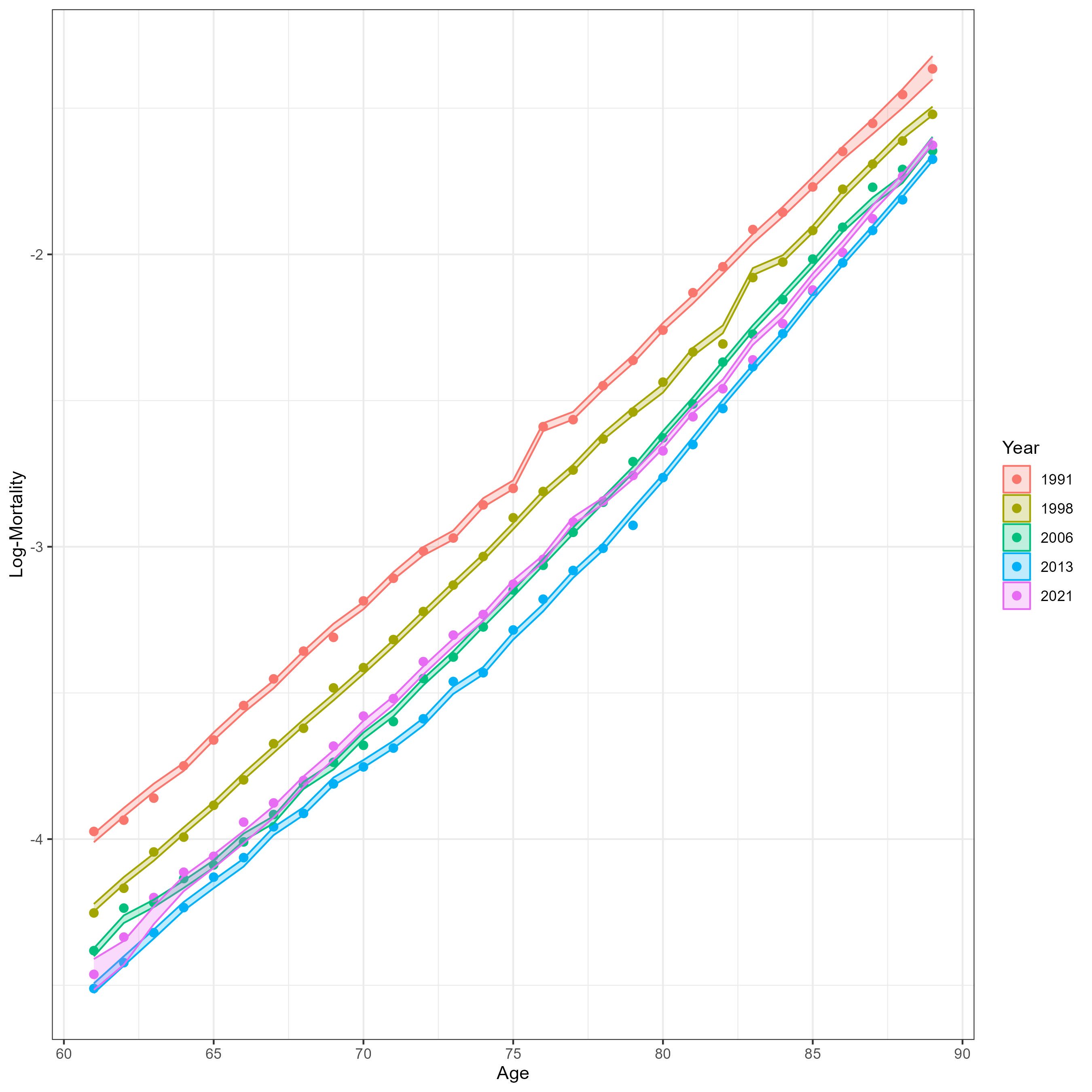} 
            \caption{Czechia}
        \end{subfigure} \\
        
        \begin{subfigure}[b]{0.3\textwidth}
            \centering
            \includegraphics[width=\textwidth]{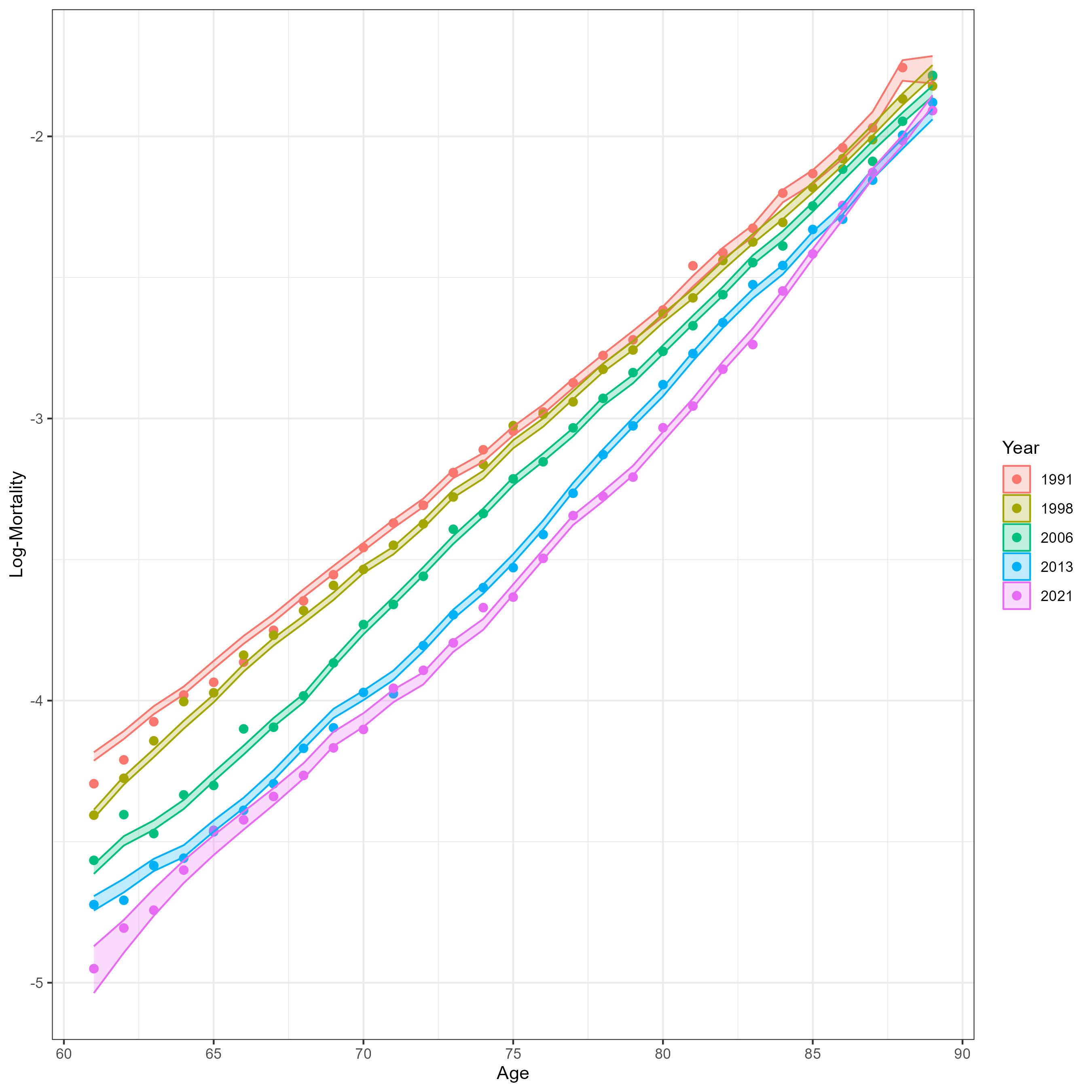} 
            \caption{Denmark}
        \end{subfigure} &
        \begin{subfigure}[b]{0.3\textwidth}
            \centering
            \includegraphics[width=\textwidth]{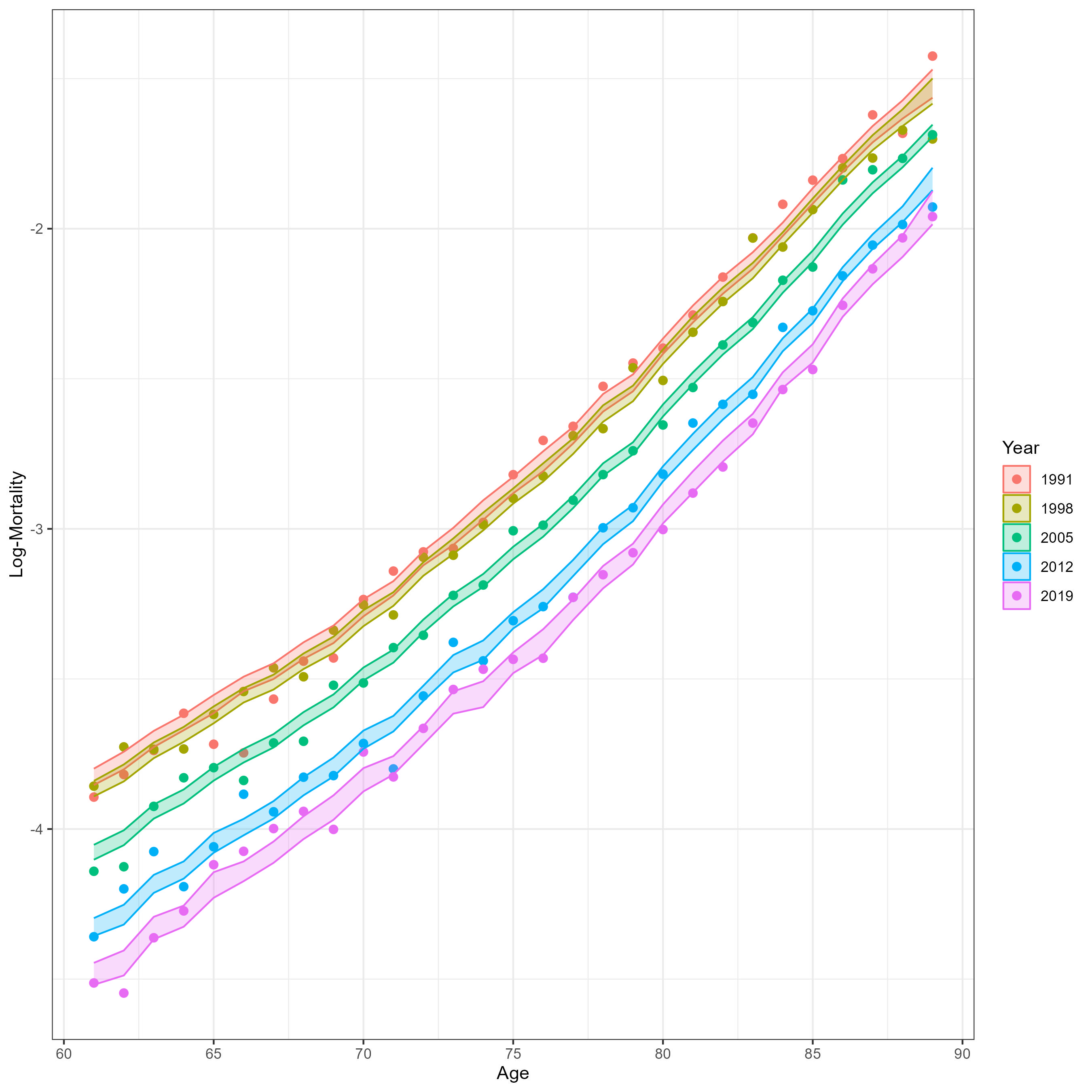} %
            \caption{Estonia}
        \end{subfigure} &
        \begin{subfigure}[b]{0.3\textwidth}
            \centering
            \includegraphics[width=\textwidth]{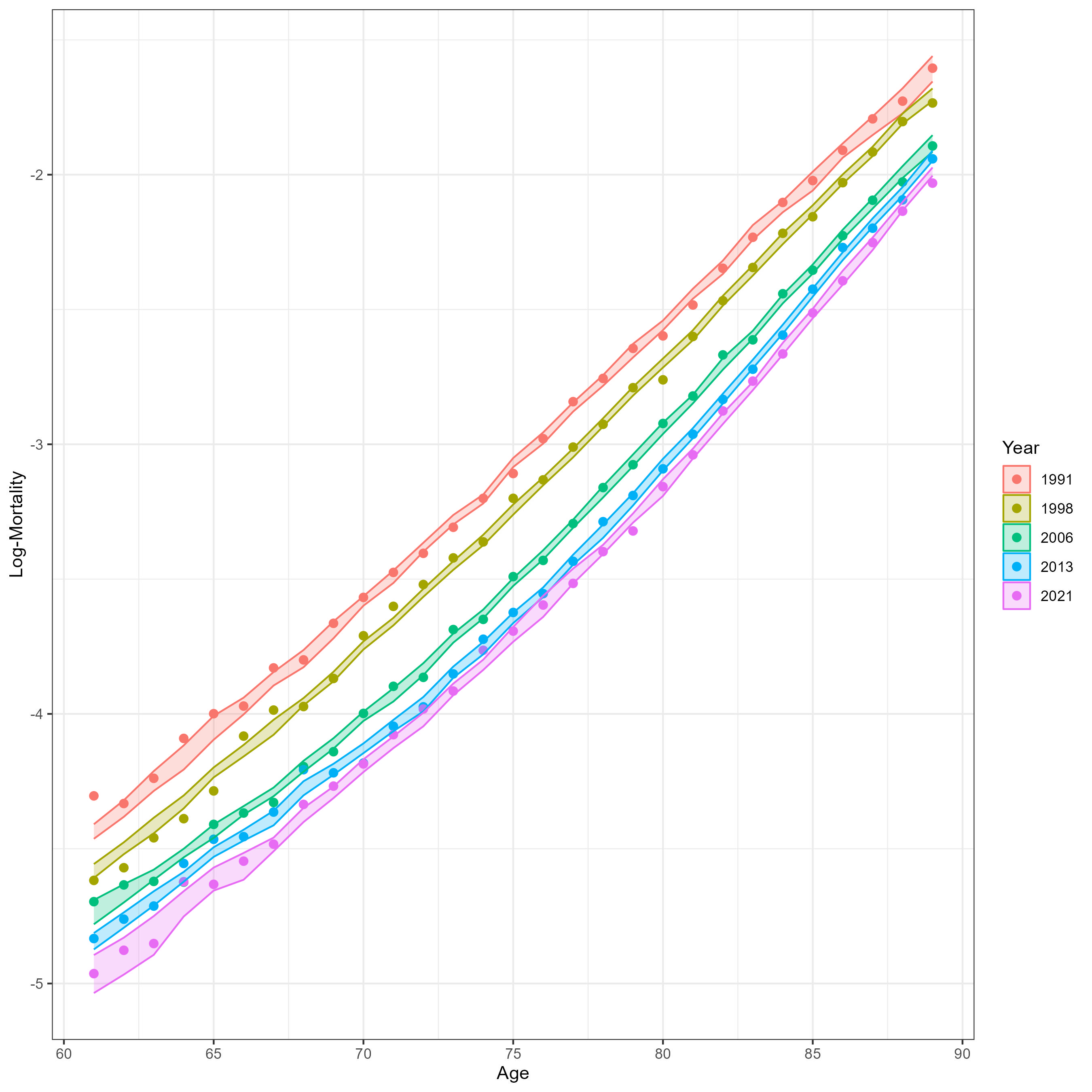} 
            \caption{Finland}
        \end{subfigure}
    \end{tabular}
    \caption{HMD data results: 95\% PCI of the log-mortality rates $m_{x,t}$. The dots represent the crude rates.}
\end{figure}

\begin{figure}[H]
\label{fig:ap_mortalityrates}
    \centering
    \begin{tabular}{ccc}
        \begin{subfigure}[b]{0.3\textwidth}
            \centering
            \includegraphics[width=\textwidth]{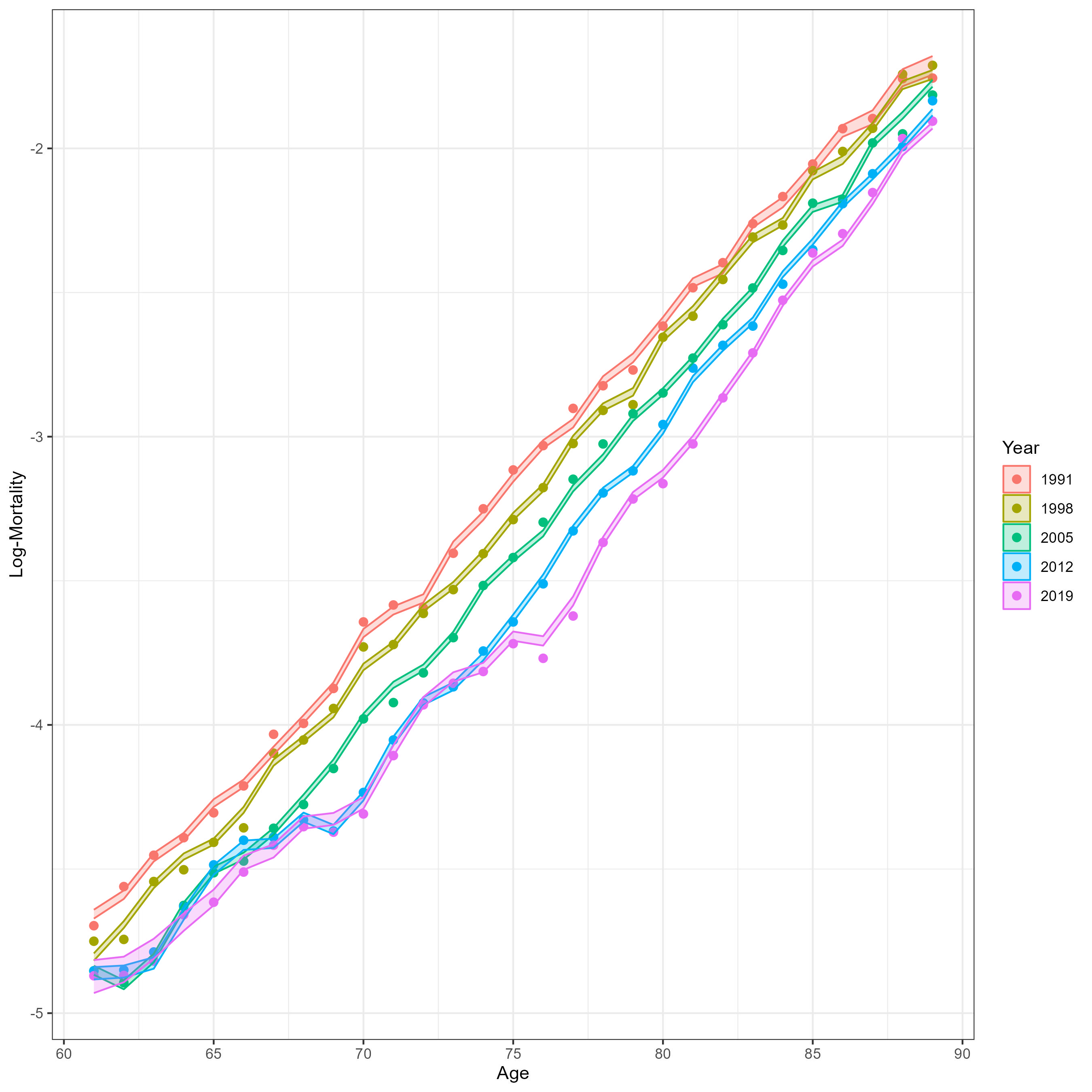} 
            \caption{Greece}
        \end{subfigure} &
        \begin{subfigure}[b]{0.3\textwidth}
            \centering
            \includegraphics[width=\textwidth]{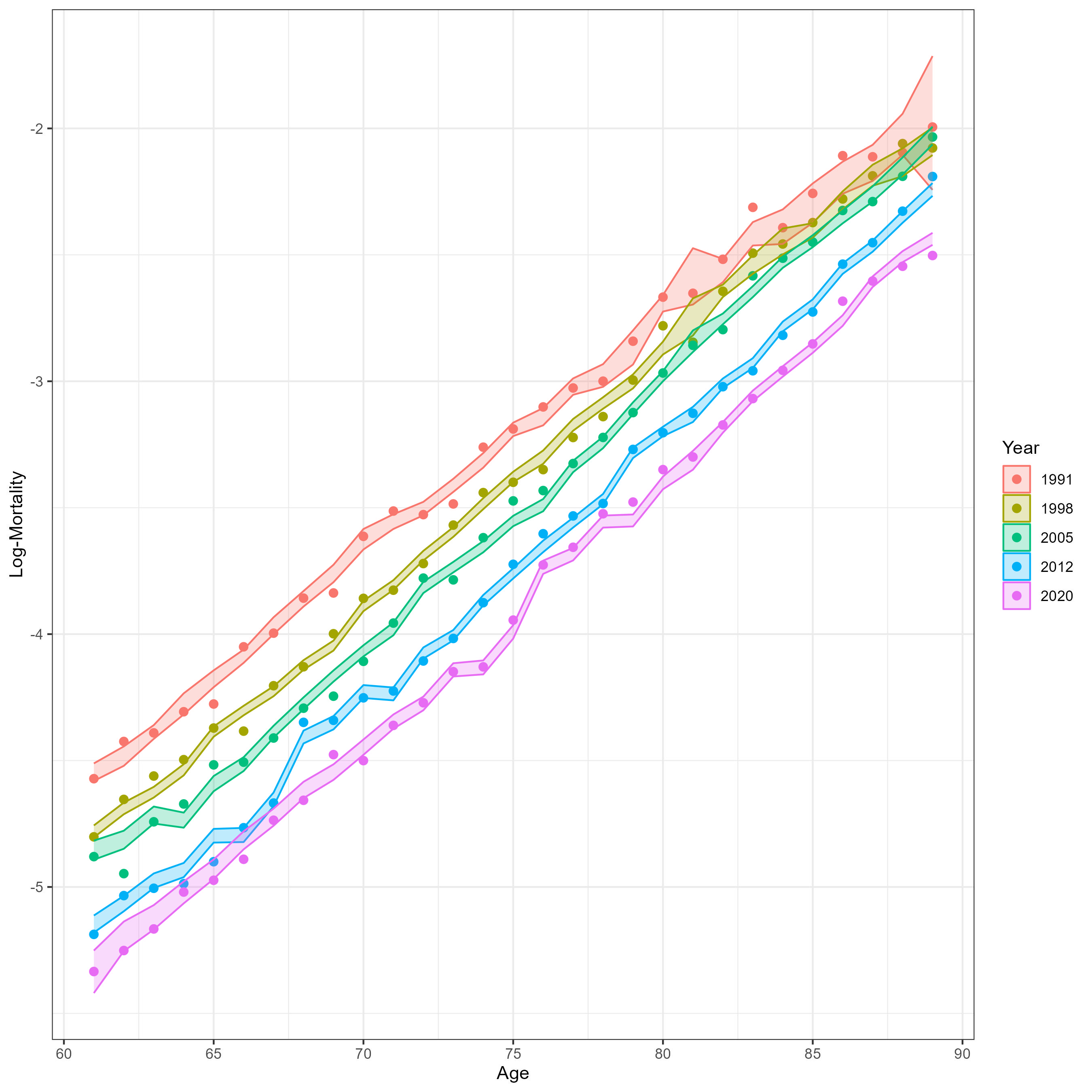} 
            \caption{HK}
        \end{subfigure} &
        \begin{subfigure}[b]{0.3\textwidth}
            \centering
            \includegraphics[width=\textwidth]{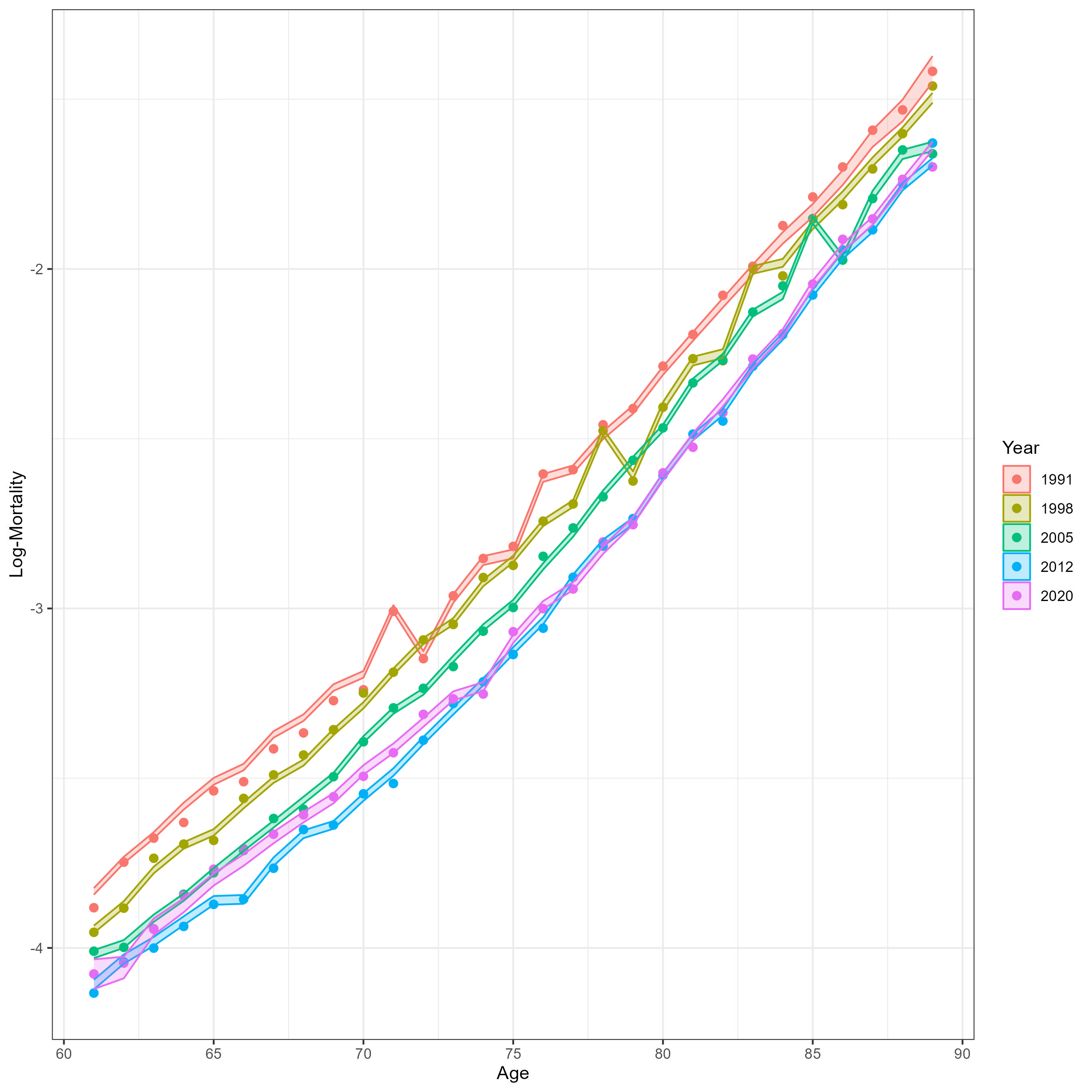} 
            \caption{Hungary}
        \end{subfigure} \\
        
        \begin{subfigure}[b]{0.3\textwidth}
            \centering
            \includegraphics[width=\textwidth]{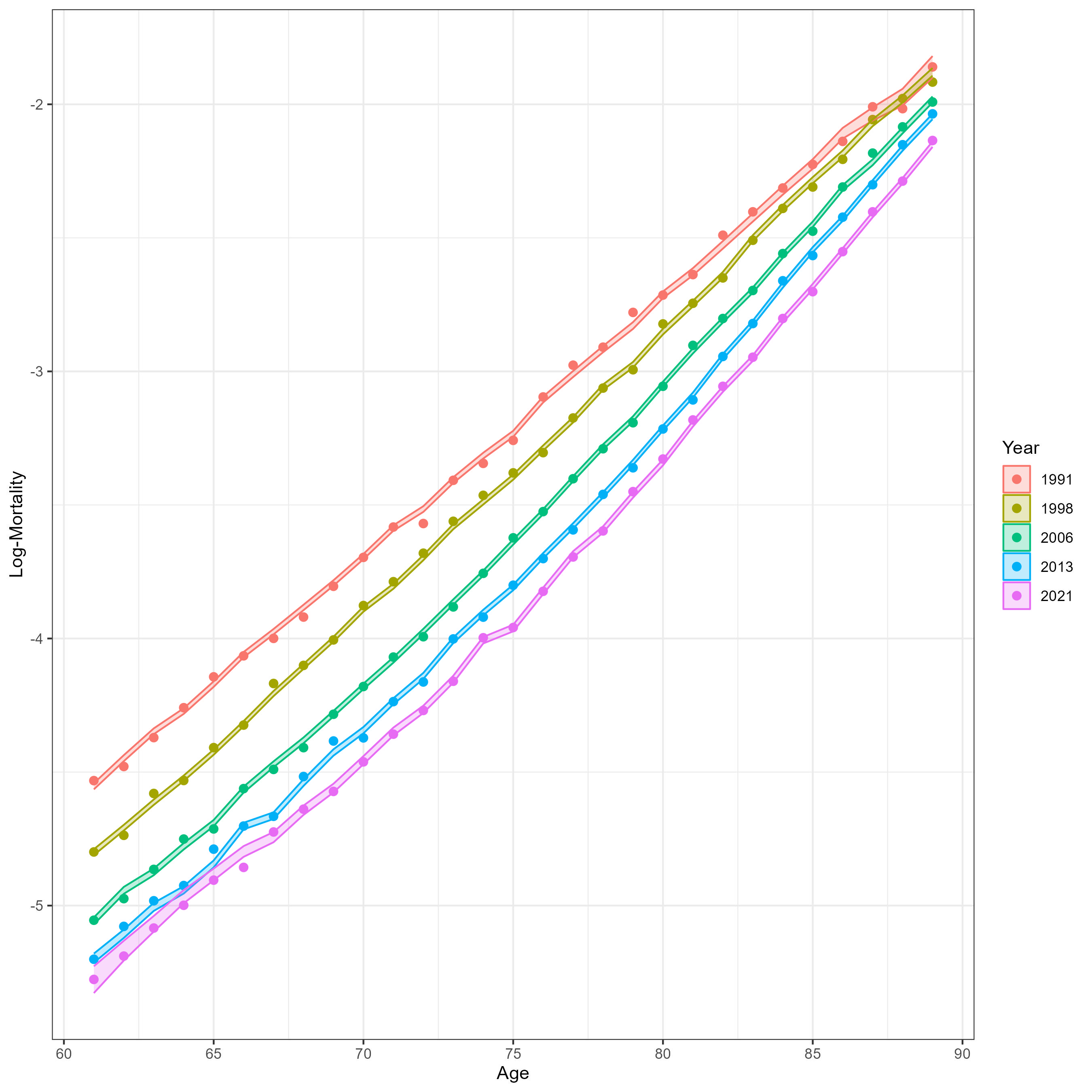} 
            \caption{Australia}
        \end{subfigure} &
        \begin{subfigure}[b]{0.3\textwidth}
            \centering
            \includegraphics[width=\textwidth]{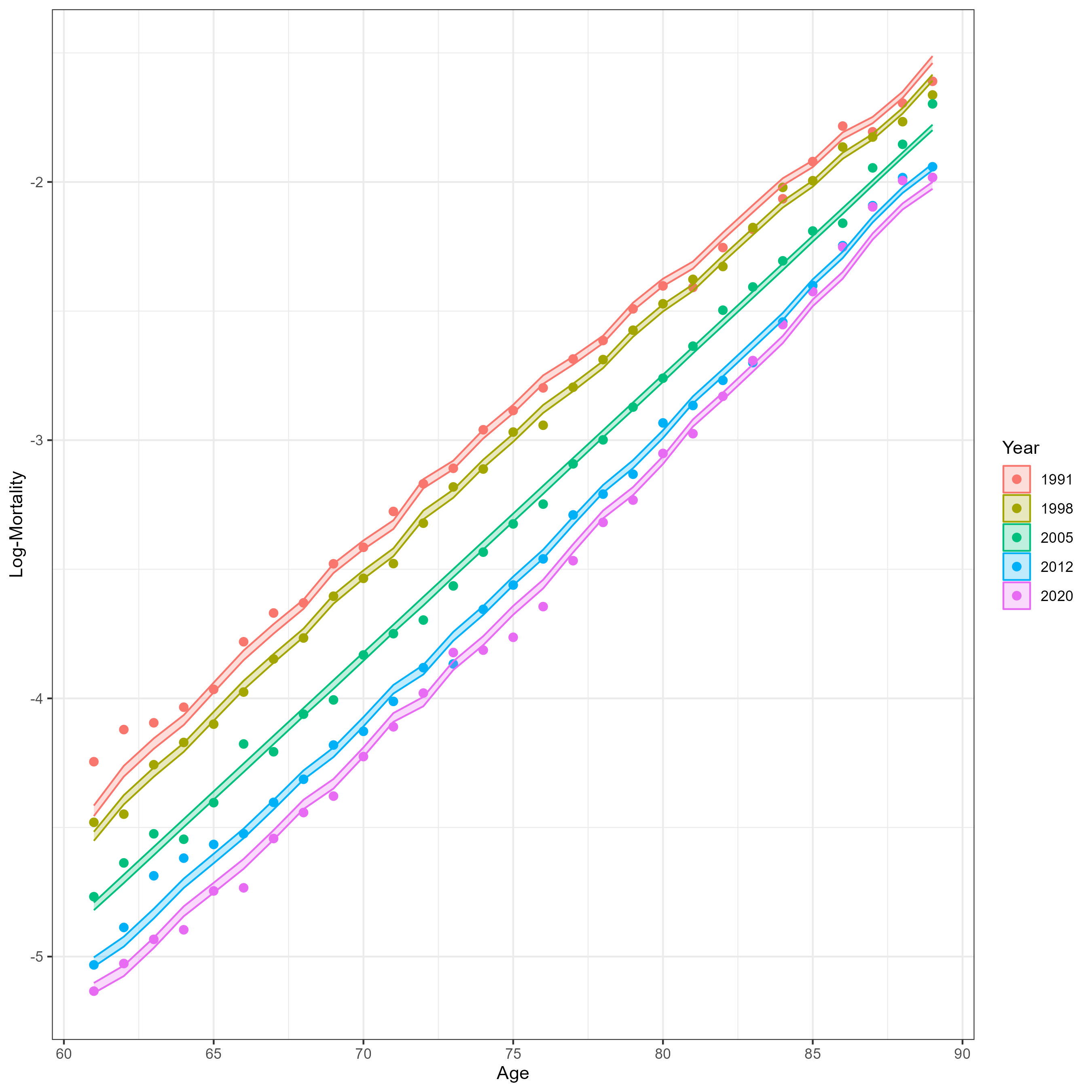} 
            \caption{Ireland}
        \end{subfigure} &
        \begin{subfigure}[b]{0.3\textwidth}
            \centering
            \includegraphics[width=\textwidth]{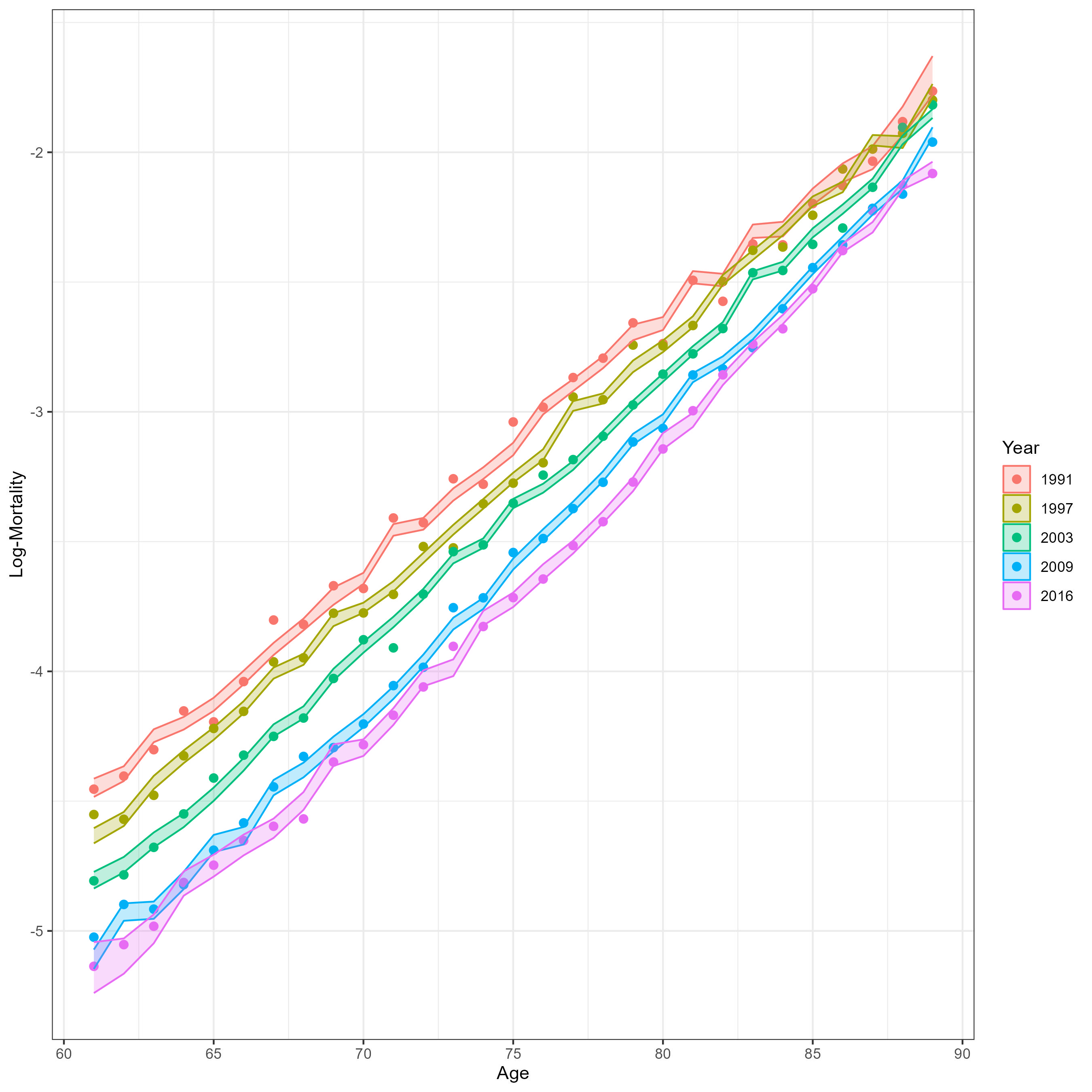} 
            \caption{Israel}
        \end{subfigure} \\
        
        \begin{subfigure}[b]{0.3\textwidth}
            \centering
            \includegraphics[width=\textwidth]{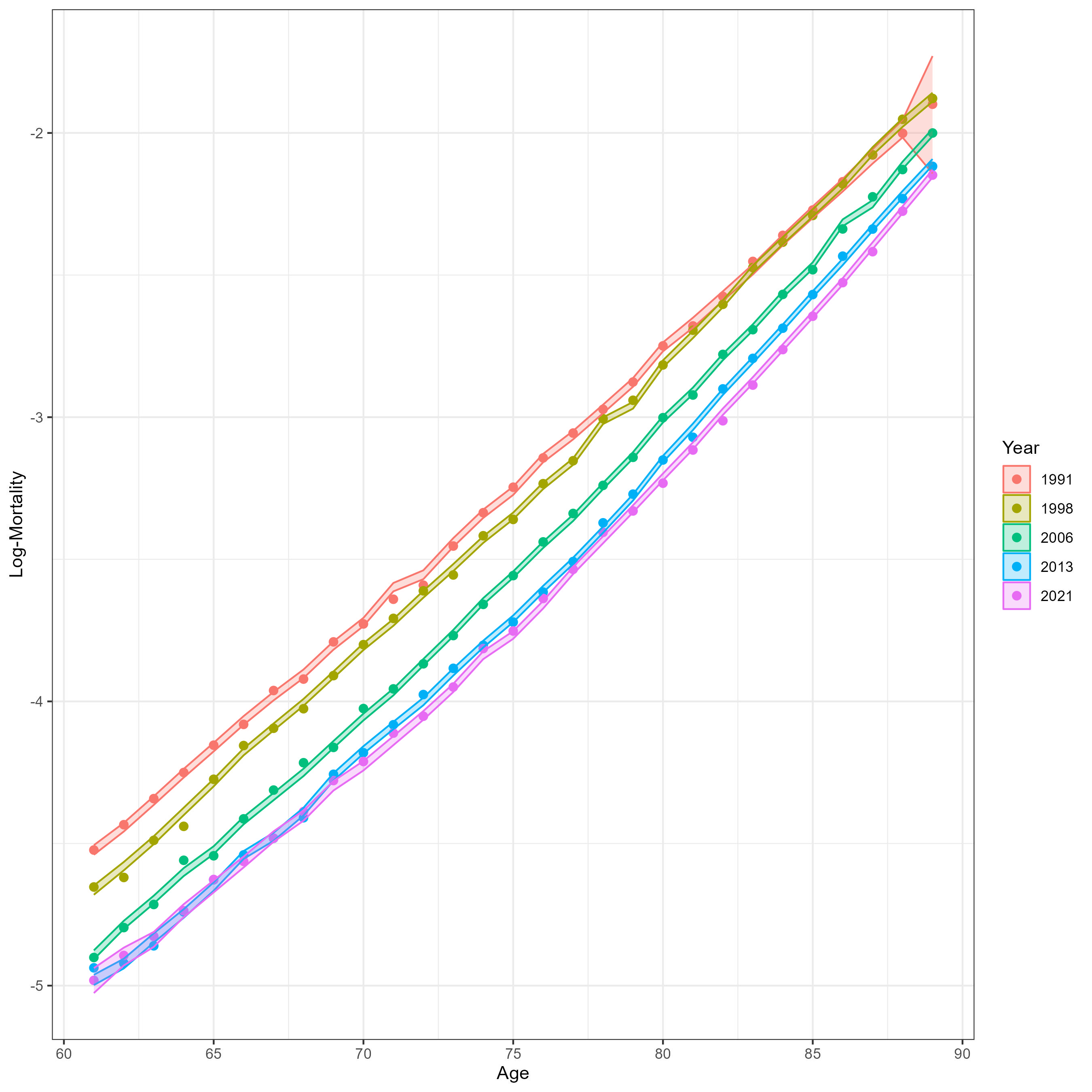} 
            \caption{Canada}
        \end{subfigure} &
        \begin{subfigure}[b]{0.3\textwidth}
            \centering
            \includegraphics[width=\textwidth]{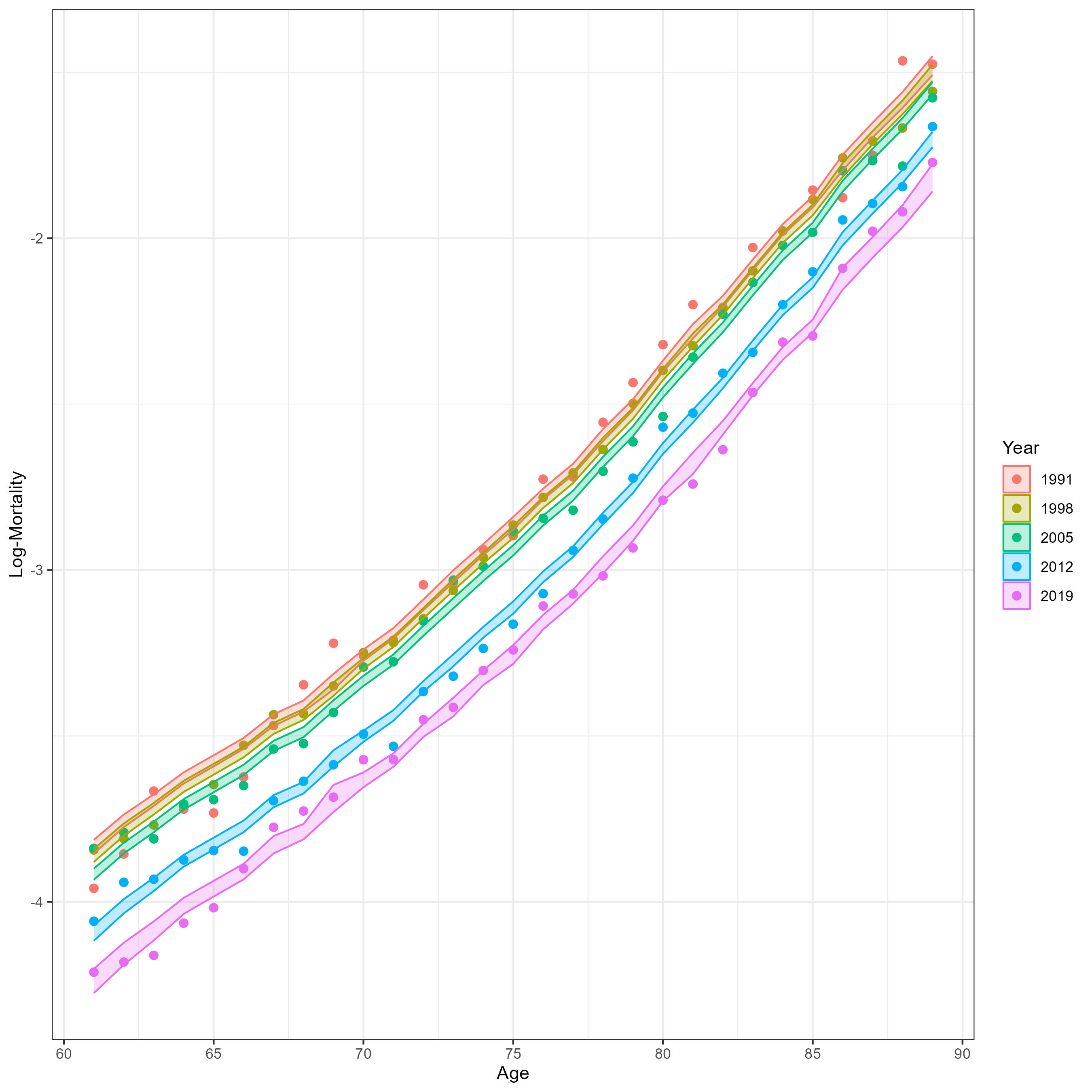} %
            \caption{Latvia}
        \end{subfigure} &
        \begin{subfigure}[b]{0.3\textwidth}
            \centering
            \includegraphics[width=\textwidth]{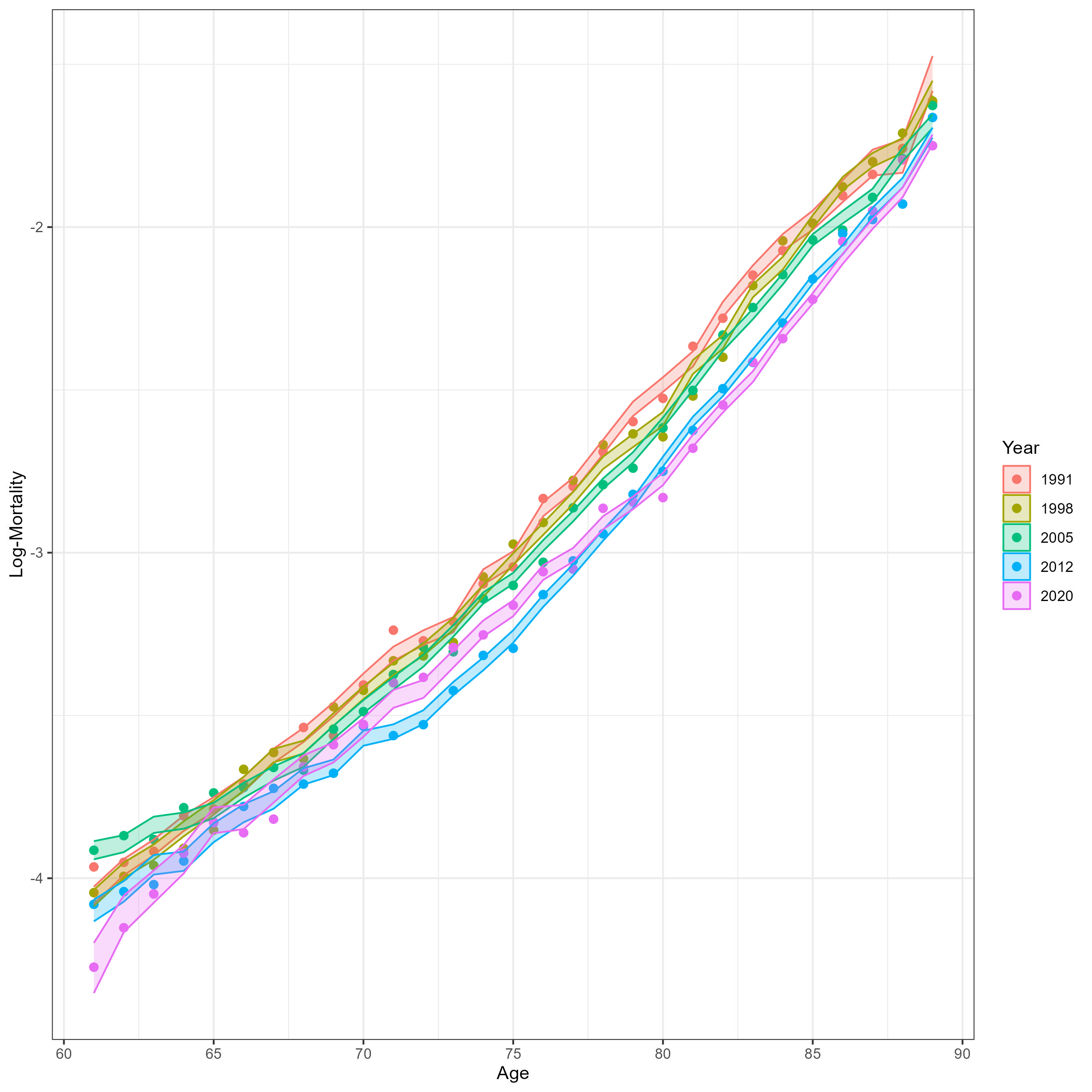} 
            \caption{Lithuania}
        \end{subfigure}
    \end{tabular}
    \caption{HMD data results: 95\% PCI of the log-mortality rates $m_{x,t}$. The dots represent the crude rates.}
\end{figure}

\begin{figure}[H]
\label{fig:ap_mortalityrates}
    \centering
    \begin{tabular}{ccc}
        \begin{subfigure}[b]{0.3\textwidth}
            \centering
            \includegraphics[width=\textwidth]{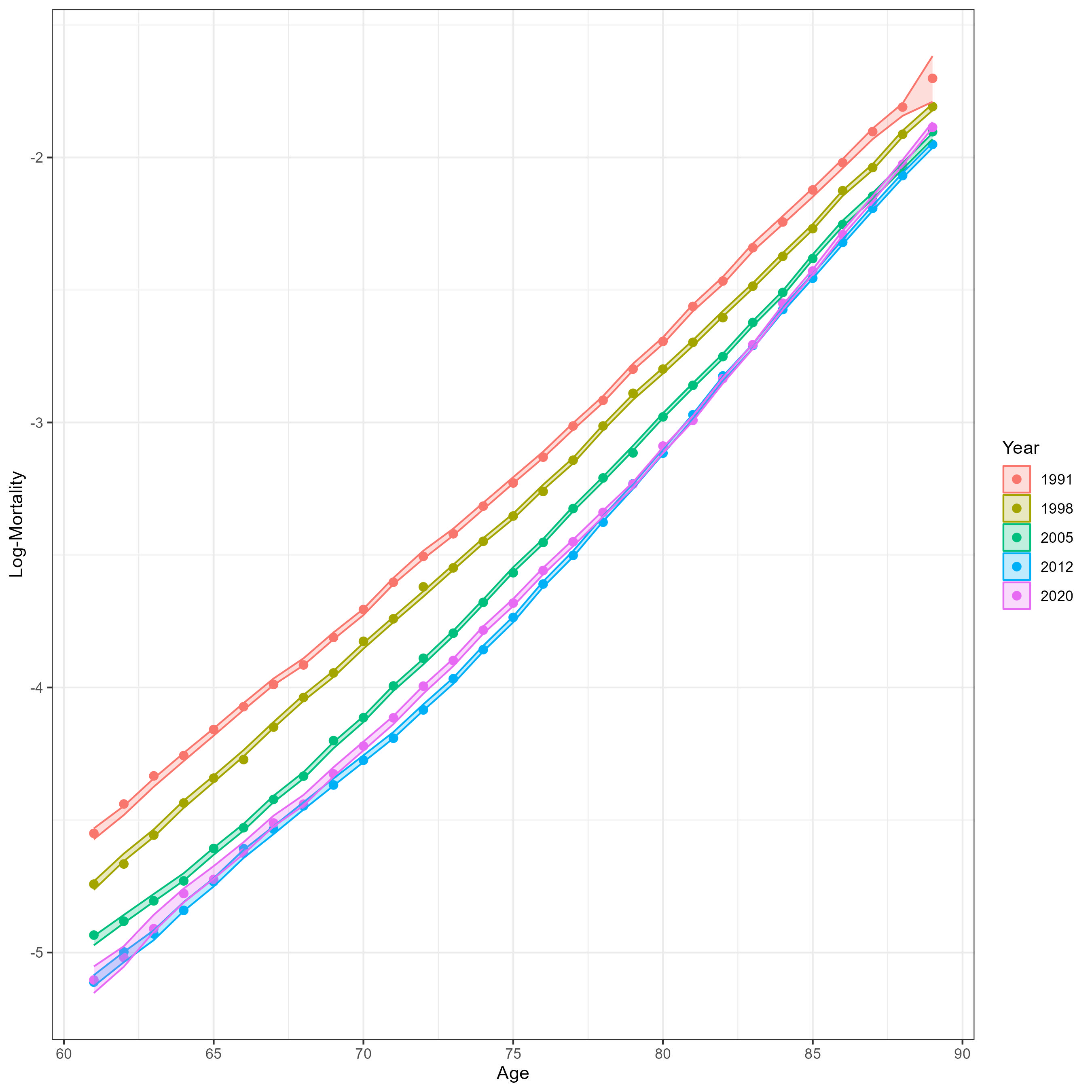} 
            \caption{Italy}
        \end{subfigure} &
        \begin{subfigure}[b]{0.3\textwidth}
            \centering
            \includegraphics[width=\textwidth]{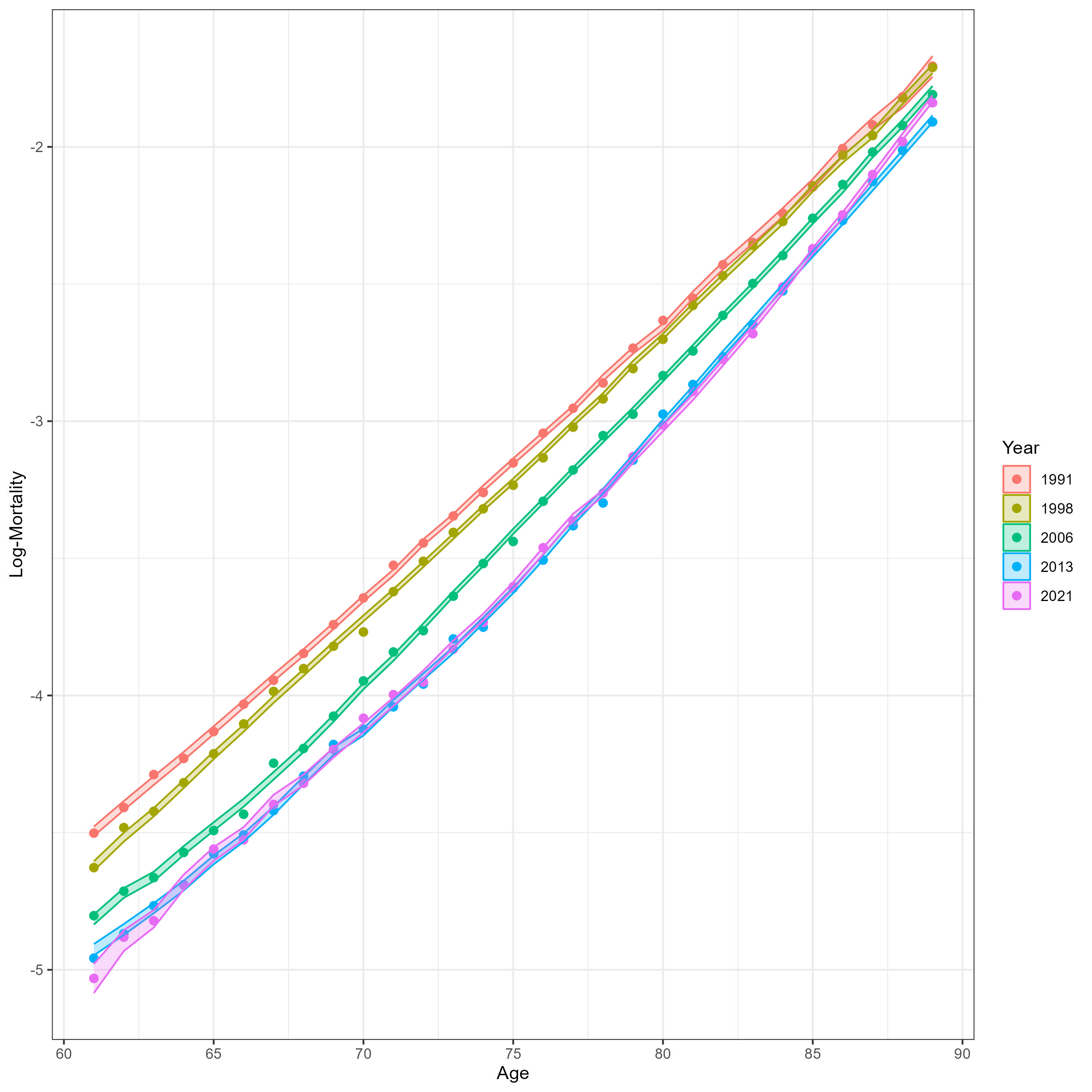} 
            \caption{Netherlands}
        \end{subfigure} &
        \begin{subfigure}[b]{0.3\textwidth}
            \centering
            \includegraphics[width=\textwidth]{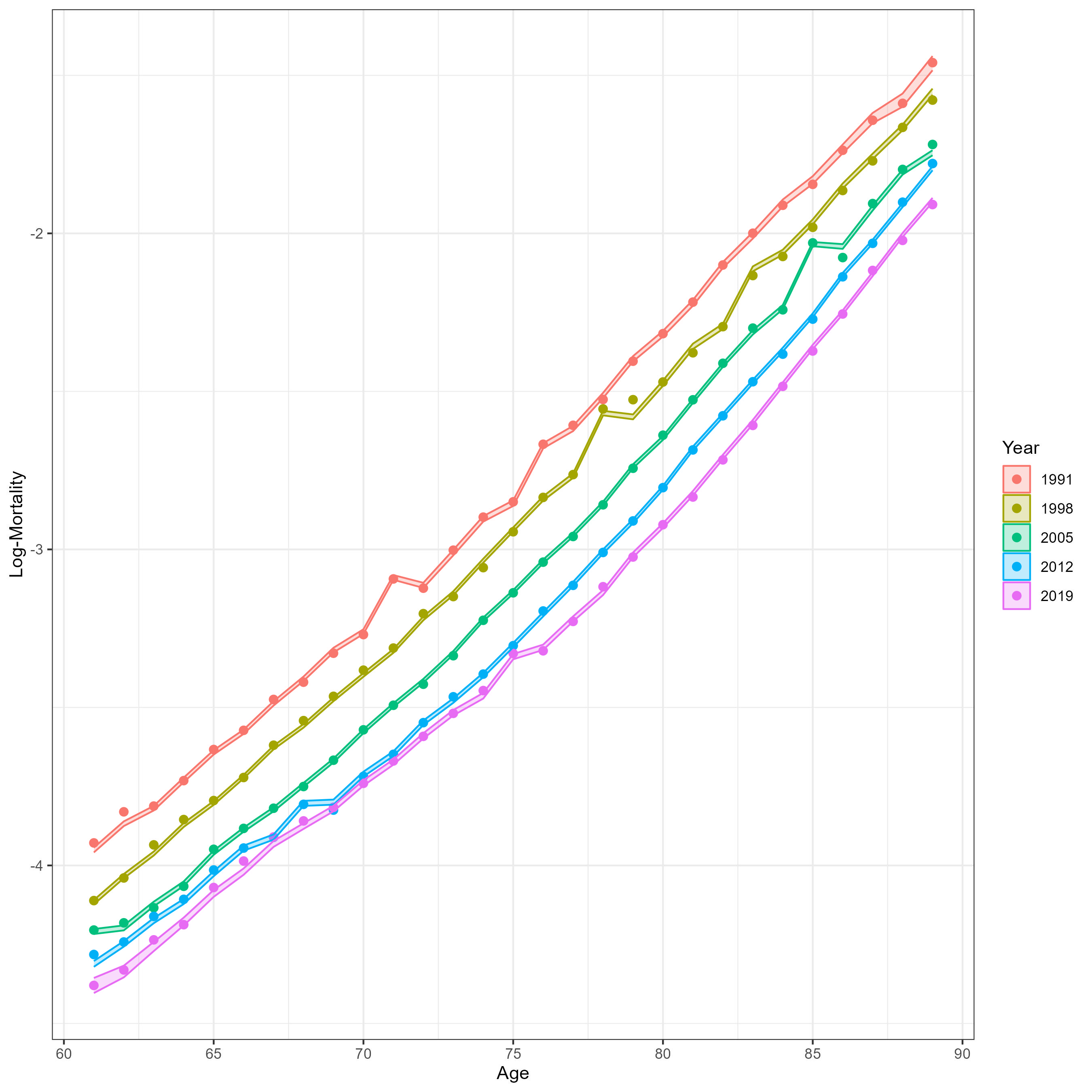} 
            \caption{Poland}
        \end{subfigure} \\
        
        \begin{subfigure}[b]{0.3\textwidth}
            \centering
            \includegraphics[width=\textwidth]{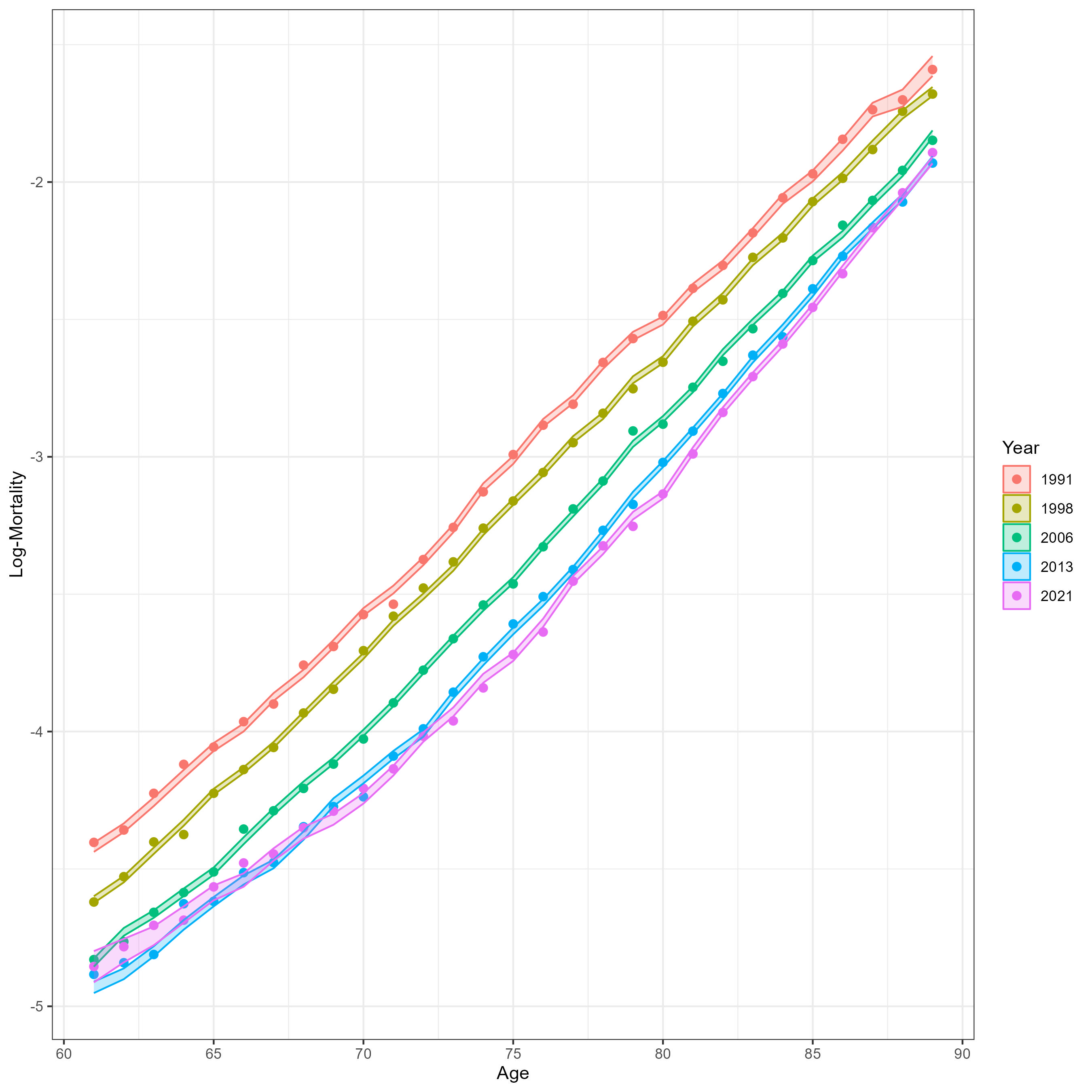} 
            \caption{Portugal}
        \end{subfigure} &
        \begin{subfigure}[b]{0.3\textwidth}
            \centering
            \includegraphics[width=\textwidth]{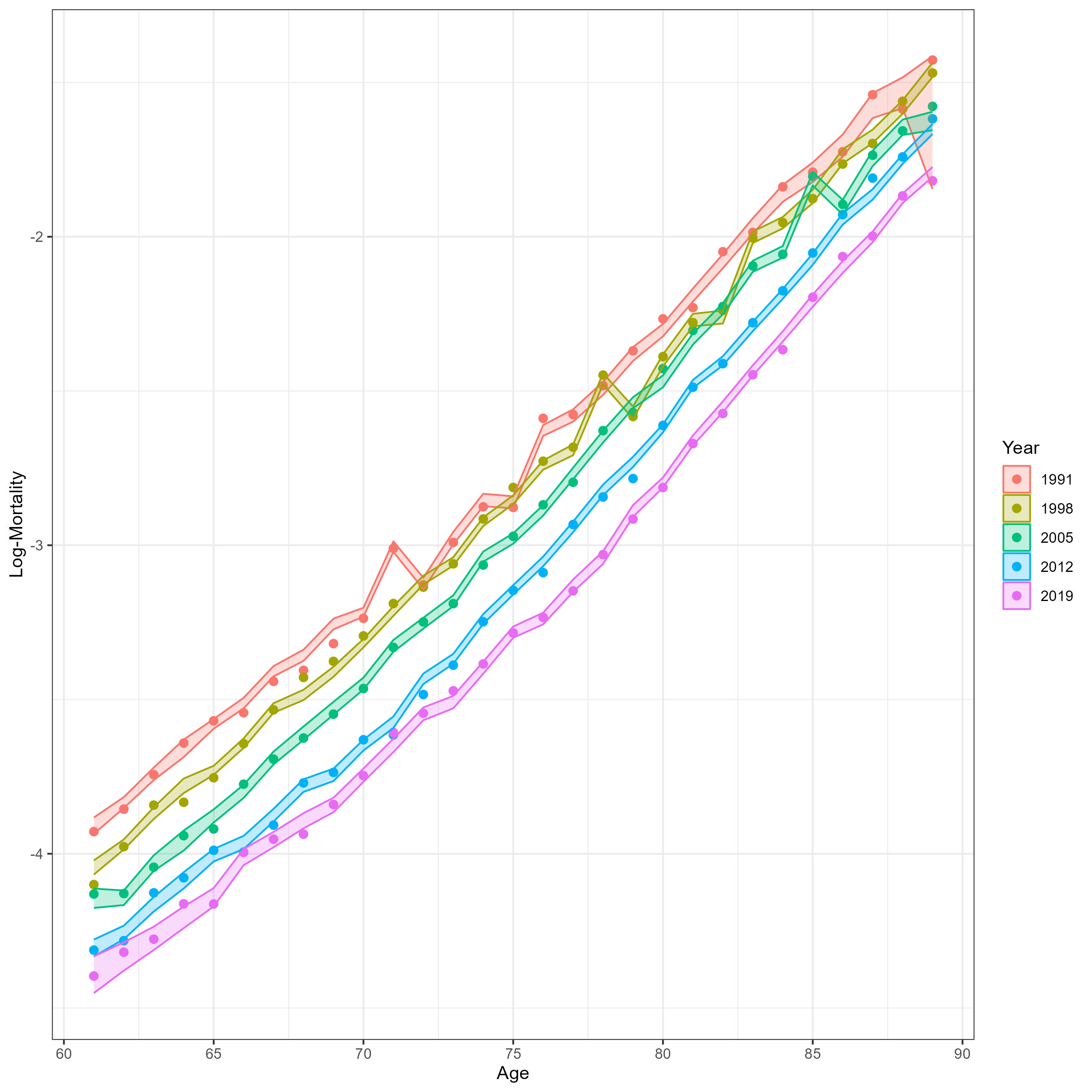} 
            \caption{Slovakia}
        \end{subfigure} &
        \begin{subfigure}[b]{0.3\textwidth}
            \centering
            \includegraphics[width=\textwidth]{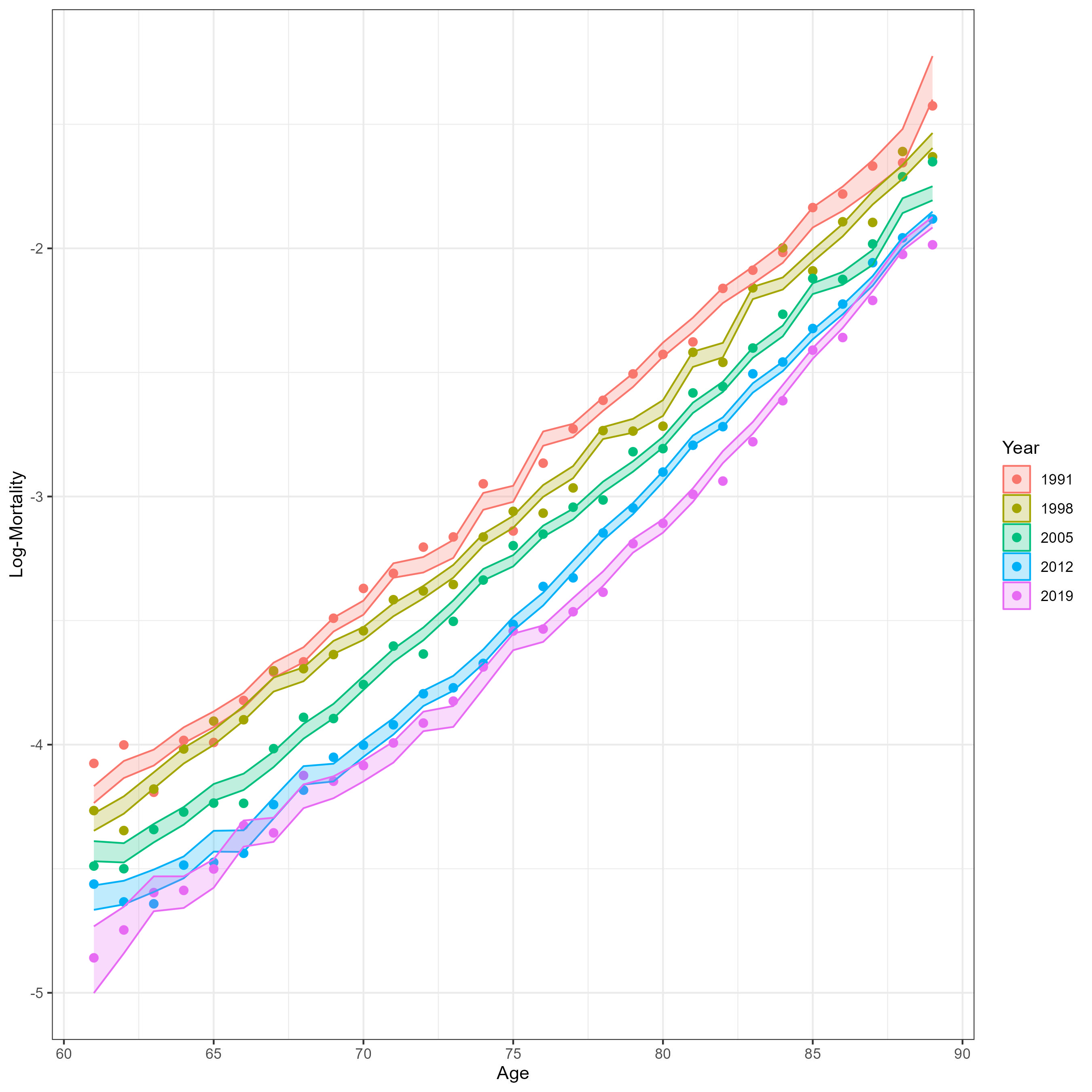} 
            \caption{Slovenia}
        \end{subfigure} \\
        
        \begin{subfigure}[b]{0.3\textwidth}
            \centering
            \includegraphics[width=\textwidth]{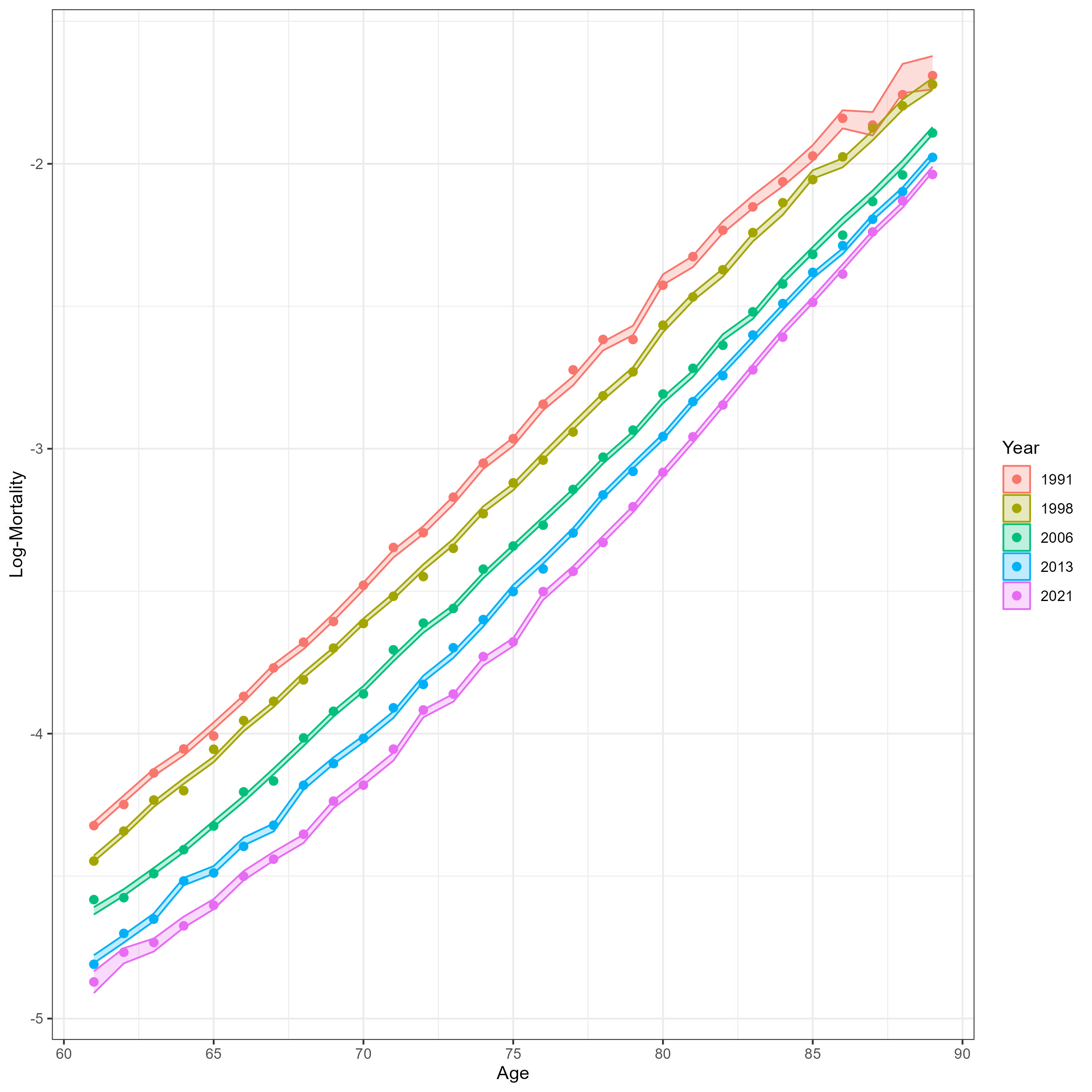} 
            \caption{Taiwan}
        \end{subfigure} &
        \begin{subfigure}[b]{0.3\textwidth}
            \centering
            \includegraphics[width=\textwidth]{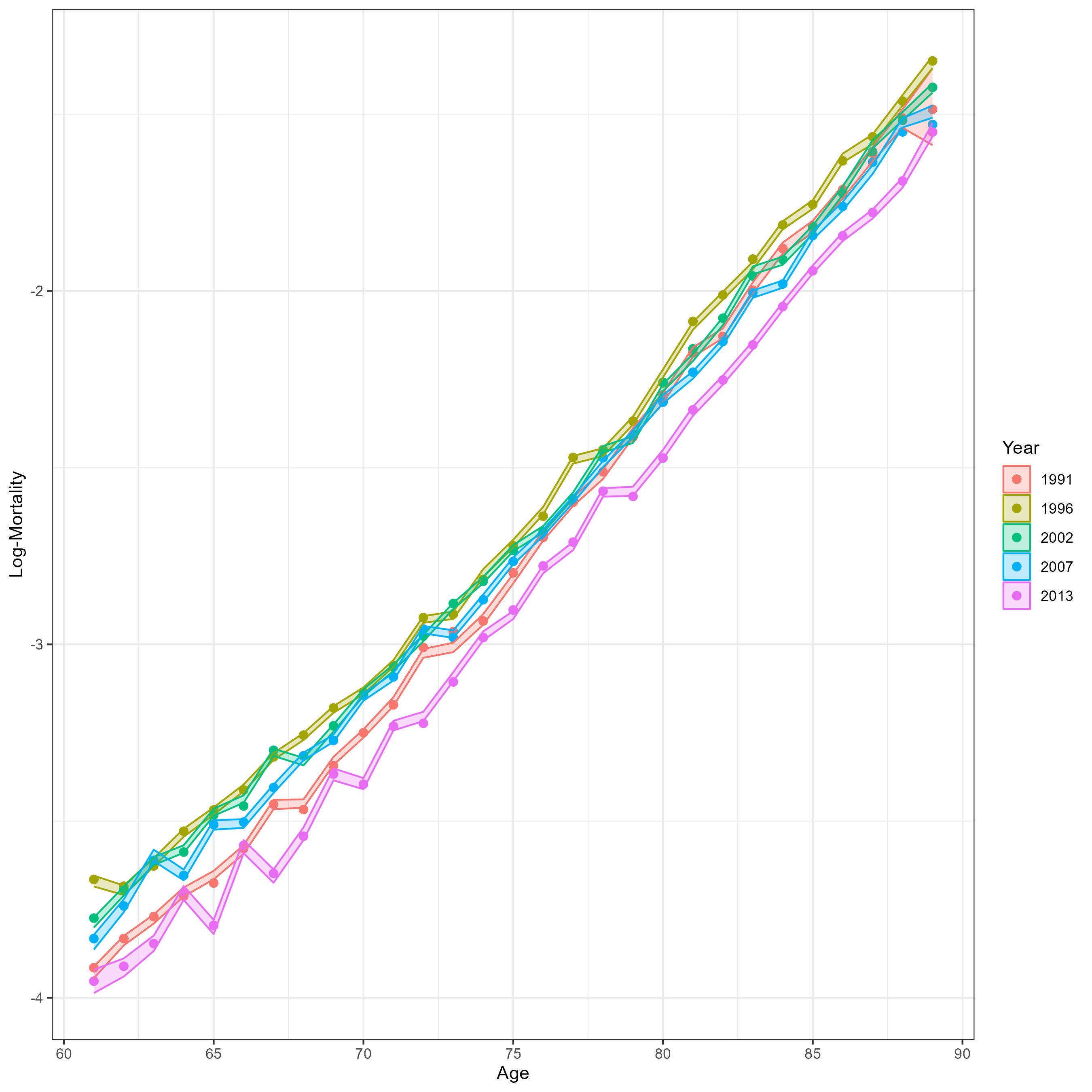} %
            \caption{Ukraine}
        \end{subfigure} &
        \begin{subfigure}[b]{0.3\textwidth}
            \centering
            
        \end{subfigure}
    \end{tabular}
    \caption{HMD data results: 95\% PCI of the log-mortality rates $m_{x,t}$. The dots represent the crude rates.}
\end{figure}

\section{Additional plots for product data}

\section{MCMC of the model on the age-period data based model }

We sample from the posterior distributions of the parameters $(\delta_1, \theta_1, \delta_2, \theta_2, \delta_3, \theta_3, \delta_4, \theta_4)$, where $\theta_i$ are the parameters corresponding to $\delta_i$, using the following Gibbs sampler.

\begin{itemize}

\item Sample $\theta_1$

\begin{itemize}

\item $\delta_1 = 1$ (unconstrained case):

We update each $a_x$ independently with a Metropolis-Hastings sampler, where the proposal distribution is taken to be the Laplace approximation to the posterior distribution.

\item $\delta_1 = 2$ (linear case):

We update jointly $(a,b)$ with a Metropolis-Hastings sampler, where the proposal distribution is taken to be the Laplace approximation to the posterior distribution.

\end{itemize}

\item Sample $\delta_1$

If $\delta_1 = 1$, we choose $q(a,b)$ as the Laplace approximation to the posterior of $(a,b)$.

If $\delta_1 = 2$, we choose $q(a_x)$ as the Laplace approximation to the posterior of $a_x$




\item Sample $\delta_2$


We distinguish four cases

\begin{itemize}
    \item \textbf{$\delta_2 = 1 \rightarrow \delta_2 = 2$}

    Since we also need to choose a value for $\delta_3$, we take for simplicity $\delta_3 = 1$.

    We propose $q(k_t)$ from its Laplace approximation to the posterior and similar for $\bar{b}$.

    \item \textbf{$\delta_2 = 2 \rightarrow \delta_2 = 1$}

    We propose $q(k_t)$ from its Laplace approximation to the posterior.

Note that dimension matching condition is satisfied.

In the following two cases, we need to further distinguish two cases, $\delta_3 = 1$ and $\delta_3 = 2$.

\item \textbf{$\delta_2 = 2 \rightarrow \delta_2 = 3$} in the case $\delta_3 = 1$

We propose as auxiliary variable $\epsilon_t$, $t=1,\dots,T$, such that $k^1_t = k_t - \epsilon_t$ and $k^2_t = k_t + \epsilon_t$, with the constraints $\sum_{t} \epsilon_t = 0$ and $\frac{1}{Y} \sum_{t} \frac{k_t + \epsilon_t}{k_t - \epsilon_t} = \bar{b}$.

    \item \textbf{$\delta_2 = 3 \rightarrow \delta_2 = 2$} in the case $\delta_3 = 1$

    The reverse move is deterministic and it is given by

$$
\begin{cases}
    \bar{b} = \sum_{t} \frac{k_t + \epsilon_t}{k_t - \epsilon_t} \\
    k_t = \frac{\left( k^1_t + \frac{k^2_t}{\bar{b}} \right)}{2} 
\end{cases}
$$

    \item \textbf{ $\delta_2 = 2 \rightarrow \delta_2 = 3$} in the case $\delta_3 = 2$

We note that we can reparametrise the model $f_1(x) + k^2_t (1 + b_x) + f_4(t-x)$ with constraints $\sum {k}^2_t = \sum {b}_x = 0$ to the model $f_1(x) + k^2_t (1 + \bar{b} b_x) + f_4(t-x)$ with the same constraints as before plus the additional constraint $b_1 = -1$.

Once this reparametrisation is given, it can be seen that the same transformation used in the case $\delta_3 = 1$ can be used here.

    \item \textbf{ $\delta_2 = 3 \rightarrow \delta_2 = 2$} in the case $\delta_3 = 2$

This proposal is deterministic.
    
\end{itemize}

\item Sample $\delta_3$

If $\delta_3 = 1$, we choose $q(b_x)$ to be the Laplace approximation to the posterior of $b_x$.


The reverse move is deterministic if $\delta_2 = 3$, while if $\delta_2 = 2$ we propose \bar{b} from the  Laplace approximation to the posterior distribution of $\bar{b}$.

\item Sample $\delta_4$

If $\delta_4 = 1$, we choose $q(g_{t-x})$ to be the Laplace approximation to the posterior of $g_{t-x}$, using the specific constraints relative to each case. 

The reverse move is deterministic

\end{itemize}

\section{MCMC of the model on the age-period-product data based model}

We sample from the posterior distributions of the parameters $(\delta_1, \theta_1, \delta_2, \theta_2, \delta_3, \theta_3, \delta_4, \theta_4)$, where $\theta_i$ are the parameters corresponding to $\delta_i$, using the following Gibbs sampler:

\begin{itemize}
    \item $Sample \theta_1$

    \item $Sample \delta_1$

If $\delta_1 = 1$, we propose $\delta_2$ by proposing $c^1_p$ from its posterior distributions.

If $\delta_1 = 2$, we either propose the move to $\delta_1 = 1$, which is deterministic, or we propose $\delta_2 = $
    
    \item $Sample \theta_2$
    \item $Sample \delta_2$
    \item $Sample \delta_3$
    \item $Sample \delta_3$
    \item $Sample \delta_4$ 
\end{itemize}

\printbibliography